\documentclass[12pt]{article}
\usepackage{amsmath,amssymb,latexsym}
\usepackage[all]{xy}

\def\sqr#1#2{{\vcenter{\vbox{\hrule height.#2pt
            \hbox{\vrule width.#2pt height#1pt \kern#1pt
                  \vrule width.#2pt}\hrule height.#2pt}}}}

\def\square
 {\mathop{\mathchoice{\sqr{12}{15}}{\sqr{9}{12}}{\sqr{6.3}{9}}{\sqr{4.5}{9}}}}

\begin{document}

\hfill hep-th/0502044

\vspace{0.5in}

\begin{center}

{\large\bf String compactifications on Calabi-Yau stacks}

\vspace{0.25in}

Tony Pantev$^1$ and Eric Sharpe$^{2}$ \\
$^1$ Department of Mathematics \\
University of Pennsylvania \\
David Rittenhouse Lab.\\
209 South 33rd Street \\
Philadelphia, PA  19104-6395\\
$^2$ Departments of Physics, Mathematics \\
University of Utah \\
Salt Lake City, UT  84112 \\
{\tt tpantev@math.upenn.edu},
{\tt ersharpe@math.utah.edu} \\

$\,$

\end{center}

In this paper we study string compactifications on Deligne-Mumford stacks.
The basic idea is that all such stacks have presentations to which
one can associate gauged sigma models, where the group gauged need be
neither finite nor effectively-acting.  Such presentations are not unique,
and lead to physically distinct gauged sigma models;
stacks classify universality classes of gauged sigma models,
not gauged sigma models themselves.
We begin by
defining and justifying a notion of ``Calabi-Yau stack,''
recall how one defines sigma models on (presentations of)
stacks, and 
calculate of physical properties of such sigma models,
such as closed and open string spectra.  We describe how
the boundary states in the open string B model on a Calabi-Yau stack
are counted by derived categories of coherent sheaves on the stack.
Along the way, we describe numerous tests that IR physics is
presentation-independent, justifying the claim that stacks classify
universality classes.
String orbifolds are one special case of these compactifications,
a subject which has proven controversial in the past; however
we resolve the objections to this description of which we are aware.
In particular, we discuss the apparent mismatch between
stack moduli and physical moduli, and how that discrepancy is
resolved.

\begin{flushleft}
January 2005
\end{flushleft}

\newpage

\tableofcontents

\newpage

\section{Introduction}

In this paper we work out the basic aspects of the physics of string
compactifications on stacks.  In a nutshell, under mild conditions
\cite{totaro,ehkv},  
every stack has a presentation as a global quotient of a space by some
group, $[X/G]$, where $G$ need not be finite and need not act
effectively (though $G$ will always have finite stabilizers).  
A sigma model on such a presentation is simply a
$G$-gauged sigma model on $X$.  Now, a given stack can have many
presentations of that form, which, as we shall see in examples, often
lead to physically very different gauged sigma models.  Thus, stacks
cannot classify gauged sigma models.  Rather, we shall argue that
stacks classify universality classes, so for the notion of ``string
compactification on a stack'' to be well-defined means that all
presentations of a single stack define gauged sigma models in the same
universality\footnote{ There is a closely related issue in current
descriptions of open string B model states using derived categories
\cite{medc,mikedc,paulalb}.  This description is realized by mapping
representatives of objects by complexes of locally-free sheaves to
D-brane/anti-D-brane combinations, generating off-shell states.
Localization on quasi-isomorphisms is realized via worldsheet
renormalization group flow, which can only be checked indirectly.  In
both stacks and derived categories, we see categories being realized
physically as follows: (only) certain distinguished representatives of
equivalence classes have a physical interpretation, and remaining
equivalences between those representatives are realized physically via
renormalization group flow.  } class of worldsheet RG flow.

Unfortunately, explicitly following renormalization group flow directly is
impossible as a practical matter, making such a statement
impossible to explicitly prove in all cases.
Instead, we shall perform numerous consistency tests.

Now, if we are not careful, it is trivially easy to generate
contradictions in the program above, that would seem to imply that 
something is amiss -- perhaps IR physics does depend upon the choice
of presentation after all.  For example:
\begin{itemize}
\item A mere moment's reflection generates an easy contradiction. 
In a sigma model on an ordinary space,
the massless spectrum of the theory is the cohomology of the target.
If we were to pretend that sigma models on stacks behave exactly as
sigma models on spaces, then we would be led to conclude that
the massless spectrum is just the cohomology of the stack.
However, this is not the
case -- massless spectra are the cohomology of a different stack,
the associated inertia stack, and not the target.
This leads to better-behaved mathematics, but apparently contradicts
the physical assertion.  This problem was resolved in \cite{meqs},
and we review its resolution (along with massless spectra computations
for more general Deligne-Mumford Calabi-Yau stacks) in
section~\ref{closedspectra}.
\item Other problems are not so easy to resolve.  
For example, in \cite{dgm} it was observed that the classical Higgs moduli space
of D-branes on orbifolds is a resolution of the quotient space,
leading many physicists to claim that string orbifolds describe
strings on resolutions of quotient spaces.
This is the single most commonly cited objection to the claim that
string orbifolds are strings propagating on stacks.
There is, however, an alternative interpretation of that Higgs moduli space
calculation,
discussed previously in \cite{tomasme,dks}, and mentioned in \cite{meqs}
as a way to explain this apparent problem.  We review the solution of this
puzzle in this paper in section~\ref{dbranemckay}.
In short, quotient stacks do {\it not} contradict \cite{dgm},
but rather give an alternative understanding of
\cite{dgm}, which is relevant to physical realizations of
various versions of the McKay correspondence.
\item Many quotient stacks do not
have deformations corresponding to CFT marginal operators,
not even infinitesimally.  For example, the standard ${\bf Z}_2$ string orbifold
of ${\bf C}^2$ has physical deformations to sigma models on both deformations
and resolutions of the quotient space ${\bf C}^2/{\bf Z}_2$,
whereas the stack $[{\bf C}^2/{\bf Z}_2]$, by contrast,
is rigid, and admits neither any complex structure nor K\"ahler deformations.
This apparent contradiction was not resolved
in \cite{meqs}, and in fact was emphasized there (and repeatedly since then)
as one of the 
significant obstructions
to this entire program of understanding strings propagating on stacks.  
In section~\ref{defthy}, we finally resolve this issue.
In particular, this problem motivated
a careful study of what the right notion of ``Calabi-Yau'' should
be for stacks, as some partial progress towards resolving this
contradiction can be made if one alters the naive definition.
We speak at length in section~\ref{cystxdefn} on what we believe
is the correct definition of ``Calabi-Yau stack.''
This problem also makes an appearance when studying closed string
massless spectra,
as the massless spectrum defines the infinitesimal physical moduli.
In particular, massless spectra of gauged sigma models are only known
for finite, effectively-acting groups; for noneffectively-acting or
nonfinite groups, the massless spectrum has been an open problem,
and one way to try to partially solve this deformation theory mismatch
in such cases is to play with the massless spectrum.
Because of the connection between this deformation theory issue
and massless spectra, resolving this issue was one of the primary hurdles
in writing this paper.
We shall discuss massless spectra and the problem posed by
deformation theory in section~\ref{closedspectra}.
\end{itemize}
Thus, it is not immediately obvious that the notion of strings propagating
on stacks is necessarily well-defined, {\it i.e.} that IR physics
is necessarily presentation-independent; however, we shall resolve the problems
outlined above in this paper.
Furthermore, we shall extensively test the presentation-independence of
IR physics, here and in \cite{glsm}, by {\it e.g.} showing that
open and closed string massless spectra are always presentation-independent,
that partition functions in examples of noneffective orbifolds are
presentation-independent,
as well as how other properties of gauged sigma models such as fractional branes
are realized between different presentations of the same stack.
In a few examples involving free fields, we can even explicitly, 
completely check
that the field theories defined by distinct presentations are
equivalent in the IR.
See section~\ref{multpres:samecft} for more details,
as well as \cite{glsm} where more tests are performed using gauged linear
sigma models, involving checking that quantum cohomology and Toda duals
are independent of presentation.

One application of this work is to local orbifolds.
Such orbifolds have an old description in terms of ``V-manifolds,''
that is, in terms of local charts each of which is an orbifold.
Unfortunately it has never been clear whether that V-manifold description
could be used to define a CFT.  Existing physical
orbifolds are all global orbifolds,
and the underlying intellectual basis, the standard consistency tests
(modular invariance, multiloop factorization / unitarity),
are only currently understood for global quotients.  
Even if one could make sense physically of a string on
a V-manifold described by a single set of orbifold coordinate charts,
which is itself not obvious,
demonstrating that physics is independent of the choice of 
orbifold coordinate atlas is well beyond the reach of existing technology.
We will side-step these issues by
using the fact that local orbifolds can be re-expressed as global
quotients of (larger) spaces by (bigger) groups.
Such global quotients can be concretely translated into physics,
unlike V-manifolds.  The price we pay is that such re-expressions are not
unique, and different re-expressions can sometimes yield very different
quantum field theories.  We argue in this paper and \cite{glsm} that
universality classes of gauged sigma models are classified by
stacks, which finally gives us a concrete handle on local orbifolds
and their relation to physics.

Another application is to massless spectra of gauged sigma models.
Such spectra are known in effective orbifolds, {\it i.e.} gauged
sigma models in which a finite effectively-acting group has been gauged.
Noneffectively-acting finite groups are a little more subtle,
and were discussed in \cite{nr}.  However, nondiscrete groups,
effectively-acting or otherwise, pose a particular problem,
as typically a $G$-gauged sigma model for $G$ nondiscrete 
is a massive theory, not a CFT,
giving us no direct Lagrangian access to the
conformal fixed point, and hence typically no direct way to
work out the vertex operators describing the massless spectrum.
We work around this issue by finding several indirect ways to
calculate massless spectra in special cases of $G$ nondiscrete, 
which allows us to
(indirectly) check a proposal for massless spectra of IR fixed points of
gauged sigma models for any group $G$.

Curiously, another outcome of the work described here and in
\cite{nr,glsm} are fields valued in roots of unity.
We shall encounter such fields from several independent lines of reasoning,
involving moduli and mirror symmetry for noneffective gaugings.
In particular, such fields give us a completely explicit description
of what seems to be a new class of abstract CFT's described by
Landau-Ginzburg models with superpotentials deformed by fields
valued in roots of unity.

In \cite{glsm} we will also extensively discuss mirror symmetry for stacks,
and will even describe a conjectured generalization of
Batyrev's mirror conjecture to Calabi-Yau stacks,
realized as hypersurfaces in toric stacks.

We begin in section~\ref{briefrev} by giving a brief review
of stacks.  Such a review is almost unnecessary, partly
because we will only ever discuss quotient stack presentations
$[X/G]$, and also on the general principle that mathematically stacks
are the next best thing to spaces, and mathematically can often be
treated formally just like spaces.  Nevertheless, to help build the
reader's confidence in this language, we provide a very short review.

In section~\ref{cystxdefn} we introduce the mathematical definition
of a Calabi-Yau stack, and describe basic facts about their presentations.
In a nutshell, a Calabi-Yau stack is a stack that can be presented
as a quotient of a Calabi-Yau by a group that preserves the holomorphic
top form.  To be careful, however, we also list several checks
that that really is the desired notion of Calabi-Yau stack.

In section~\ref{cystxexs} we discuss numerous examples of Calabi-Yau stacks
and their (multiple) presentations.  Other examples are discussed
in \cite{nr,glsm}.  For purposes of comparison, we also discuss
a non-Calabi-Yau stack, which happens to be closely related to
Calabi-Yau's.

In section~\ref{sigmamodels} we discuss precisely why one associates
a $G$-gauged sigma model to a presentation $[X/G]$:  the gauged sigma model
can be interpreted as a sigma model on $[X/G]$.  This phenomenon was discussed
previously in \cite{meqs} for finite effectively-acting quotients,
but the basic point, that the `maps' one sums over in a path integral
description of a gauged sigma model coincide with the 
`maps' that define the presentation, applies equally well here.
The general yoga is that to define a sigma model on a stack,
one must specify a presentation.  Then, mathematical equivalences 
between presentations are realized physically via worldsheet
renormalization group flow.

Checking that IR physics is presentation-independent cannot be done
directly, as it is not currently possible to explicitly follow
renormalization group flow in all cases, but there are a variety of
independent tests that one can perform.
In section~\ref{sigmamodels} we also study several examples of 
multiple presentations of stacks
and check that IR physics does seem to be presentation-independent.
Further tests of such equivalences are performed later in this paper
and in \cite{glsm}, where we shall see in examples of gauged linear
sigma models describing toric stacks that quantum cohomology and
Toda duals are presentation-independent.

In section~\ref{closedspectra} we discuss closed string massless spectra,
and make a (presentation-independent) conjecture for the general answer.
For presentations as global quotients by finite effectively-acting groups,
the answer is known and agrees with the conjecture, as we shall review.
For presentations as global quotients by finite noneffectively-acting groups,
the matter has not been previously studied in the physics
literature to our knowledge; however,
the calculation is doable, albeit a bit more subtle than in the
effectively-acting case, as was discussed in \cite{nr}.
We review how the answer obtained in \cite{nr} agrees with the conjecture
presented here.
For presentations as global quotients by nondiscrete groups,
it is in principle not possible to directly calculate the massless
spectrum.  Nevertheless, some strong indirect calculations are possible,
such as calculating spectra at associated Landau-Ginzburg orbifold points,
and using quantum cohomology calculations (which can be performed
without knowing the massless spectrum) to reverse-engineer the
massless spectrum, as we shall discuss in \cite{glsm}.
This gives us strong reasons to believe that the conjecture for the
closed string massless spectrum is correct.

In section~\ref{defthy} we discuss one of the most significant obstructions
to this entire program:  the mismatch between physical deformations of
gauged sigma models and mathematical deformations of stacks.
Stacks typically have fewer moduli than the corresponding physical theories.
For example, the string orbifold corresponding to $[ {\bf C}^2/{\bf Z}_2]$
has marginal operators describing deformations to physical theories 
describing sigma models on complex structure deformations and
K\"ahler resolutions of quotient spaces.  The stack $[ {\bf C}^2/{\bf Z}_2]$,
by contrast, is rigid, and admits neither complex structure deformations
nor any K\"ahler resolutions to Calabi-Yau's.
In noneffective orbifolds, according to the results of the massless
spectrum calculation above, there are physical moduli that do not
have any obvious correspondence at all to any geometric moduli.
Such difficulties suggest that there may be some fundamental problem
with this program -- perhaps IR physics is presentation-dependent after
all.  We resolve these difficulties in section~\ref{defthy}.

In section~\ref{Dbranegen} we collect some useful general facts
about D-branes on stacks.  One direction involves D-branes in noneffective
orbifolds.  Specifically, even if a group acts trivially on the space,
it is nevertheless consistent with the Cardy condition for the group
to act non-trivially on the Chan-Paton factors.  Mathematically,
the resulting D-branes can be understood as sheaves on gerbes,
which are the same as (twisted) sheaves on the underlying space.
Another matter discussed in section~\ref{Dbranegen} is the 
often-quoted objection to describing gauged sigma models with stacks,
that D-brane probes see resolutions of quotient spaces, and so
gauged sigma models should describe strings on resolutions, not
some sort of stack.  We speak to this objection in section~\ref{Dbranegen}.

Finally, in section~\ref{openB}, we discuss the open string B model on
stacks, and how the technology developed to understand the physical
relevance of derived categories when the target is a space
can be
adapted to target stacks.

In an appendix we describe circumstances under which the total
space of a principal ${\bf C}^{\times}$ bundle is a Calabi-Yau manifold,
which turns out to be rather useful.

In the sequel \cite{glsm} we discuss gauged linear sigma models for
toric stacks, and closed string A model calculations in that context.
We describe more tests of presentation-independence, as well as quantum
cohomology calculations and tests of the closed string massless spectrum
conjecture.  We also describe mirror symmetry for stacks in detail there.
We will see that physical fields valued in roots of unity and the
resulting new abstract CFT's defined by such fields play an important
role in understanding mirror symmetry for stacks.

\section{Brief review of stacks}   \label{briefrev}

In this section, we shall give an extremely brief overview
of some relevant properties of stacks.

At one level, this review is almost unnecessary, on the grounds
that stacks can be treated almost the same as spaces.
One can do differential geometry on stacks; one can define
tangent spaces, normal bundles, differential forms, and metrics
on stacks, all of which have properties essentially identical to
those defined on ordinary spaces.  Thus, one approach the reader
can take to this paper is to read ``space'' whenever he or she
reads the word ``stack.''

Furthermore, for the purposes of this paper and \cite{nr,glsm},
we will use the fact that all\footnote{Note first that by `stack' we typically
mean a smooth complex algebraic Deligne-Mumford stack.
Now, there are mild conditions for such stacks to have presentations as global
quotients -- strictly speaking, one requires that the underlying space
(historically known as the `moduli space' of the stack,
despite the fact that there is no moduli problem here, because of the historical
roots of stacks in moduli problems) be quasi-projective, and then
the result follows from a universally-believed conjecture of Grothendieck
for which a proof has recently been announced.  Strictly speaking,
there do exist smooth Deligne-Mumford stacks which do not admit presentations
as global quotients, but these stacks live over exotic objects such as
non-separated schemes, and so are irrelevant for our purposes in this paper.} 
stacks have presentations as
quotient stacks $[X/G]$, so we do not require the general machinery
of stacks.  

For completeness, however, let us very briefly review some basic
aspects of the story.

Technically, we generalize spaces to stacks by first defining spaces
in terms of their categories of incoming maps, and then slightly
generalizing the allowed categories.  A suitably well-behaved
category of incoming maps, that does not describe a space, describes
a stack.  That may seem at first an obscure approach,
but a moment's reflection upon how sigma models are defined physically
should make it clear that this approach has utility in physics.

This language is in many ways closely analogous to the typical
approach to noncommutative geometry.  There, one defines noncommutative
spaces by first defining spaces in terms of an algebra of functions.
Then, by allowing more general algebras that do not come from
ordinary spaces, one creates a notion of a noncommutative space.

The stacks we shall consider in this paper all have the property
of possessing an ``atlas,'' which is an ordinary space together with
a well-behaved map into the stack. 

As a result of possessing an atlas, 
we can define functions, bundles, differential forms, metrics,
{\it etc} on the stacks appearing in this paper in two completely
equivalent ways:
\begin{enumerate}
\item First, we can formally define such objects in terms of their
pullbacks to all other spaces.  If we know what all possible pullbacks
look like, then we can reproduce the original object.
This definition is very useful for defining sigma models
on stacks \cite{meqs}.
\item Second, instead of describing all possible pullbacks,
we can equivalently just describe the function, bundle, differential
form, {\it etc} on the atlas.  For example, if ${\cal X}$ is a stack
with atlas $X$, then a function $f$ on the stack ${\cal X}$
is a function $\tilde{f}$ on the space $X$ such that  
$p_1^* \tilde{f} = p_2^* \tilde{f}$,
where $p_{1,2}: X \times_{ {\cal X} } X \rightarrow X$
are the two projection maps, and (by definition of atlas) $X \times_{ {\cal X}
} X$ is a space, not a stack.  This is a generalization of the
notion of $G$-invariance in orbifolds, as we shall see explicitly below.
Differential forms, bundles, {\it etc}
are defined similarly.
We shall see that this definition is very useful for
generalizing the ansatz of \cite{dougmoore} to compute
open string spectra on stacks.
\end{enumerate}
We shall not attempt to give a detailed treatment of why these two
definitions are equivalent, though at least one direction should
be approximately clear.  Since there is a map from the atlas into
the stack, if we know all possible pullbacks, then we necessarily
know an object on the atlas.

Since we will be using the second description extensively in
this paper, let us illustrate how a few more objects are defined
on atlases:
\begin{itemize}
\item A metric on the stack ${\cal X}$ is a metric $g_{\mu \nu}$
on the space $X$, subject to the constraint that $p_1^* g = p_2^* g$,
where $p_{1,2}: X \times_{ {\cal X} } X \rightarrow X$ are the
projection maps.
\item A differential form on the stack ${\cal X}$ is a differential
form $\omega$ on the space $X$, such that $p_1^* \omega = p_2^* \omega$.
\item A holomorphic nowhere-zero top form on the stack ${\cal X}$
is a holomorphic nowhere-zero top form $\omega$ on the space $X$ such that
$p_1^* \omega = p_2^* \omega$.
\item A bundle on the stack ${\cal X}$ is a bundle $E$ on the space
$X$, together with isomorphisms
\begin{displaymath}
\psi: \: p_1^* E \: \stackrel{\sim}{\longrightarrow} \: p_2^* E
\end{displaymath}
obeying the constraint that
\begin{displaymath}
p_{13}^* \psi \: = \: \left( p_{23}^* \psi \right) \circ 
\left( p_{12}^* \psi \right)
\end{displaymath}
where $p_{ij}$ are the projections $X \times_{ {\cal X} }
X \times_{ {\cal X} } X \rightarrow X \times_{ {\cal X} } X$. 
\end{itemize}

Let us try to illustrate this technology in three basic examples.
\begin{enumerate}
\item First, consider the special case that the stack is an ordinary
space.  (Spaces are special cases of stacks, after all).
Denote the space by $X$.  We can take the atlas on $X$ to be
$X$ itself, and the map from the atlas into the stack is given by
the identity map.  In this case, $X \times_X X = X$, and the
projection maps are the identity map, so in terms of the definition
above, the constraint $p_1^* f = p_2^* f$ is trivial,
so a function on the `stack' $X$ is the same as a function
on the atlas, which is also $X$.
\item The next example is a quotient stack $[X/G]$, where $G$ is a finite
group acting on $X$ by diffeomorphisms.  Maps $Y \rightarrow [X/G]$ into
the quotient stack are defined by\footnote{As algebraic stacks,
we have a holomorphic bundle and a holomorphic map.
As differentiable stacks, we would have a smooth bundle with connection
and a smooth map.} pairs $\left( E \rightarrow Y,
E \rightarrow X\right)$, where $E \rightarrow Y$ is a principal $G$-bundle
over $Y$, and $E \rightarrow X$ is a $G$-equivariant map from the total
space of the bundle into $X$.  
That description might sound obscure at first, but a moment's reflection
reveals that physicists have seen such a description before -- this data
is equivalent to a twisted sector map!  Summing over twisted sector maps
in string orbifolds is equivalent to summing over maps into a quotient
stack.
An atlas for $[X/G]$ is given by $X$,
and there is a canonical map $X \rightarrow [X/G]$ defined by the
trivial principal $G$-bundle $X \times G \rightarrow X$,
together with the (canonically $G$-equivariant) evaluation map
$X \times G \rightarrow X$.
It is straightforward to check, using the
definition given above, that functions on $[X/G]$ are the 
same as $G$-invariant functions on the
atlas $X$, and similarly for bundles, differential forms, metrics, 
{\it etc}.  For example, a function $f$ on $X$ defines a function
on $[X/G]$ if, as mentioned earlier, $p_1^* f = p_2^* f$,
where $p_{1,2}: X \times_{ [X/G] } X \rightarrow X$ are the
projection maps.  
Now, it can be shown that $X \times_{[X/G]} X =  X \times G$,
and the two projection maps are the projection onto the first factor and the
evaluation map, respectively.  
Hence, the constraint $p_1^* f = p_2^* f$ is the
same as demanding that $f$ be $G$-invariant. 
\item The third example we shall consider is that of a gerbe on
a space $X$.  Gerbes are to $B$ fields what principal bundles are 
to gauge fields -- gerbes play a role analogous to the total space
of a bundle.  The canonical trivial $G$-gerbe on a space $X$ is
the quotient stack $[X/G]$ where $G$ acts trivially on $X$.
In particular, even when $G$ acts trivially, the stack $[X/G]$ is 
still not the same as the space $X$, since maps into $[X/G]$ are
partially indexed by principal $G$-bundles.  (Physically, trivial
group actions will yield distinct physics because of nonperturbative effects,
due ultimately to those same principal $G$-bundles.)
More general $G$-gerbes still
look locally like the quotient stack $[X/G]$, just as general
vector bundles look locally like a trivial vector bundle.
The most important feature of gerbes for this discussion is that
sheaves on gerbes are equivalent to twisted sheaves on the underlying space,
twisted in the sense of \cite{cks}.
In particular, the stack $[\mbox{point}/G]$ is denoted $BG$, and the
trivial $G$-gerbe $[X/G]$, where $G$ acts trivially on $X$,
is the same stack as $X \times BG$ -- the stack $BG$ is analogous to the
fiber of a vector bundle.
\end{enumerate}

A description of a stack in terms of an atlas with relations is known
as a presentation.  Every stack has a presentation of the form of a 
global quotient stack $[X/G]$, for some $X$ and some $G$.
However, presentations are not unique -- a given stack can have many
different presentations of the form $[X/G]$.
For example, any space $X$ is the same as the stack \linebreak 
$[X \times {\bf C}^{\times} / {\bf C}^{\times}]$ where the ${\bf C}^{\times}$
acts only, effectively, on the ${\bf C}^{\times}$ factor, which amounts to the
physics statement that if I gauge $U(1)$ rotations of a sigma model
on $X \times S^1$, then the resulting gauged sigma model is in the
same universality class as $X$.

Physically, we will associate a $G$-gauged sigma model on $X$
to a presentation $[X/G]$, and then we shall argue that stacks
classify universality classes, which gives a very precise meaning
to standard intuitions about relations between different gauged sigma
models.  (For example, the statement that the A-twist of the
nonlinear sigma model on
${\bf P}^{N-1}$ and the A-twist of a $U(1)$ gauge theory with $N$ chiral
superfields lie in the same universality class is just a special case
of our claim regarding stacks and universality classes.)

In the course of this paper as well as \cite{nr,glsm}, we shall see
many more examples of stacks and (multiple) presentations.
See \cite{vistoli,gomez} for more information.

\section{Calabi-Yau stacks}    \label{cystxdefn}

\subsection{Basic definition}

Under what circumstances is a stack Calabi-Yau?
We shall define a stack to be Calabi-Yau when its canonical
bundle is trivial.

On the one hand, this definition seems like an obvious
extension of the definition of Calabi-Yau for spaces.
Since a space is a special kind of stack, our definition of
Calabi-Yau stack must specialize to the usual notion of Calabi-Yau
for spaces.

However, the correctness of this definition is slightly clouded
by deformation theory arguments.  We shall see explicitly that
a string orbifold is just a sigma model on a stack of the form
$[X/G]$ where $G$ is finite and acts effectively.
Thus, one would expect marginal operators in the string orbifold
CFT to correspond to deformations of the stack $[X/G]$.
Unfortunately, as we explain in more detail in section~\ref{defthy},
stacks do not always have enough complex structure deformations to match
CFT marginal operators.  Also, if one tries to blow up a stack,
then typical attempts give stacks which have nontrivial
canonical bundle (essentially for the same reason that blowing up
a smooth point of a space makes the canonical bundle nontrivial).  
(See section~\ref{ex:nonCY} for such an attempted
blowup of $[{\bf C}^2/{\bf Z}_2]$.)
If a ``Calabi-Yau stack'' somehow did {\it not} necessarily
have trivial canonical bundle, then perhaps we could begin to
match up some of the CFT marginal operators.  Yet on the
other hand, in section~\ref{ex:nonCYphysics} we study sigma models
on such blowups with nontrivial canonical bundle, and see that
they are not consistent with supersymmetry.  We will 
argue in section~\ref{defthy}
that we can can resolve this deformation theory puzzle without having
to demand that stacks with nontrivial canonical bundle
be Calabi-Yau.

In any event, as the preceding discussion illustrates, some
nontrivial checks of our definition are called for.

In a few pages we shall see explicitly that our definition
of Calabi-Yau stack does include all the obvious special cases,
namely, Calabi-Yau spaces, and orbifolds of Calabi-Yau spaces
by finite effectively-acting groups that preserve the holomorphic
top form.

In fact, for stacks presented as $[X/G]$, with $G$ an affine algebraic
group, we will shortly argue that our definition of ``Calabi-Yau,'' in
terms of triviality of the canonical bundle, holds if and only if $X$
is Calabi-Yau and $G$ preserves the holomorphic top form, which is
precisely the condition one would expect in a gauged sigma model.

Another test of our definition comes from worldsheet beta functions.
In sigma models on ordinary manifolds, after all, the reason for the
Ricci-flatness condition is that it makes worldsheet beta functions
vanish.  In the next section, we shall associate gauged sigma models
to presentations of stacks of the form $[X/G]$.  In such a gauged sigma
model, locally in patches on $X$ we can write $X = Y \times G$,
and gauge fix down to $Y$.  Vanishing of beta functions in such
gauge-fixed patches gives a transverse Calabi-Yau condition,
that $X$ must be Calabi-Yau perpendicular to the action of $G$.
This is not quite a sufficient condition for triviality of the canonical
bundle of the stack, as we shall see in section~\ref{transverseCY}, 
but is a necessary
condition.  (To get a sufficient condition, we need global information
on the $G$ action.)  In any event, although this is only a necessary
condition, it does help justify our definition.

Another check of this definition appears in section~\ref{serre},
concerning open string spectra.  We argue in section~\ref{openB}
that open string spectra in the B model are counted by
Ext groups on stacks.  In order for the resulting spectra to
be well-behaved under Serre duality (a property
of all known spectrum calculations) the canonical bundle must be trivial.

Another check of our definition will appear in section~\ref{ex:nonCYphysics},
where we consider the physics of a non-Calabi-Yau stack,
in fact the stack corresponding to the obvious notion of
blowup of $[ {\bf C}^2/{\bf Z}_2 ]$.  We will argue that the physics
of that stack does not appear to be consistent with what we would
like to call ``Calabi-Yau.''

Further checks of this definition of Calabi-Yau for stacks appear
in examples throughout this paper and \cite{glsm}.
For example, in \cite{glsm} we shall see that this notion of Calabi-Yau
leads to superpotentials in gauged linear sigma models with the behavior
one would expect for a Calabi-Yau.

Our potential objections to this notion of ``Calabi-Yau'' revolved
around reconciling mathematical deformation theory with physical
deformation theory.  In section~\ref{defthy} we will resolve
this discrepancy in a fashion that does not involve using an
alternative notion of Calabi-Yau.

\subsection{Presentations as global quotients}

A given stack can be presented in many different forms.  However,
under mild conditions, smooth Deligne-Mumford stacks can be described
in the form of a global quotient $[X/G]$ for some space $X$ and some
(typically non-finite) group $G$.  A typical sufficient condition for
this is that the stack in question has enough vector bundles,
i.e. that every coherent sheaf on the stack can be resolved by vector
bundles \cite{totaro}. This is always the case for smooth orbifolds
with quasi-projective moduli spaces. Another sufficient condition is
that the moduli space of the stack is quasi-projective and of
dimension for which the Brauer map is known to be surjective
\cite{ehkv}, \cite{kv}. From now on we will tacitly assume that all
the stacks we are dealing with satisfy one of these conditions.  The
reader should be careful to notice we are {\it not} assuming that all
smooth Deligne-Mumford stacks are orbifolds in the sense traditionally
used in the physics literature.  An orbifold, as the term is
traditionally used, refers to a quotient in which $G$ is finite and
acts effectively.  Here, $G$ need neither be finite nor act
effectively.  For technical reasons, however, we will restrict the types
of $G$ apperaing in the constructions and we will either work with $G$
reductive algebraic or compact Lie; for $[X/G]$ with $G$ one of the
above, there exist $X'$ and $G'$ of the other type such that $[X/G] =
[X'/G']$.

There are more possible presentations
of a given stack than just as global quotients,
as we outline in the next section; 
however, it is presentations as global quotients
that seem most useful for describing\footnote{Something analogous
happens when describing open strings with derived categories.
The objects in a derived category can typically be represented
as complexes of many different forms.  However, for states which
do not have an 
on-shell description, the cleanest dictionary
between off-shell open string B model states and objects in the derived
category is obtained by picking a representative of a given object
as a complex of locally-free sheaves.  Perhaps later people will
better understand sigma models on stacks with presentations not of the
form above; but for the moment, we will only consider presentations of
this more restrictive form, which are already adequate to describe
all Deligne-Mumford stacks.} string sigma models. 
In particular, to define a sigma model on a stack requires that
one specify a presentation of the stack, and only for presentations
of the form $[X/G]$ for $G$ a compact Lie group do we understand
how to define sigma models.  (See section~\ref{sigmamodels} for
more details.)  For this reason, we shall concentrate
on presentations of the form of global quotients in this paper.
It would be extremely interesting to understand how to define
sigma models using presentations of other forms; we outline
the implications of the Calabi-Yau condition above for alternative
presentations in section~\ref{transverseCY}.

Given a presentation of the form $[X/G]$, with $G$ an affine algebraic
group, it is instructive to note that the corresponding stack has
trivial canonical bundle (and hence is a Calabi-Yau stack) if and only
if $X$ is a Calabi-Yau space, and $G$ preserves the holomorphic
top form. Indeed, the natural projection $X \to [X/G]$ realizes $X$ as
a principal $G$ bundle over $[X/G]$. Since $G$ is parallelizable, the
relative canonical class of this principal bundle is trivial and so if
$[X/G]$ has a trivial canonical class, then so does $X$. The converse
is obvious since a $G$-equivariant nowhere vanishing top form
descends to $[X/G]$ by definition.

\subsection{Alternate presentations and transverse Calabi-Yau atlases}
\label{transverseCY}

Although we assume that our algebraic stacks have at least one presentation of
the form $[X/G]$, such stacks typically also have presentations in
different forms.  In principle, a map into an alternate presentation
can be described in terms of maps into an atlas subject to constraints.
However, for general presentations, we do not understand how to realize
such constraints.  For presentations as global quotients, the
relevant constraints can be realized in terms of introducing a 
gauge field (and then Faddeev-Popov gauge fixing, and so forth).
We do not understand the physical analogue of this procedure for
presentations of alternate forms.

Thus, understanding the physical relevance of presentations of stacks
not in the form of global quotients will have to wait for future work.
However, although the mathematics is not of central importance to this
paper, we shall briefly review some alternative presentations.

One particular case of an alternate presentation has atlases of the
form of transverse Calabi-Yau's. Transverse Calabi-Yau structures have
been around for quite some time in foliation theory but were recently
brought center stage by Hitchin \cite{hitchin-transverse} because of
their relation to generalized complex geometry and the branes in
$(2,2)$ sigma models. Informally, a transverse Calabi-Yau structure on
a manifold $X$ is a foliation $\mathcal{F}$ on $X$ which is equipped
with a transverse complex structure which has a trivial canonical
class. To spell this out, recall that instead of thinking of a
Calabi-Yau manifold $X$ as a complex manifold with a trivial canonical
bundle, we can alternatively say that $X$ is a real manifold of
dimension $2n$, equipped with a complex locally decomposable $n$-form
$\Omega = \theta_{1}\wedge \ldots \wedge \theta_{n}$ such that
$\Omega\wedge \overline{\Omega}$ is nowhere zero and $d\Omega =
0$. Indeed, given such a form $\Omega$, we can declare that a complex
vector field $\xi$ is of type $(0,1)$ if and only if $i_{\xi}\Omega =
0$. The integrability condition $d\Omega = 0$ then implies that all
such vector fields are closed under the Lie bracket and so by the
Newlander-Nirenberg theorem we get a complex structure on $X$ for
which $\Omega$ is a holomorphic volume form. Guided by this
interpretation of the Calabi-Yau condition Hitchin defines a {\em
transverse Calabi-Yau structure} on a manifold $X$ to be a complex
locally decomposable $n$-form 
$\Omega = \theta_{1}\wedge \ldots \wedge \theta_{n}$ such that
$\Omega\wedge \overline{\Omega}$ is nowhere zero and $d\Omega =
0$. The difference is that now $\dim X$ is only greater than or equal
to $2n$. The easiest example of a transverse Calabi-Yau is the product
of a Calabi-Yau with any manifold. More generally, we can take any
Calabi-Yau manifold $Y$ of complex dimenion $n$ and any manifold $X$
equipped with a differentiable map $X \to Y$. Then the pullback of the
holomorphic volume form on $Y$ is a transverse Calabi-Yau structure on
$X$. 

Note that if $X$ is equipped with a transverse Calabi-Yau structure
$\Omega$, then the distribution 
$\mathcal{F} = \{ \xi \in TX | i_{\xi}\Omega = 0 \}$, is automatically
integrable and so gives a foliation on $X$ of  codimension $2n$. The
stack $[X/\mathcal{F}]$ is naturally a Calabi-Yau stack in the
sense we explained before. In fact there is a transverse version of
Yau's theorem which asserts that for a compact $X$ which is
transversally Calabi-Yau and transversally K\"{a}hler, one can find a
transverse K\"{a}hler metric which is Ricci flat \cite{kacimi}. This
suggests that stacks obtained from transverse Calabi-Yau structures are
interesting candidates for targets of supersymmetric sigma models.

We can build simple examples of Calabi-Yau stacks (and spaces) with
presentations involving transverse Calabi-Yau atlases as follows.  Let
$X$ be any projective Calabi-Yau manifold and $L$ a line bundle over
$X$ which is not holomorphically trivial.  Let $Y$ be the total space
of $L$.  Then $Y$ is not a Calabi-Yau, but it is a transverse
Calabi-Yau.  More generally, for any Calabi-Yau space $X$ and any
space $Y$ mapping to $X$, the groupoid with atlas $Y$ and relations $Y
\times_X Y$ is a transverse CY.  More complicated examples can be
obtained by orbifolding K3 surfaces but we will not discuss these
here. In fact we shall not consider transverse Calabi-Yau's further in
this paper but we would like to keep them in mind.

\section{Examples of Calabi-Yau stacks}  \label{cystxexs}

In this section, we shall describe several classes of examples
of Calabi-Yau stacks, to which we shall repeatedly refer
throughout the rest of this paper.

\subsection{Ordinary Calabi-Yau spaces}

As spaces are special cases of stacks, and from our definition of
Calabi-Yau stack, it should be clear that Calabi-Yau spaces
are examples of Calabi-Yau stacks.

\subsection{Global orbifolds (and multiple presentations)}  \label{multpres}

Global orbifolds of Calabi-Yau spaces by finite groups
that preserve the holomorphic top form certainly define
examples of Calabi-Yau stacks.

One of the easiest examples of a nontrivial stack is the
quotient stack $[{\bf C}^2/{\bf Z}_2]$, where the ${\bf Z}_2$ acts
on the ${\bf C}^2$ by inverting the signs of the coordinates.
The corresponding orbifold is supersymmetric, and since the ${\bf Z}_2$
preserves the holomorphic form $dx \wedge dy$ on ${\bf C}^2$,
this stack is a Calabi-Yau stack.

Now, as described earlier, stacks can have many different presentations,
even multiple presentations of the form $[X/G]$.  Physically these
different presentations should yield sigma models in the same universality
class, hence defining the same CFT, which we shall explicitly check
later in examples.

In particular, let us describe an alternative presentation of
the stack $[{\bf C}^2/{\bf Z}_2]$, also of the form $[X/G]$ for 
some $G$.  (In the next section we will describe
local orbifolds using the same ideas we use here to define alternative
presentations of global orbifolds.)

First, define the three-fold
\begin{displaymath}
X \: = \: \frac{ {\bf C}^2 \times {\bf C}^{\times} }{ {\bf Z}_2 }
\end{displaymath}
where the generator of ${\bf Z}_2$ acts by sending
\begin{displaymath}
\left( x, y, t \right) \: \mapsto \: \left( - x, - y, 
- t \right)
\end{displaymath}
Since the ${\bf Z}_2$ acts freely, $X$ is smooth.
Also note that $X$ is Calabi-Yau.  That statement might surprise readers
at first, since the quotient $[ {\bf C}^3 / {\bf Z}_2 ]$ is not Calabi-Yau.
Here, however, instead of ${\bf C}^3$, we have ${\bf C}^2 \times 
{\bf C}^{\times}$, with holomorphic top form given by
\begin{displaymath}
dx \wedge dy \wedge d \log t
\end{displaymath}
Because the last factor is $d \log t$, instead of $dt$, this holomorphic form
is invariant under the ${\bf Z}_2$, and so $X$ is a smooth Calabi-Yau manifold.

Now, let ${\bf C}^{\times}$ act on $X$ as follows.
A given $\lambda \in {\bf C}^{\times}$ maps
\begin{displaymath}
\lambda: \: \left( x, y, t \right) \: \mapsto \:
\left( x, y, \lambda t \right)
\end{displaymath}
Again, because the holomorphic top form contains a factor $d \log t$ rather
than $dt$, this ${\bf C}^{\times}$ action preserves the
holomorphic top form, and so, since $X$ is a Calabi-Yau,
the stack $[X / {\bf C}^{\times}]$ is a Calabi-Yau stack.
Also note that this ${\bf C}^{\times}$ does not quite act freely -- 
the points $[0,0,t] \in X$ have nontrivial stabilizer, given by
$\pm 1 \in {\bf C}^{\times}$.

Finally, one can show that
\begin{displaymath}
[ {\bf C}^2 / {\bf Z}_2 ] \: \cong \: [ X / {\bf C}^{\times} ]
\end{displaymath}
for the $X$ and ${\bf C}^{\times}$ action above.
To see this explicitly, define $Y \subset X$ to be the image of
${\bf C}^2 \hookrightarrow X$ under the inclusion map
\begin{displaymath}
\left(x, y \right) \: \mapsto \: [ x, y, 1 ]
\end{displaymath}
If the ${\bf C}^{\times}$ acted freely, then the image of $Y$ would
simply be translated through $X$, and the quotient $[X/{\bf C}^{\times}]$
would be isomorphic to $Y$.  However, the ${\bf C}^{\times}$ does not
quite act freely -- the points  $[0,0,t] \in X$
have nontrivial stabilizer in ${\bf C}^{\times}$, given by
$\pm 1 \in {\bf C}^{\times}$, 
and so $[X/{\bf C}^{\times}] \cong [Y/{\bf Z}_2]$.

This alternative description of $[{\bf C}^2/{\bf Z}_2]$ can be
easily generalized to other orbifolds.  For example,
we can write $[ {\bf C}^2/{\bf Z}_n] = [X/{\bf C}^{\times}]$
where
\begin{displaymath}
X \: = \: \frac{ {\bf C}^2 \times {\bf C}^{\times} }{ {\bf Z}_n }
\end{displaymath}
where the generator of ${\bf Z}_n$ acts on ${\bf C}^2 \times
{\bf C}^{\times}$ as
\begin{displaymath}
\left(x, y, t \right) \: \mapsto \:
\left( \alpha x, \alpha^{-1} y, \alpha t \right)
\end{displaymath}
for $\alpha = \exp(2 \pi i / n)$, and ${\bf C}^{\times}$ acts
on $X$ by rotating the ${\bf C}^{\times}$ factor, just as for
$[ {\bf C}^2 / {\bf Z}_2 ]$.

We shall compare the physics of the orbifold with the $U(1)$ gauged
sigma model of the example above in section~\ref{ex:finvscont}.

\subsection{Local orbifolds}    \label{locorb:mathexs}

Let us discuss some examples of local orbifolds explicitly.
For our first example, start with the quotient space $T^4/{\bf Z}_2$,
and resolve a few singularities.  The resulting singular space
still has some quotient singularities, but can no longer be described
as a global quotient by a finite group.

The stack we will describe next looks like the space above,
except that each of the local quotient singularities has been
replaced by a stack structure.  The specific construction will
be closely analogous to the description of $[{\bf C}^2/{\bf Z}_2]$
in the form $[X/{\bf C}^{\times}]$, for $X = \left(
{\bf C}^2 \times {\bf C}^{\times} \right) / {\bf Z}_2$.

Begin by defining $S$ to be the blowup of the cover $T^4$ at the preimages
of those points which will be resolved.  Note that we can still extend
the ${\bf Z}_2$ action on $T^4$ across the exceptional divisors of the blowups,
and then take the global quotient stack $[S/{\bf Z}_2]$, but unfortunately,
in addition to the stacky structures over the quotient singularities,
this stack also has ${\bf Z}_2$-gerbes over the exceptional divisors,
which we do not desire, and also is not Calabi-Yau.
The space $S$ will be useful momentarily, but we shall have to
work harder.

Define $L = {\cal O}_S\left( - \sum E_i \right)$,
{\it i.e.} the line bundle on $S$ whose divisor is minus the sum
of the exceptional divisors of the blowups of $T^4$.
$L$ possesses a canonical ${\bf Z}_2$-equivariant structure, with respect to
the ${\bf Z}_2$ obtained by extending the action on $T^4$ across the
exceptional divisors.  
(Put another way, since $L$ is an integral multiple of the canonical
bundle $K$, and $K$ has a canonical ${\bf Z}_2$-equivariant structure,
so does $L$.)

Define $L^{\times}$ to be the total space of the principal ${\bf C}^{\times}$
bundle associated to $L$, or equivalently the space obtained from the total
space of $L$ by omitting the zero section.  
The only fixed points of the 
${\bf Z}_2$ action lie within the zero section of $L$,
so ${\bf Z}_2$ acts freely on $L^{\times}$.

Define $Y = L^{\times}/{\bf Z}_2$.  Since the ${\bf Z}_2$ only fixes the
zero section of $L$, $Y$ is a smooth space.

Finally, define ${\cal Y} = [ Y / {\bf C}^{\times} ]$,
where the ${\bf C}^{\times}$ action rotates the fibers of $L^{\times}$ once.

The stack ${\cal Y}$ is the desired local orbifold,
describing quotient stack structures locally over the singularities
but not imposing any sort of extra gerbe structures over exceptional divisors.

As a check, note that if we did not blow up any points at all,
then $S = T^4$, $L = T^4 \times {\bf C}$,
$Y = \left( T^4 \times {\bf C}^{\times} \right) / {\bf Z}_2$,
and the stack ${\cal Y} = [ Y / {\bf C}^{\times}]$.
From our previous discussion of how $[ {\bf C}^2 / {\bf Z}_2 ]$
can be alternately presented as $[ X/{\bf C}^{\times}]$
for $X = \left( {\bf C}^2 \times {\bf C}^{\times} \right)/{\bf Z}_2$,
it should be clear that ${\cal Y} = [ T^4/{\bf Z}_2]$, exactly
as should be in the case no points are resolved.

Also note that this stack is a Calabi-Yau stack.
First, note that since $L$ is an integral multiple of the canonical
bundle of the blowup of $T^4$, the total space of $L^{\times}$
is a Calabi-Yau manifold (see appendix~\ref{ppalcxCY} for details).
Since the holomorphic top form on $L^{\times}$ is of the form
$\omega \wedge d \log t$ where $\omega$ is a holomorphic top form
on $T^4$, and the ${\bf Z}_2$ preserves $\omega$ and merely translates
$\log t$, preserving $d \log t$,
we see that the ${\bf Z}_2$ preserves the holomorphic top form
on $L^{\times}$, hence, $Y = L^{\times} / {\bf Z}_2$ is also a Calabi-Yau
manifold.  Finally, since the ${\bf C}^{\times}$ action
on the Calabi-Yau $Y$ also preserves the holomorphic top form, 
as it just translates
the image of $\log t$, leaving the image of $d \log t$ invariant,
we see that the stack ${\cal Y} = [Y/{\bf C}^{\times}]$ is a 
Calabi-Yau stack.

The same ideas apply to other local orbifolds obtained from
global orbifolds by blowing up a few points.
Consider, for example, the local orbifold obtained by starting with
the global quotient $T^6/{\bf Z}_3$, blowing up some of the quotient 
singularities, and replacing the remaining quotient singularities
with stacky points.

As before, define $S$ to be the blowup of points on $T^6$ that 
are the preimages of the points on $T^6/{\bf Z}_3$ to be blown up.

Define $L$ to be the total space of the line bundle ${\cal O}_S\left(
- \sum E_i \right)$, where the $E_i$ are the exceptional divisors,
and let $L^{\times}$ be the total space of the corresponding principal
${\bf C}^{\times}$ bundle.

Finally, define $Y = L^{\times} / {\bf Z}_3$, and
${\cal Y} = [ Y / {\bf C}^{\times} ]$.
This stack ${\cal Y}$ is a Calabi-Yau stack, for the same reasons as in
the previous example.

As before, the stack ${\cal Y}$ describes a partial blowup of
$T^6/{\bf Z}_3$ with the singularities replaced by local quotient stack
structures.

\subsection{Flat gerbes}  \label{flatgerbes}

\subsubsection{Generalities}

A `gerbe' is the analogue of a principal bundle for $B$ fields,
rather than gauge fields.  The total `space' is no longer an
honest space, but rather is a stack.  Gerbes correspond to quotients
by noneffectively-acting groups.

The canonical trivial $G$-gerbe on a space $X$ 
is the quotient stack $[X/G]$, where $G$ acts trivially on $X$.
Nontrivial gerbes look locally like the trivial gerbe.

Let us work out descriptions of more general gerbes as global
quotients of auxiliary spaces.
Recall $G$-(banded-)gerbes on a space $X$, for abelian $G$,
are classified by $H^2(X, C^{\infty}(G))$.
Suppose there exists a group $K$, and a central extension $A$ of $K$ by $G$
\begin{displaymath}
0 \: \longrightarrow \: G \: \longrightarrow \: A \:
\longrightarrow \: K \: \longrightarrow \: 0
\end{displaymath}
such that the characteristic class in $H^2(X, C^{\infty}(G))$
is the image of an element of $H^1(X, C^{\infty}(K))$
under the coboundary map in the long\footnote{If any of the groups
are nonabelian, this `long' exact sequence is only defined
for a few degrees, but enough so as to make sense of this statement.} 
exact sequence associated
to the short exact sequence above.
That element of $H^1(X, C^{\infty}(K))$ defines a principal $K$-bundle
on $X$, call it $P$.  The bundle $P$ has a non-faithful action of
the extension $A$, and in fact, it can be shown that the
quotient stack $[P/A]$ is the same as the $G$-gerbe classified
by the element of $H^2(X, C^{\infty}(G))$ that we started with.
Note the quotient stack $[P/A]$ is a stack over $[P/K] = P/K = X$.

For example, let us consider the nontrivial ${\bf Z}_2$-gerbe
on ${\bf P}^1$.  (Since 
\begin{displaymath}
H^2({\bf P}^1, C^{\infty}({\bf Z}_2) \: = \:
H^2({\bf P}^1,{\bf Z}_2) \: = \: {\bf Z}_2,
\end{displaymath} 
there is precisely
one nontrivial ${\bf Z}_2$-gerbe over ${\bf P}^1$.)
Let us make use of the fact that ${\bf P}^1 = ( {\bf C}^2-0)/{\bf C}^{
\times}$, and that the nontrivial extension
\begin{equation}  \label{ext1}
0 \: \longrightarrow \: {\bf Z}_2 \: \longrightarrow \:
{\bf C}^{\times} \: \longrightarrow \: {\bf C}^{\times} \:
\longrightarrow \: 0
\end{equation}
has the property that the nontrivial element of $H^2({\bf P}^1,
{\bf Z}_2)$ is the image\footnote{
We can check this by noting that the ${\bf C}^{\times}$ bundle
$({\bf C}^2-0)$ over ${\bf P}^1$ is precisely realizing the Hopf
construction -- if we restricted to an $S^3$, with a $U(1)$ action,
then this would be precisely the Hopf construction.
The characteristic class of the Hopf fibration is odd.
The long exact sequence 
\begin{displaymath}
H^1({\bf P}^1, C^{\infty}({\bf Z}_2)) \: \longrightarrow \:
H^1( {\bf P}^1, C^{\infty}( {\bf C}^{\times} )) \: \longrightarrow \:
H^1( {\bf P}^1, C^{\infty}( {\bf C}^{\times} )) \: \longrightarrow \:
H^2( {\bf P}^1, C^{\infty}( {\bf Z}_2)) \: \longrightarrow \:
H^2( {\bf P}^1, C^{\infty}( {\bf C}^{\times} ))
\end{displaymath}
associated to the extension~(\ref{ext1})
simplifies to
\begin{displaymath}
0 \: \longrightarrow \: {\bf Z} \: \stackrel{2 \times}{\longrightarrow} \:
{\bf Z} \: \longrightarrow \: {\bf Z}_2 \: \longrightarrow \: 0
\end{displaymath}
where the second ${\bf Z}$ classifies ${\bf C}^{\times}$ bundles on
${\bf P}^1$ and the ${\bf Z}_2$ is $H^2({\bf P}^1, C^{\infty}({\bf Z}_2))$.
Since the Hopf fibration has odd characteristic class,
and the first nontrivial map is multiplication by two,
the Hopf fibration cannot be in the image, so it must map to the
nontrivial element of ${\bf Z}_2$.
} of the element of $H^1({\bf P}^1,C^{\infty}({\bf C}^{\times}))$
corresponding to the principal ${\bf C}^{\times}$-bundle
$\{ {\bf C}^2-0 \} \rightarrow {\bf P}^1$.
In particular, the quotient stack $[ ( {\bf C}^2-0)/ {\bf C}^{\times}]$,
where the ${\bf C}^{\times}$ in the quotient is the nontrivial
extension, {\it i.e.} acts on ${\bf C}^2-0$ as
$(x,y) \mapsto (t^2 x, t^2 y)$, is the same as the nontrivial ${\bf Z}_2$-gerbe
on ${\bf P}^1$, and so this gerbe has atlas $\{ {\bf C}^2-0 \}$.

We can build ${\bf Z}_n$ gerbes on other spaces similarly.
Let $P$ be a principal ${\bf C}^{\times}$ bundle on $X$ such that
the mod $n$ reduction of its first Chern class is the element of
$H^2(X, {\bf Z}_n)$ classifying the desired ${\bf Z}_n$ gerbe.  
If we extend ${\bf C}^{\times}$ by ${\bf Z}_n$,
so as to get a ${\bf C}^{\times}$ action that rotates the fibers
of $P$ $n$ times, then the quotient of $P$ by that extension,
$[P / {\bf C}^{\times}]$, is the gerbe in question.

We can also often describe ${\bf Z}_n$ gerbes as global quotients
by finite groups, following the same general procedure as above,
a fact we shall use later in this paper.

We can build flat nontrivial $U(1)$ gerbes using the same general
ideas.
Let $M$ be a space such that $H^3_{tors}(M, {\bf Z})$ is nonempty.
Let $\alpha$ be an $r$-torsion element of $H^3(M, {\bf Z})$,
which will classify our gerbe.
Just as previously we built $G$-gerbes by finding a suitable
extension of some group $K$ by $G$, and using a principal $K$-bundle as
an atlas, here we use the extension
\begin{equation}  \label{flatgerbeext}
0 \: \longrightarrow \: {\cal O}^* \: \longrightarrow \:
GL(r, {\bf C}) \: \longrightarrow \: PGL(r, {\bf C}) \: \longrightarrow
\: 0
\end{equation}
and will construct the desired flat gerbe by quotienting a
principal $PGL(r, {\bf C})$ bundle by $GL(r, {\bf C})$.

Let $P'$ be a principal ${\bf P}^{r-1}$ bundle on $M$,
such that the obstruction\footnote{From the short exact 
sequence~(\ref{flatgerbeext})
we get a long exact sequence whose relevant part is
\begin{displaymath}
H^1(M, {\cal O}^* ) \: \longrightarrow \: H^1(M, GL(r,{\bf C})) \: 
\longrightarrow \: H^1(M, PGL(r, {\bf C})) \: \longrightarrow \:
H^2(M, {\cal O}^*) 
\end{displaymath}
which makes it manifestly clear that the obstruction to lifting a
${\bf P}^{r-1}$ bundle to a rank $r$ vector bundle is given by
an element of $H^2(M, {\cal O}^*)$.}
to lifting $P'$ to a $GL(n, {\bf C})$ vector
bundle is $\alpha$.

Let $P$ denote the frame bundle of $P'$, {\it i.e.}
$\mbox{Isom}\left( P, {\bf P}^{r-1} \times M\right)$.
$P$ will be the atlas of our gerbe.

Now, $PGL(r, {\bf C})$ acts on $P$ freely, and $GL(r, {\bf C})$ acts
on $P$ through the projection map into $PGL(r, {\bf C})$.

As a result, we can describe the flat $U(1)$ gerbe on $M$ classified by
$\alpha$ as $[P/GL(r,{\bf C})]$.

In this paper, we will only study Deligne-Mumford stacks,
meaning, quotients in which all points have finite stabilizers.
Clearly, a $U(1)$ gerbe, in which stabilizers are copies of $U(1)$,
cannot be a Deligne-Mumford stack.

\subsubsection{Non-banded gerbes}

This section is more technical, and can be omitted on a first reading.

Technically, the $G$-gerbes of the form described above
are `banded' gerbes, and on a space $X$ are classified by elements of
$H^2(X, C^{\infty}(G))$.  There is a slightly more general
class of gerbes that we shall sometimes consider, that can be described
analogously.  This more general class of gerbes,
known as $G$-gerbes (whereas the previous class is, strictly
speaking, $G$-banded-gerbes) can also be described as global quotients.

To describe a $G$-gerbe as a global quotient, as before we require
a short exact sequence
\begin{displaymath}
0 \: \longrightarrow \: G \: \longrightarrow \: A \:
\longrightarrow \: K \: \longrightarrow \: 0
\end{displaymath}
except that now $A$ no longer need be a {\it central} extension
of $K$, merely some extension.  As before, Let $P$ be a 
principal $K$-bundle on $X$, then the $G$-gerbe can
(for certain $P$) be described as $[P/A]$, where $A$ acts by
first projecting to $K$.

Non-banded $G$-gerbes are no longer classified merely by
$H^2(X, C^{\infty}(G))$, but rather are classified by
the hypercohomology group
\begin{displaymath}
H^2\left( X, G \rightarrow \mbox{Aut}(G) \right)
\end{displaymath}
({\it i.e.} sheaf cohomology valued in a complex), which fits into
an exact sequence
\begin{displaymath}
H^2(X, C^{\infty}(G) ) \: \longrightarrow \:
H^2\left(X, G \rightarrow \mbox{Aut}(G) \right) \: \longrightarrow \:
H^1(X, \mbox{Out}(G))
\end{displaymath}
where $\mbox{Out}(G)$ denotes the outer automorphisms of $G$,
and $\mbox{Aut}(G)$ denotes all automorphisms of $G$.
Intuitively, $H^1(X, \mbox{Out}(G))$ encodes how
$K$ acts on $G$ globally, which for a central extension is trivial.

Both banded and non-banded $G$-gerbes can be Calabi-Yau.
For most of this paper, we shall ignore the distinction.

Later we shall see several examples of non-banded gerbes,
and study the physics of strings propagating on such gerbes.
Since we have already discussed some examples of banded gerbes,
presented as global quotients by $U(1)$, let us do the same for
an example of a non-banded gerbe.
Define
\begin{displaymath}
\widetilde{ {\bf H} } \: = \: \frac{ {\bf C}^{\times} \times {\bf H} }{
{\bf Z}_4 }
\end{displaymath}
where ${\bf H}$ is the eight-element group formed from the quaternions
\begin{displaymath}
{\bf H} \: = \: \{ \pm 1, \pm i, \pm j, \pm k \}
\end{displaymath}
and the ${\bf Z}_4$ acts on ${\bf C}^{\times}$ in the obvious way,
and acts on ${\bf H}$ as $<i>$, so that we have a pushout diagram
\begin{displaymath}
\begin{array}{ccc}
{\bf Z}_4 & \longrightarrow & {\bf H} \\
\downarrow & & \downarrow \\
{\bf C}^{\times} & \longrightarrow & \widetilde{ {\bf H} }
\end{array}
\end{displaymath}
which gives us a commuting diagram
\begin{displaymath}
\begin{array}{ccccccccc}
& & 0 & & 0 & & 0 & & \\
& & \downarrow & & \downarrow & & \downarrow & & \\
0 & \longrightarrow & {\bf Z}_4 & \longrightarrow & {\bf H} &
\longrightarrow & {\bf Z}_2 & \longrightarrow & 0 \\
& & \downarrow & & \downarrow & & \downarrow & & \\
0 & \longrightarrow & {\bf C}^{\times} & \longrightarrow &
\widetilde{ {\bf H} } & \longrightarrow & {\bf Z}_2 &
\longrightarrow & 0 \\
& & \downarrow & & \downarrow & & \downarrow & & \\
0 & \longrightarrow & {\bf C}^{\times} & \longrightarrow &
{\bf C}^{\times} & \longrightarrow & 0 & & \\
& & \downarrow & & \downarrow & &    & & \\
 & & 0 & & 0 & & & &
\end{array}
\end{displaymath}
We also have a short exact sequence
\begin{displaymath}
0 \: \longrightarrow \: {\bf Z}_4 \: \longrightarrow \:
\widetilde{ {\bf H} } \: \longrightarrow \:
{\bf C}^{\times} \times {\bf Z}_2 \: \longrightarrow \: 0
\end{displaymath}
Now, if we are given a principal ${\bf C}^{\times} \times {\bf Z}_2$-bundle
$L$ over $X$, a non-banded gerbe can be constructed as
the quotient $[L/\widetilde{ {\bf H} } ]$.

\subsubsection{Calabi-Yau gerbes}    \label{cygerbes}

From our definition of Calabi-Yau stack,
a Calabi-Yau gerbe is a stack with trivial canonical bundle.
It turns out that this statement has a very trivial consequence:
a Calabi-Yau gerbe over a space or stack $X$ is Calabi-Yau
if and only if $X$ is Calabi-Yau.
(The canonical bundle of the gerbe turns out to be given by the pullback
of the canonical bundle on $X$.)

Let us check that statement for presentations of several different forms.
Consider a presentation of the form $[X/G]$, where $G$ is finite
and acts noneffectively.
Write
\begin{displaymath}
1 \: \longrightarrow \: K \: \longrightarrow \: G \: \longrightarrow \:
H \: \longrightarrow \: 1
\end{displaymath}
where $K$ is the trivially-acting normal subgroup, and $H = G/K$ acts
effectively, and both $K$ and $H$ are finite.
In this language, $G$ acts on $X$ by first projecting to $H$,
which has an effective action on $X$.
From our previous general analysis, we said that a presentation
$[X/G]$ defines a Calabi-Yau stack if and only if $X$ is Calabi-Yau
and $G$ preserves the holomorphic top form.
In the present case, $[X/G]$ is a $K$-gerbe over $[X/H]$.
Thus, $[X/G]$ is Calabi-Yau if and only if $X$ is Calabi-Yau and
$G$ preserves the holomorphic top form, but since $G$ acts by first
projecting to $H$, we see that $G$ preserves the holomorphic top form
if and only if $H$ preserves the holomorphic top form.
Thus, for a $K$-gerbe $[X/G]$ over $[X/H]$, the gerbe $[X/G]$ is
Calabi-Yau if and only if $[X/H]$ is Calabi-Yau.

For presentations as quotients by nonfinite groups,
the same analysis applies, but there is a subtlety that may confuse the
reader.  Consider for example the presentations of banded
${\bf Z}_k$ gerbes discussed earlier, as ${\bf C}^{\times}$ quotients
of total spaces of principal ${\bf C}^{\times}$ bundles,
of the form $[L/{\bf C}^{\times}]$.
From our general claim at the beginning of this section,
if this gerbe lives over a manifold $M$, {\it i.e.}
$L$ is a bundle over $M$, then the gerbe
$[L/{\bf C}^{\times}]$ should be Calabi-Yau if and only if $M$
is Calabi-Yau.  But for $[L/{\bf C}^{\times}]$ to be Calabi-Yau
means that $L$ is Calabi-Yau.

Now,
it is well-known that total spaces of line bundles over Calabi-Yau's
are not typically Calabi-Yau, but what is less well-known is that
the total space of a principal ${\bf C}^{\times}$ bundle over a Calabi-Yau,
unlike a line bundle, is always Calabi-Yau.
A proof of this somewhat counterintuitive fact, along with a generalization,
is given in appendix~\ref{ppalcxCY}.

Thus, although the details are more subtle,
we see also that for presentations as quotients by nonfinite groups,
again gerbes over Calabi-Yau's are themselves Calabi-Yau stacks.

An alternative way to understand the same result is, briefly, as follows.
One presentation involves picking a good cover of the Calabi-Yau,
and putting a trivial ${\bf Z}_n$-gerbe on each element of the cover,
with relations that encode the gerbe structure.
In this presentation, over any element of the good cover,
the atlas looks like $n$ disjoint copies of the underlying open set,
together with relations that preserve the Calabi-Yau structure.
Some readers might find that this
this particular presentation makes the fact that this stack has
trivial canonical bundle more clear.

In \cite{glsm}, using gauged linear sigma models to describe
gerbes on ambient toric varieties, we shall get another check
of the statement that Calabi-Yau gerbes are merely gerbes
on Calabi-Yau's.

\subsection{Some non-Calabi-Yau stacks}
\label{ex:nonCY}

In the previous several subsections, we have discussed a variety
of Calabi-Yau stacks, and multiple presentations thereof.
In this section, we shall briefly discuss some examples of
non-Calabi-Yau stacks.

One easy way to construct some non-Calabi-Yau stacks is to start
with a Calabi-Yau, and orbifold by a finite group that does not
preserve the holomorphic top form.  The physics of such examples
was recently considered in \cite{evaallan,hkmm}.

One particular example of that form merits attention:
an example in which the canonical class of the stack is torsion,
but the canonical class of the underlying variety is nonzero
and non-torsion.
Consider the stack $[T^2/{\bf Z}_2]$, where the ${\bf Z}_2$ inverts
the sign of the complex coordinate.  This stack is not Calabi-Yau
(after all, the ${\bf Z}_2$ does not preserve the holomorphic
top form), and the canonical bundle is two-torsion.
However, the underlying variety is $T^2/{\bf Z}_2 = {\bf P}^1$.
This is not only not Calabi-Yau, but fails to be Calabi-Yau in a worse
fashion:  the canonical bundle is not even torsion,
rather it is nonzero and nontorsion.

A slightly more devious example of a non-Calabi-Yau stack
can be constructed as follows.
Begin by considering $\mbox{Bl}_1 {\bf C}^2$, the blowup of
${\bf C}^2$ at a single point.  Consider a ${\bf Z}_2$ action on
${\bf C}^2$ that leaves the location of the blowup fixed and inverts
coordinates, and extend that ${\bf Z}_2$ (trivially) over the
exceptional divisor in $\mbox{Bl}_1 {\bf C}^2$.
As an algebraic variety,
\begin{displaymath}
\left( \mbox{Bl}_1 {\bf C}^2 \right) / {\bf Z}_2 \: \cong \:
\widetilde{ {\bf C}^2 / {\bf Z}_2 }
\end{displaymath}
(The ${\bf Z}_2$ quotient of $\mbox{Bl}_1 {\bf C}^2$ has complex
codimension one fixed points, but fixed points in codimension one
are invisible algebraically.)

Finally, we are ready to describe the stack.
The stack in question is $[ \left( \mbox{Bl}_1 {\bf C}^2 \right) / {\bf Z}_2 ]$.
Since the corresponding space is Calabi-Yau, the reader might
incorrectly guess that this stack is also Calabi-Yau.
Because the ${\bf Z}_2$ has
a codimension one locus of fixed points, the stack
$[ \left( \mbox{Bl}_1 {\bf C}^2 \right) / {\bf Z}_2 ]$ 
looks like a ${\bf Z}_2$-gerbe
over the exceptional divisor of $\widetilde{ {\bf C}^2/{\bf Z}_2 }$.
Away from the exceptional divisor, however, the stack
looks like the space $\widetilde{ {\bf C}^2/{\bf Z}_2 }$.

Because of the close relationship between the stack and the space,
the reader might incorrectly guess that the stack
$[ \left( \mbox{Bl}_1 {\bf C}^2 \right) / {\bf Z}_2 ]$ is also Calabi-Yau.
However, this is not the case.  
Metrically, a Ricci-flat metric on this stack would be a ${\bf Z}_2$-invariant
Ricci-flat metric on $\mbox{Bl}_1 {\bf C}^2$, but that space
admits no Ricci-flat metrics, much less a ${\bf Z}_2$-invariant
one.  More formally, the canonical bundle of 
$[ \left( \mbox{Bl}_1 {\bf C}^2 \right) / {\bf Z}_2 ]$
can be shown to be nontrivial.

\section{Sigma models on (presentations of) Calabi-Yau stacks}  
\label{sigmamodels}

In the previous section, we described Calabi-Yau stacks
and their various presentations.  As we shall describe in more
detail in this section, each presentation of a stack defines
a sigma model, and the sigma models arising from different presentations
all seem to be in the same universality class.  Moreover, the
Calabi-Yau condition for stacks seems to correctly distinguish
supersymmetric cases from non-supersymmetric cases -- we shall
see some interesting examples.

In section~\ref{sigmacoupling} we discuss sigma model perturbation theory,
and explicitly describe the version of this perturbation theory in
which gauged sigma models are weakly coupled ({\it i.e.} an expansion
in the curvature of the covering space, rather than a resolution of the
quotient space).  In section~\ref{basicdefn} we review how a gauged
sigma model is naturally a sigma model on a quotient stack,
as was previously discussed in \cite{meqs}.  We correlate qualitative
features of such models with mathematical properties of
stacks in section~\ref{wellbehave}, and in sections~\ref{gerbefinpres} and
\ref{nonmincharge} outline examples with noneffective gaugings, 
{\it i.e.} gerbes,
whose physics is discussed more extensively in \cite{nr,glsm}.
In section~\ref{ex:nonCYphysics} we discuss a non-Calabi-Yau example.

Now, a given stack can have many different presentations of the form
$[X/G]$; one of the most important parts of this proposal is that
different presentations should define gauged sigma models lying in the
same universality class, so that stacks classify universality classes.
In section~\ref{multpres:samecft} we check that statement in some
examples of Calabi-Yau stacks with multiple very different presentations.
We will do further tests later in this paper and also in
\cite{glsm}, where we shall discuss examples of toric stacks
with multiple presentations in the form of gauged linear sigma models,
and there we will see explicitly in examples that quantum cohomology and
Toda duals are presentation-independent.

Finally, in section~\ref{tftstx} we discuss the topological A and B models
on stacks.  We will discuss the open string B model more extensively
later in this paper in section~\ref{openB}, and the closed string A model
will be discussed in greater detail in \cite{glsm}.

\subsection{A word on sigma model coupling constants}
\label{sigmacoupling}

Before beginning the description of sigma models on (presentations of)
stacks, let us take a moment to speak to the issue of coupling constants
and sigma model perturbation theory.

In quotients of a Calabi-Yau $X$ by finite effectively-acting groups $G$,
people have often done sigma model perturbation theory
in the radius of a resolution of the quotient space:  $\widetilde{X/G}$.
In such an expansion, the orbifold CFT lies at strong coupling,
and hence, in such an expansion, no geometric picture of string orbifolds
can or should be expected.

However, in any $G$-gauged sigma model on $X$, there is another,
more natural, parameter in which to perform sigma model
perturbation theory, namely, the radius of the cover $X$.
With respect to {\it this} coupling constant,
sigma model perturbation theory can be made weakly coupled at the orbifold 
point, merely by taking $X$ large,
and so a geometric description can be expected.

Throughout this paper, whenever sigma model perturbation theory is called
for, we will be expanding about large radius on the cover, not large radius
on a resolution of a quotient.  This is why it is sensible for us to speak
of geometry in string orbifolds in the first place.

\subsection{Basic definition}   \label{basicdefn}

Let us first recall the path-integral description of a $G$-gauged sigma
model on a space $X$.
We begin by representing the worldsheet as a polygon with sides identified.
When we gauge the action of $G$, maps from the worldsheet into $X$
are allowed to have branch cuts, meaning that at the edges of the polygon,
a map into $X$ is only well-defined up to a transformation defined
by a map from that edge into $G$.

When $G$ is discrete, this leads to the usual picture of
picking $2g$ elements of $G$, for a genus $g$ worldsheet, obeying a constraint
that is simply the condition that the image of the edges in $X$ close
as you walk around the boundary of the polygon.

When $G$ is not discrete, the same picture applies, though one can
have nonconstant maps into $G$ along the boundaries.

In both cases, one must sum over equivalence classes of principal
$G$ bundles on the worldsheet (we are gauging a group $G$ after all),
which are represented by the maps from the edges into $G$.

Then, the action is defined as a ($G$-invariant) sigma model action
of the usual form, defined on the interior of the polygon,
with ordinary derivatives replaced by covariant
derivatives.

The combination of edge maps and maps into $X$ with $G$-branch cuts
outlined above can be described much more efficiently.
Let $E$ denote the total space of a principal $G$-bundle,
defined by the edge maps into $G$.
The `maps' into $X$ can be equivalently described as maps from
the total space of $E$ into $X$, together with a $G$-equivariance
condition that can be stated as $f(g \cdot x) = g \cdot f(x)$
for any such map $f$.  If we restrict $f$ to any section of $E$
defined over the interior of the polygon, then we get the
maps into $X$ described above, with $G$ actions at the boundaries.
The $G$-equivariance condition ensures that the map from the total
space of $E$ contains no more information than the map on a 
local section of $E$; the two maps are equivalent.

Thus, the `maps' that are summed over in a gauged sigma model are
pairs consisting of
\begin{enumerate}
\item a principal $G$-bundle $E$ over the worldsheet,
with connection
\item a $G$-equivariant map $f: E \rightarrow X$
\end{enumerate}

The first, and perhaps most important, point in defining sigma models
on stacks is that a map into $[X/G]$ is defined by the same\footnote{
After translation from the algebraic category to the differentiable
category.  See the next paragraph for details.}
pair of
data as above.
Thus, summing over maps into $[X/G]$ is precisely the same operation
that the path integral measure performs in a gauged sigma model -- in both
cases we sum over principal $G$ bundles, together with maps defined
with suitable cuts.

Strictly speaking, as we have been implicitly discussing algebraic
stacks up to this point, there are two important issues that bear
mentioning, though their resolutions are trivial.
\begin{enumerate}
\item Technically, in much of this paper we implicitly speak of 
algebraic stacks,
just as physicists often talk about Calabi-Yau manifolds as
algebraic varieties.  To define sigma models, there is always implicitly
a technical step in which one replaces smooth algebraic varieties with
differentiable manifolds, in which the defining data changes from
algebraic function rings and Zariski topologies to coverings by
smaller open sets and metrics.
The analogous step here involves replacing algebraic stacks,
defined by holomorphic maps, with differentiable
stacks, defined by smooth maps.  Thus, as an algebraic stack $[X/G]$
is defined by a reductive $G$, a holomorphic principal $G$ bundle,
and a holomorphic map from the total space of that bundle into $X$,
but after replacing that by the corresponding differentiable stack,
one speaks of smooth bundles with connection and smooth maps.
As this step is obvious and trivial, we will not belabor the point
further.
\item As algebraic stacks we
think of $G$ as a 
reductive algebraic group, {\it i.e.} something noncompact such as
${\bf C}^{\times}$.  Strictly speaking, to relate to a physical
gauged sigma model, we want to replace any reductive algebraic
group $G$ with a corresponding compact Lie group, as part of the
implicit transition from the algebraic category to the differentiable
category, which also means modifying the atlas.  
In a $(2,2)$ supersymmetric worldsheet theory, 
D-terms implicitly take care of this sort of matter.
The counting of $G$-bundles on the worldsheet
is the same for $G$ reductive
algebraic or compact Lie, so, we do not gain or lose any
twisted sectors in this process.
\end{enumerate}
As these issues are trivially easy to resolve, we shall not discuss
them further.  We will merely speak of `stacks,'
and will freely interchange algebraic stacks $[X/G]$
with the differentiable stacks $[Y/H]$ where $H$ is the Lie group
associated to the reductive algebraic group $G$, and $Y$ is the 
differentiable manifold obtained from $X$ by D-term fixing.

So far we have seen that the path integral sum in a gauged sigma model
is the same thing as a sum over maps into a quotient stack $[X/G]$.
Formulating an action is equally straightforward.
In \cite{meqs} it was observed that for $G$ finite, this is completely
trivial -- just use the usual nonlinear sigma model action,
on a local section of the principal bundle that partially defines
a given map into $X$.  Since the action is $G$-invariant
(otherwise, one could not gauge $G$), the action is well-defined across
the boundaries of the polygon.

For $G$ nondiscrete, we have to work very slightly harder, but not much.
For bosonic strings, the only complication beyond the above is that
ordinary derivatives must be replaced with covariant derivatives.
Ultimately this is because in the algebraic category,
the $\overline{\partial}$ acting on a $G$-equivariant map from
the total space of the principal $G$-bundle into $X$
becomes a covariant derivative when reinterpreted in the smooth category.
For supersymmetric strings, one also has D-terms.
We do not see how D-terms can be directly derived from these considerations,
but rather take the attitude that they are obtained by supersymmetrizing
a bosonic action, and so need not be directly describable as
terms in an ordinary nonlinear sigma model, so long as the bosonic action
can be obtained directly.

Given the observation that the path integral of a gauged sigma model
precisely sums over maps into $[X/G]$, describing gauged
sigma models as `ordinary' sigma models on a stack seems trivial.
What is far more tricky is the proposed presentation-independence of IR physics;
if universality classes of gauged sigma models depend upon the presentation
of a stack, not just the stack, then this whole program of understanding
string compactifications on stacks fails.

In a few sections, we will study some
specific examples of stacks with multiple presentations of the form
of global quotients $[X/G]$, and will check in those examples that
the IR physics really is presentation-independent.
Before that, however, we will pause to note how
certain mathematical properties of stacks seem related to certain 
physical properties of string orbifolds,
and then we shall briefly review the highlights of the physics of
gauged sigma models with noneffective group actions,
as studied more carefully in \cite{nr,glsm}.
Gauged sigma models with noneffective group actions form a 
strong testing ground for the ideas of these papers,
as they have physical properties which seem extremely obscure at first,
but which have a very simple understanding in terms 
of stacks.

In passing, it would be entertaining if the mathematical dimension
of the stack always matched the central charge of the corresponding
conformal field theory.  For example, the dimension of the
trivial gerbe $[\mbox{point}/G]$ is negative, given by $- \mbox{dim } G$,
and certainly the central charges of pure ghost systems are negative.
However, we only consider Deligne-Mumford stacks, meaning all
fixed points have finite stabilizers, so the only examples of
the form $[\mbox{point}/G]$ that we consider have $G$ finite,
for which the claim is not very interesting.

\subsection{Well-behavedness of sigma models on stacks}
\label{wellbehave}

As is well-known, string orbifolds are well-behaved CFT's,
at least when the covering spaces are smooth.  Unlike sigma models
on singular spaces, string orbifold CFT's do not have divergent
correlation functions or other such properties.

We have previously said that all Deligne-Mumford stacks can be
presented in the form $[X/G]$, for some not necessarily finite,
not necessarily effectively acting $G$.  However, a stronger statement
can be made.  A Deligne-Mumford stack $[X/G]$ has the further property
that the stabilizers of all fixed points are finite\footnote{
Technicality lovers should note that everywhere in this paper
we are working in characteristic zero.} subgroups of $G$.
(A stack $[X/G]$ in which not all stabilizers are finite is called
Artin.)

In other words, a Deligne-Mumford stack looks locally like
a quotient by a finite (albeit not-necessarily-effectively-acting) group.
Globally, we may have to resort to non-finite groups, but locally,
we can describe these stacks as quotients by finite groups.
This is part of the reason why we only consider Deligne-Mumford stacks
in this paper, and not more general Artin stacks.

In \cite{evaed} it was argued, among other things, that the
bad behavior of a sigma model on a singular space is essentially
a local feature of the target space.  This argument was made in the
context of linear sigma models, but if we assume that essentially the
same statement holds true for nonlinear sigma models,
then sigma models on Deligne-Mumford stacks, expressed as
$[X/G]$ for $X$ smooth and $G$ acting by diffeomorphisms,
should be well-behaved, and not ever suffer from any of the 
undesirable properties
of sigma models on singular spaces.

We shall not use this conjecture anywhere in the remainder of this paper,
but given the amount of confusion that the well-behavedness property
of string orbifold CFT's has generated over the years, and the number
of technical convolutions that have been used to 
try explain that well-behavedness
feature, we think it important to observe that such well-behavedness
naturally correlates with the geometry of stacks.

\subsection{Example:  Gerbes presented as global quotients by finite
groups}   \label{gerbefinpres}

A gerbe is a local orbifold by a trivially-acting group.
As with all local orbifolds, it can be described as a global quotient
by a noneffectively-acting group, {\it i.e.} a group in which some
of the elements act trivially.
Now, the reader might at first think that such a physical theory
should be identical to the physical theory on the underlying space,
but examples of such theories, presented as global quotients
by finite noneffectively-acting groups, were described in \cite{nr},
and there we saw explicitly in several examples
that such theories are distinct from theories 
on the underlying spaces.

In fact, in \cite{nr} we did quite a bit more:  we checked the
consistency of noneffective orbifolds, calculated massless spectra of
noneffective orbifolds, and began to address deformation theory issues.

As we discussed many such examples in detail in \cite{nr},
for brevity we refer the reader to that reference.

\subsection{Example:  Gerbes presented as quotients by nonfinite groups}
\label{nonmincharge}

As discussed earlier in section~\ref{flatgerbes},
another typical presentation of gerbes is as a quotient by a 
nonfinite group.  For example, we constructed ${\bf Z}_k$ gerbes
as $U(1)$ quotients of total spaces of principal $U(1)$ bundles,
where the $U(1)$ rotates the fibers $k$ times.

Physically a sigma model on such a presentation is a $U(1)$ gauged
sigma model on the total space of the bundle, in which the
fields have nonminimal $U(1)$ charges, since the $U(1)$ rotates the fibers
multiple times.

From the discussion of presentations of gerbes as quotients by
finite groups, it is clear that the physical theory of a string
on a gerbe is not the same as the physical theory of a string
on the underlying space (not even with a $B$ field).  
For our conjecture that stacks, not presentations thereof, classify
universality classes, it had better be the case that a two-dimensional
gauge theory with nonminimal charges must be different from a two-dimensional
gauge theory with minimal charges.

Indeed, that is the case.  Although such two-dimensional gauge
theories are the same perturbatively, they are very different
nonperturbatively.

This fact will play a crucial role in \cite{glsm}, where we will extensively
study gauged linear sigma models for toric stacks, which typically
look like ordinary gauged linear sigma models, but with nonminimal charges.
There, we will explicitly calculate some of the many ways in which
the theories differ -- from different correlation functions to
different R-symmetry anomalies.

Since this physical effect is obscure, let us take a moment to
describe
more carefully the general reasons why these theories are
distinct.  (We would like to thank J.~Distler and R.~Plesser for providing
the detailed argument that we review in this section.)
For a different discussion of two dimensional
gauge theories with fermions of nonminimal charges, 
see \cite{edold}[section 4].  (The discussion there is most
applicable to the present situation when $m \ll M$, in the notation
of that reference.)

To be specific, consider a gauged linear sigma model with a single
$U(1)$ gauge field, and with chiral superfields, all of charge $k$,
with $k>1$.  (Mathematically, this corresponds to a ${\bf Z}_k$ gerbe
on a projective space, as we shall review in the next section.)
One might argue that this theory should be the same as a theory
with chiral superfields of charge $1$, as follows.
Since instanton number is essentially monopole number,
from Dirac quantization since the electrons have charges a multiple
of $k$, the instantons must have charge a multiple of $1/k$,
and so zero modes of the Higgs fields in a minimal nonzero instanton
background would be sections of ${\cal O}(k/k) = {\cal O}(1)$,
just as in a minimal charge GLSM.  Making the charges nonminimal
has not changed the physics.
In order to recover the physics we have described,
we require the Higgs fields to have charge $k$ while the
instanton numbers are integral, not fractional.

Closer analysis reveals subtleties.
Let us break up the analysis into two separate cases:  first,
the case that the worldsheet is noncompact, second,
that the worldsheet is compact.  For both cases, it will be important
that the worldsheet theory is two-dimensional.

First, the noncompact case.
Since the $\theta$ angle couples to $\mbox{Tr }F$,
we can determine the instanton numbers through the periodicity of
$\theta$.  Suppose we have the physical theory described above,
namely a GLSM with Higgs fields of charge $k$,
plus two more massive fields, of charge $+1$ and $-1$.
In a two-dimensional theory, the $\theta$ angle acts as an electric
field, which can be screened by pair production, and that screening
determines the periodicity of $\theta$.
If the only objects we could pair produce were the Higgs fields
of charge $k$, then the theta angle would have periodicity
$2 \pi k$, and so the instanton numbers would be multiples
of $1/k$.  However, since the space is noncompact, and the
electric field fills the entire space, we can also pair produce
arbitrary numbers of the massive fields, which have charges
$\pm 1$, and so the $\theta$ angle has periodicity $2 \pi$,
so the instantons have integral charges.

We can phrase this more simply as follows.
In a theory with only Higgs fields of charge $k$,
the instanton numbers are multiples of $1/k$, and so the resulting
physics is equivalent to that of a GLSM with minimal charges.
However, if we add other fields of charge $\pm 1$,
then the instanton numbers are integral,
and if those fields become massive, and we work at an energy scale
below that of the masses of the fields, then we have a theory
with Higgs fields of charge $k$, and integral instanton numbers,
giving us the physics that corresponds to a gerbe target.

Thus, we see in the noncompact case that there are two
possible physical theories described by Higgs fields of charge $k$:
one is equivalent to the GLSM with minimal charges,
and the other describes the gerbe.

The analysis for the compact worldsheet case is much shorter.
Strictly speaking, to define the theory nonperturbatively on a
compact space, we must specify, by hand, the bundles that the
Higgs fields couple to.  If the gauge field is described by
a line bundle $L$, then coupling all of the Higgs fields to
$L^{\otimes k}$ is a different prescription from coupling all
of the Higgs fields to $L$.  As a result, the spectrum of zero modes
differs between the two theories, hence correlation functions and
anomalies differ between the two theories,
and so the two physical theories are very different,
as we shall see in examples later.

We shall assume throughout this paper that the worldsheet is
compact, though as we have argued the same subtlety shows up
for noncompact worldsheets.

We are interested in these physical theories
because they often crop up in describing stacks.
Since stacks are defined via their incoming maps, a precise definition
of which bundles the superfields couple to is part of the definition
of the stack, and so there is no ambiguity in which physical
theory to associate to a given (presentation of a) stack.

\subsection{Example:  Sigma model on a non-Calabi-Yau stack}
\label{ex:nonCYphysics}

To help check that our definition of ``Calabi-Yau stack'' is
a sensible definition, let us consider some sigma models on
non-Calabi-Yau stacks.

As mentioned in section~\ref{ex:nonCY}, some easy examples
of non-Calabi-Yau stacks are orbifolds of Calabi-Yau's by
finite groups that do not preserve the holomorphic top form.
Sigma models on such stacks have been extensively studied
in \cite{evaallan,hkmm} (though neither paper explicitly
used the language of stacks).  They exhibit a number of interesting
properties, but what is most relevant for the present discussion
is that they are not spacetime supersymmetric.
In these examples, our definition of ``Calabi-Yau stack''
correctly distinguishes spacetime supersymmetric cases from
non-spacetime-supersymmetric cases.

Let us next consider the other example discussed in
section~\ref{ex:nonCY}.
Recall this example in question is the non-Calabi-Yau stack
$[ \left( \mbox{Bl}_1 {\bf C}^2 \right) / {\bf Z}_2 ]$.
Because of the relationship between the algebraic spaces
\begin{displaymath}
\left( \mbox{Bl}_1 {\bf C}^2 \right) / {\bf Z}_2 \: \cong \:
\widetilde{ {\bf C}^2 / {\bf Z}_2 }
\end{displaymath}
we can interpret the stack $[ \mbox{Bl}_1 {\bf C}^2 / {\bf Z}_2 ]$
as a naive blowup of the stack $[ {\bf C}^2/{\bf Z}_2]$.
(``Naive'' because it does not have trivial canonical bundle.)

How does the physics of a sigma model on the stack
$[ \left( \mbox{Bl}_1 {\bf C}^2 \right) / {\bf Z}_2 ]$
compare to the physics of a sigma model on the space
\begin{displaymath}
\left( \mbox{Bl}_1 {\bf C}^2 \right) / {\bf Z}_2 \: \cong \:
\widetilde{ {\bf C}^2 / {\bf Z}_2 }  ?
\end{displaymath}

Because of the close relationship between the stack and the space,
the reader might (incorrectly) suspect that the sigma model
on the stack, {\it i.e.} the ${\bf Z}_2$ orbifold
of $\mbox{Bl}_1 {\bf C}^2$, is supersymmetric, and is in the same
universality class as $\widetilde{ {\bf C}^2 / {\bf Z}_2 }$.
However, it is easy to check that the orbifold is not supersymmetric,
and it seems unlikely that it can be in the same universality class.
To see that it cannot be supersymmetric\footnote{In nonlinear sigma
models, it is possible to explicitly compute the spectrum of
spacetime fermions, but much more difficult to compute the spectrum
of spacetime bosons.  A direct test of spacetime susy is therefore out
of the question.}, first note that the vertex operator for a spacetime
supercharge cannot come from a twisted sector, as the supercharge
must be able to exist everywhere on the target space, not just 
in a fixed-point locus.  The untwisted sector is, as a physical theory,
equivalent to the unorbifolded theory.  Now, when the target space
is K\"ahler but not Calabi-Yau, we have a massive theory
with worldsheet $(2,2)$ supersymmetry, but the left- and right- R-symmetries
are both anomalous.  Hence, there can be no spectral flow in such
models; thus, no supersymmetry.

\subsection{Multiple presentations define same CFT}
\label{multpres:samecft}

Our description of sigma models on stacks revolves around a choice
of presentation of the stack.  In order to speak consistently
about strings on the stack, rather than strings on a presentation
of the stack, the physics must be independent of the choice of
presentation.  Technically, sigma models associated to various presentations
of a single stack must flow in the IR to the same CFT.

We shall check that statement in a number of ways in this paper.
First, in this section we shall describe several explicit examples
of different-looking presentations of the same stack, and shall
check explicitly in each case that the different presentations
actually describe the same theory.  Trivial examples of such
equivalences are easy to produce:
\begin{itemize}
\item One of the simplest such examples is the stack
$[X \times U(1) / U(1)]$, where the $U(1)$ acts only on the $U(1)$,
by effective rotations.
The corresponding sigma model is a gauged sigma model on $X \times U(1)$,
where rotations of the $U(1)$ are gauged.  Such a sigma model is
obviously equivalent to a sigma model on $X$,
and indeed, the stack $[X \times U(1) / U(1) ]$ is the same stack
as $X$.  
\item 
Another trivial example is the stack 
$[\left( X \times {\bf Z}_2 \right) / {\bf Z}_2]$,
where the ${\bf Z}_2$ acts solely on the ${\bf Z}_2$ factor
in $X \times {\bf Z}_2$.  
Here the atlas consists of two disjoint copies of $X$,
which are interchanged by the group action.
Clearly, a ${\bf Z}_2$-gauged sigma model on two copies of $X$
is equivalent to a sigma model on $X$,
and indeed, this stack is the same stack as $X$.
\item A well-known example involves the A-twist of the ${\bf C}
{\bf P}^{N-1}$ model.  The A-twist of a nonlinear sigma model
on ${\bf C} {\bf P}^{N-1}$ is believed to lie in the same worldsheet
universality class as the A-twist of a gauge theory with
$N$ chiral superfields, each of charge one under a single $U(1)$,
corresponding to the presentation of ${\bf C} {\bf P}^{N-1}$
as the quotient $[ ( {\bf C}^N - 0 ) / {\bf C}^{\times}]$.
Although there is not a CFT here, nevertheless this is 
an easy example in which multiple presentations of the same stack
lie in the same universality class.
\end{itemize}
We shall discuss some much more nontrivial equivalences
in more detail below.

Second, in later sections we shall argue that physical quantities
can be calculated in a presentation-independent fashion.
In section~\ref{closedspectra} we shall show that the massless closed string
spectra can be expressed in terms of the stack, not merely in terms
of a presentation, and in section~\ref{openB} we shall argue that
the massless states of the open string B model can similarly be
expressed in terms of the stack, and not in terms of some presentation
thereof.  Also, in our companion work \cite{glsm} we shall again
see multiple presentations of the same stack, all giving rise to
gauged linear sigma models, giving the same RG-invariant physics.

\subsubsection{Example:  finite quotient versus continuous quotient}
\label{ex:finvscont}

Let us work through an example to substantiate our claim that
different presentations of a given stack should describe the
same CFT.  In particular, consider the alternate description of
the orbifold $[ {\bf C}^2 / {\bf Z}_n ]$ in the form $[X/U(1)]$,
where $X = \left( {\bf C}^2 \times U(1) \right) / {\bf Z}_n$,
as described in section~\ref{multpres}.
As stacks, $[ {\bf C}^2/{\bf Z}_n ] = [X/U(1)]$.
Let us outline why the corresponding physical theories should
be identified.

Begin with a bosonic sigma model on ${\bf C}^2 \times U(1)$,
{\it i.e.} four free real bosons, plus a boson compactified on a circle.
Now, consider the ${\bf Z}_n$ quotient that acts as
\begin{displaymath}
\left(x,y,t\right) \: \mapsto \: \left( \alpha x,\: \alpha^{-1} y, \: t+
\frac{2 \pi R }{n} \right)
\end{displaymath}
where $x$ and $y$ are complex coordinates on ${\bf C}^2$,
$t$ is a real coordinate on the circle, of radius $R$,
and $\alpha = \exp( 2 \pi i / n)$.
This ${\bf Z}_n$ acts freely, so the ${\bf Z}_n$ orbifold of the sigma
model on ${\bf C}^2 \times U(1)$ is identical to a sigma model
on $X$.  The description as a ${\bf Z}_n$ orbifold of a free field
theory is more amenable to analysis than the description as a nonlinear
sigma model, so we shall work with the former.

As a ${\bf Z}_n$ orbifold of a free field theory, the sigma model on $X$
has $n$ sectors:  an untwisted sector, describing bosons propagating
freely on ${\bf C}^2 \times U(1)$, plus $n-1$ twisted sectors,
in which the bosons obey boundary conditions of the form
\begin{eqnarray*}
x\left(\sigma + 2 \pi \right) & = &  \alpha^k x(\sigma) \\
y\left(\sigma + 2 \pi \right) & = &  \alpha^{-k} y(\sigma) \\
t\left(\sigma + 2 \pi \right) & = &  t(\sigma) + 2 k \pi R / n
\end{eqnarray*}
where $x$, $y$, $t$ are as above, $k \in \{ 0, 1, ..., n-1\}$,
$\sigma$ is a worldsheet coordinate,
and $R$ is the radius of the target-space $S^1$.
Since the ${\bf Z}_n$ acts freely, this twisted sector does not generate
any new massless states -- all the twisted sector states have
strictly positive mass.  The lowest mass states in the twisted sector
must wind halfway around the circle, and so can never shrink to
zero length.  The spectrum of massless states in the sigma
model on $X$ 
consists only of ${\bf Z}_n$-invariant states on ${\bf C}^2 \times U(1)$.

So far we have described a sigma model on $X$ as a ${\bf Z}_n$ orbifold
of a free field theory.  Next, let us gauge rotations of the $U(1)$
in $X$, {\it i.e.} gauge the action that sends
\begin{displaymath}
[x,y,t] \: \mapsto \: [x,y, \lambda + t]
\end{displaymath}
for $\lambda \in U(1)$.
In terms of the description of the sigma model on $X$ as a 
${\bf Z}_n$ orbifold of a free field theory, we perform this gauging
in each twisted sector separately.  (Since the $U(1)$ commutes with
the ${\bf Z}_n$, the $U(1)$ action continues to be a global symmetry in
the twisted sectors -- if we set $t'(\sigma) = t(\sigma) + \kappa$,
then $t(\sigma + 2 \pi) = t(\sigma) + 2k\pi R/n$ implies
$t'(\sigma + 2 \pi) = t'(\sigma) + 2k\pi R/n$.)  
However, since we are just gauging
rotations of the $U(1)$ in ${\bf C}^2 \times U(1)$, the effect
is merely to eliminate the $t$ direction.

The technically inclined reader will note that the
total space of the principal ${\bf Z}_n$ bundle
\begin{displaymath}
{\bf C}^2 \times U(1) \: \longrightarrow \:
\frac{ {\bf C}^2 \times U(1) }{ {\bf Z}_n }
\end{displaymath}
is $U(1)$-equivariantizable, since the $U(1)$ commutes with the
${\bf Z}_n$, so there are no subtleties in the
iterated gauging.  In principle, however, in iterated gaugings of this form 
the later gaugings need not be truly equivariant, as we shall discuss
in the next section.

The resulting theory, obtained by gauging the $U(1)$ action on the
sigma model on $X$, contains $n$ sectors:  an untwisted sector,
consisting of four free real bosons, and $n-1$ twisted sectors,
in which the bosons (expressed in complex coordinates) obey the
boundary conditions
\begin{eqnarray*}
x\left(\sigma + 2 \pi \right) & = &  \alpha^k x(\sigma) \\
y\left(\sigma + 2 \pi \right) & = &  \alpha^{-k} y(\sigma)
\end{eqnarray*}

The reader will immediately recognize this, however, as being the same
thing as the $[ {\bf C}^2 / {\bf Z}_n ]$ orbifold.
Massless states in twisted sectors arise in the $[X/U(1)]$ presentation
from massive states winding noncontractible loops in $X$, that shrink to
zero size when the $U(1)$ is gauged.

Thus, we see in this example that multiple presentations of the same
stack define the same CFT, albeit through different-looking sigma
models.

Note that if we supersymmetrize, it becomes more difficult to test
this equivalence directly.  This is because when we gauge a $U(1)$,
we get bosonic potentials such as D-terms, which give a manifestly
non-conformal theory.  In the supersymmetric case, the equivalence
of universality classes is actually more difficult to test than
in the bosonic case.

In section~\ref{check:openspectra}, we check that fractional branes
on $[{\bf C}^2/{\bf Z}_n]$ are also all realized on the alternate
presentation $[X/{\bf C}^{\times}]$ discussed in section~\ref{multpres}.
In a nutshell, fractional branes that in the presentation
$[{\bf C}^2/{\bf Z}_n]$ are supported at the fixed point of ${\bf Z}_n$
are realized in the alternative presentation
as D-branes supported on a codimension two subset of $X$,
invariant under the ${\bf C}^{\times}$.
Group actions on the Chan-Paton factors are naturally related between
the two presentations.
We show in section~\ref{openB} that open string spectra in the B model
can be expressed in terms of the stacks, not in terms of presentations,
so open string spectra are guaranteed to be presentation-independent.
As an explicit check of that statement, in section~\ref{check:openspectra}
we compute open string spectra between fractional branes on
$[ {\bf C}^2/{\bf Z}_n]$, in the two different presentations.
Although the calculational details differ, we do indeed get
the same result for the two presentations, as expected.

\subsubsection{Example:  distinct presentations of a trivial Calabi-Yau gerbe}
\label{trivgerbe}

Next, let us consider checking the presentation-independence of the physics
associated to another stack.
In particular, let us consider several different presentations of
the trivial ${\bf Z}_2$ gerbe over a space $X$.
Almost all of these presentations involve finite quotients,
and so yield CFT's immediately, so to test
presentation-independence we can simply compare partition functions.
(For multiple presentations as finite effectively-acting quotients,
presentation-independence of the partition functions is more or less
clear, but as noneffective group actions have not been carefully
studied in the physics literature, we think it useful to 
check some examples in detail.)
We will argue that each different-looking presentation ends up
defining the same partition function, at arbitrary genus,
albeit in very different ways.
Now, merely because the partition functions match does not uniquely
determine that the physical theories are the same,
but nevertheless this is an excellent test of our ideas.

One presentation of the trivial ${\bf Z}_2$ gerbe on $X$
is as a global quotient of $X$ by a trivially-acting
${\bf Z}_2$, {\it i.e.} $[X/{\bf Z}_2] = X \times B {\bf Z}_2$.

A second presentation of the trivial ${\bf Z}_2$ gerbe on $X$
is given by quotienting $X \times {\bf Z}_2
\times {\bf Z}_2$ by the eight-element group ${\bf H}$ formed
from the quaternions, with elements
\begin{displaymath}
{\bf H} \: = \: \{ \pm 1, \pm i, \pm j, \pm k \}
\end{displaymath}
This finite group projects to ${\bf Z}_2 \times {\bf Z}_2$, with kernel
${\bf Z}_2$, and can be described as a central extension:
\begin{displaymath}
1 \: \longrightarrow \: {\bf Z}_2 \: \longrightarrow \:
{\bf H} \: \longrightarrow \: {\bf Z}_2 \times {\bf Z}_2
\: \longrightarrow \: 1
\end{displaymath}
Let ${\bf H}$ act on $X \times {\bf Z}_2 \times {\bf Z}_2$ by first
projecting ${\bf H}$ to ${\bf Z}_2 \times {\bf Z}_2$, and then
act by left multiplication on the ${\bf Z}_2 \times {\bf Z}_2$.
The resulting stack $[ \left( X \times {\bf Z}_2 \times {\bf Z}_2
\right) / {\bf H} ]$ is another presentation of the trivial
${\bf Z}_2$ gerbe over $X$.

In the special case that there exists a space $W$ with a free
action of ${\bf H}$ (the finite group defined above), such that
$X = W / {\bf H}$,
there is yet another presentation of the trivial ${\bf Z}_2$ gerbe on $X$.
(An example of such a Calabi-Yau is discussed in \cite{beauville}.)
Define $Y = W / {\bf Z}_2$, where this ${\bf Z}_2$ is the center
of ${\bf H}$.
Let ${\bf H}$ act on $Y$ by first projecting ${\bf H}$ to
${\bf Z}_2 \times {\bf Z}_2$, and then let ${\bf Z}_2 \times
{\bf Z}_2$ act on $Y$ in the natural fashion.
In these circumstances, the trivial\footnote{
This gerbe is trivial because the atlas, a principal ${\bf Z}_2\times
{\bf Z}_2$-bundle over $X$, can be lifted to a principal ${\bf H}$-bundle
over $X$.  Since the characteristic class of the gerbe is defined via
the image of the characteristic class of the bundle in the part
of the long exact sequence given by
\begin{displaymath}
H^1(X, C^{\infty}({\bf H})) \: \longrightarrow \:
H^1(X, C^{\infty}({\bf Z}_2 \times {\bf Z}_2) ) \: \longrightarrow \:
H^2(X, C^{\infty}( {\bf Z}_2 ) )
\end{displaymath}
and since the characteristic class of the ${\bf Z}_2\times {\bf Z}_2$
bundle is itself the image of something, its image in $H^2$ must
vanish, so the gerbe must be trivial.
} ${\bf Z}_2$ gerbe on
$X$ can be described as $[Y/{\bf H}]$.

A fourth presentation of the trivial ${\bf Z}_2$ gerbe on $X$
can be obtained by starting with
$X \times {\bf C}^{\times}$,
and letting ${\bf C}^{\times}$ act on the ${\bf C}^{\times}$ factor
by rotating twice.
The resulting stack
$[\left( X \times {\bf C}^{\times} \right) / {\bf C}^{\times}]$
is another presentation of the trivial ${\bf Z}_2$ gerbe over $X$.

Now, let us consider how each of these different presentations
has the same partition functions.

The first presentation of the gerbe is also easy to understand.
In this presentation, we quotient $X$ by a ${\bf Z}_2$ that acts
completely trivially on $X$.  A one-loop partition function
is simply $| {\bf Z}_2 |^2/| {\bf Z}_2 | = 2$ copies of the 
one-loop partition function
for a sigma model on $X$.
Similarly, the $g$-loop partition function is
$| {\bf Z}_2 |^{2g}/| {\bf Z}_2 |^g = 2^g$ copies of the $g$-loop partition
function for a sigma model on $X$.
Since ultimately for a string compactification one wants to couple
to worldsheet gravity, such overall factors cannot be discarded
(just as cosmological constants cannot be renormalized away),
so we are led to believe that this theory is distinct from
a sigma model on $X$, as was discussed in much more detail in \cite{nr}.  
We shall check that alternate presentations give rise
to the same partition function (including numerical factors).

The second presentation is another quotient, but this time acting
on a disconnected atlas, with $| {\bf Z}_2 \times {\bf Z}_2 |$ components,
by a group whose effectively-acting part merely exchanges the
components.  First, note that a ${\bf Z}_2 \times {\bf Z}_2$ orbifold of
$X \times {\bf Z}_2 \times {\bf Z}_2$ is physically equivalent
to a sigma model on $X$.  After all, since $X$ is disconnected,
and the group action exchanges components, there can be no twisted
sector states at all (massless or massive), 
since it is not possible for the two ends of a 
connected string to lie on distinct disconnected components.  As a result,
all of the one-loop twisted sector contributions to the 
one-loop partition function necessarily vanish, except for the
\begin{displaymath}
{\scriptstyle \pm 1} \square_{ \pm 1} 
\end{displaymath}
twisted sectors.
Thus, the one-loop partition function of the orbifold is given by
\begin{eqnarray*}
Z_{orb} & = & \frac{1}{ | {\bf H} | } \sum_{g,h; gh=hg} {\scriptstyle g}
\square_h \\
& = & \frac{1}{| {\bf H} |} \sum_{g \in \{ \pm 1\} }
\sum_{ h \in \{ \pm 1\} } {\scriptstyle g} \square_h \\
& = & \frac{1}{2} \left( {\scriptstyle 1} \square_1 \right)
\end{eqnarray*}
since $\pm 1$ act trivially.  Now, since we are orbifolding not $X$,
but rather $X \times {\bf Z}_2 \times {\bf Z}_2$, the untwisted sector
above is not quite the same as the one-loop partition function of a
sigma model on $X$, but rather
\begin{displaymath}
{\scriptstyle 1} \square_1 \: = \: | {\bf Z}_2 \times {\bf Z}_2 | Z(X)
\end{displaymath}
so that we find that the one-loop partition function of the orbifold
is given by
\begin{displaymath}
\frac{1}{2} | {\bf Z}_2 \times {\bf Z}_2 | Z(X) \: = \: 2 Z(X)
\end{displaymath}
exactly the same as we obtained for the previous presentation of the
trivial gerbe.

We can repeat the analysis above for $g$-loop partition functions.
Since the cover is disconnected, and the group action exchanges components,
the only contributions to the $g$-loop partition function come from
twisted sectors with group elements in $\{ \pm 1 \}$.
Thus, the $g$-loop partition function of the orbifold is given by
\begin{displaymath}
\frac{ | {\bf Z}_2 |^{2g} }{ |{\bf H} |^g} \: = \: \frac{ 2^{2g} }{ 8^g} \: = \:
\left( \frac{1}{2} \right)^g
\end{displaymath}
times the $g$-loop partition function of the sigma model on
$X \times {\bf Z}_2 \times
{\bf Z}_2$.  Since the $g$-loop partition of the theory on
$X \times {\bf Z}_2 \times {\bf Z}_2$ is just $| {\bf Z}_2 \times
{\bf Z}_2 |^g = 4^g$ times\footnote{Consider a theory on the disjoint union of
$n$ copies of $X$.  This theory has $n$ times as many states as a sigma
model on $X$ -- one set of states for each copy of $X$.
Since in a genus $g$ partition function one has states propagating on each
of $g$ loops, the result is that a genus $g$ partition function when the
target is $n$ copies of $X$, should be $n^g$ times the genus $g$ partition
function for $X$.} the $g$-loop partition function for $X$,
the result is that the $g$-loop partition function for this orbifold
is $4^g/2^g = 2^g$ times the partition function for $X$, exactly matching
the result for the previous presentation.

The third presentation of the trivial gerbe is more interesting.
Since ${\bf H}$ acts freely on $W$, we can think of a sigma model
on $X$ as an ${\bf H}$-orbifold of a sigma model on $W$,
{\it i.e.} $X = [W/{\bf H}]$, and similarly a sigma model on $Y$
can be viewed as a ${\bf Z}_2$-orbifold of $W$,
{\it i.e.} $Y = [W/{\bf Z}_2]$. 
We want to consider the quotient $[Y/{\bf H}]$ or,
equivalently, $[ [W/{\bf Z}_2] / {\bf H}]$.
In other words, we want to orbifold an orbifold.

For simplicity,
let us first consider the ${\bf Z}_2 \times {\bf Z}_2$ orbifold
of $[W/{\bf Z}_2]$, before considering the gerbe case.
In principle, quotienting an orbifold is relatively straightforward.
One would think that one can twist each of the twisted sectors of 
the original orbifold,
so Hilbert space sectors are now labelled by two group elements instead
of one, one from each of the two groups.  
However, we need to be slightly more careful.
In order to orbifold the orbifold, we must lift the ${\bf Z}_2 \times
{\bf Z}_2$ action on $Y = W/{\bf Z}_2$ to $W$, if we want to 
express everything in terms of twisted sectors on $W$ ultimately.
Ordinarily when we define an orbifold, we want honest group actions,
hence honest equivariant structures, which in this case would
mean an honest ${\bf Z}_2 \times {\bf Z}_2$-equivariant structure
on the total space of the principal ${\bf Z}_2$-bundle $W \rightarrow Y$.

However, no such honest equivariant structure exists.
Since ${\bf H}$ is the largest group of automorphisms of $W$ that
covers the action of ${\bf Z}_2 \times {\bf Z}_2$ on $Y$ and
commutes with the structure group of the bundle $W \rightarrow Y$,
a ${\bf Z}_2 \times {\bf Z}_2$-equivariant structure on $W$ would be
a homomorphism ${\bf Z}_2 \times {\bf Z}_2 \rightarrow {\bf H}$
splitting the natural projection 
${\bf H} \rightarrow {\bf Z}_2 \times {\bf Z}_2$. 
Since no such homomorphism exists, there cannot be an honest equivariant
structure. 

Since we are orbifolding a ${\bf Z}_2$ orbifold, 
{\it i.e.} a theory with a gauged ${\bf Z}_2$,
we do not need an honest ${\bf Z}_2 \times {\bf Z}_2$-equivariant
structure:  a ${\bf Z}_2$-projectivized equivariant structure
is sufficient.

Before considering the full $[ [ W/{\bf Z}_2]/{\bf H}]$ orbifold,
let us first consider a $[ [W/{\bf Z}_2]/{\bf Z}_2 \times {\bf Z}_2]$
orbifold.  Let $i$, $j$ denote the generators of ${\bf Z}_2 \times {\bf Z}_2$.
As mentioned above, in order to define this orbifold, we must pick
some lift of ${\bf Z}_2 \times {\bf Z}_2$ to the automorphism group ${\bf H}$
of W, but our lift need not (can not, in fact) respect the group
law on ${\bf Z}_2 \times {\bf Z}_2$.  It will become clear shortly
that the specific lift does not matter, so we shall choose
to lift $i \in {\bf Z}_2 \times {\bf Z}_2$ to $i \in {\bf H}$
and $j \in {\bf Z}_2 \times {\bf Z}_2$ to $j \in {\bf H}$.
Following \cite{ginsparg}[section 8.5], denoting a twisted
sector in the $[W/{\bf Z}_2]$ orbifold by 
\begin{displaymath}
{\scriptstyle a} \square_b,
\end{displaymath}
and a twisted sector in the ${\bf Z}_2 \times {\bf Z}_2$ orbifold by
\begin{displaymath}
{\scriptstyle m} \square_n\,{'},
\end{displaymath}
we derive the relations that
\begin{eqnarray*}
{\scriptstyle 1} \square_1\,{'} & = & \frac{1}{2}
\left( {\scriptstyle +1} \square_{+1} \: + \:
{\scriptstyle -1} \square_{+1} \: + \:
{\scriptstyle +1} \square_{-1} \: + \:
{\scriptstyle -1} \square_{-1} \right) \\
{\scriptstyle i} \square_1\,{'} & = & \frac{1}{2}
\left( {\scriptstyle +i} \square_{+1} \: + \:
{\scriptstyle -i} \square_{+1} \: + \:
{\scriptstyle +i} \square_{-1} \: + \:
{\scriptstyle -i} \square_{-1} \right) \\
{\scriptstyle i} \square_j\,{'} & = & \frac{1}{2}
\left( {\scriptstyle +i} \square_{+j} \: + \:
{\scriptstyle -i} \square_{+j} \: + \:
{\scriptstyle +i} \square_{-j} \: + \:
{\scriptstyle -i} \square_{-j} \right)
\end{eqnarray*}
and so forth.
Note in particular that the one-loop twisted sector
\begin{displaymath}
{\scriptstyle i} \square_j\,{'}
\end{displaymath}
vanishes, because none of the
twisted sectors in its expansion close as twisted sectors in an
${\bf H}$ quotient of $W$, {\it i.e.} none of the pairs commute in 
${\bf H}$.
It is now easy to check that the full one-loop partition function of
the $[[W/{\bf Z}_2]/{\bf Z}_2 \times {\bf Z}_2]$ orbifold is given by
\begin{displaymath}
\frac{1}{| {\bf Z}_2 \times {\bf Z}_2|} 
\sum_{a,b \in {\bf Z}_2 \times {\bf Z}_2}
{\scriptstyle a} \square_b\,{'} \: = \:
\frac{1}{|{\bf H}|} \sum_{g,h \in {\bf H}, gh=hg} 
{\scriptstyle g} \square_h
\end{displaymath}
In other words, the one-loop partition function of
$[[W/{\bf Z}_2]/{\bf Z}_2 \times {\bf Z}_2]$ exactly
matches that of $[W/{\bf H}]$, exactly as one would expect.

Now, let us turn to the ${\bf Z}_2$ gerbe, presented as the
$[[W/{\bf Z}_2]/{\bf H}]$ orbifold.
${\bf H}$ acts on the $[W/{\bf Z}_2]$ by first projecting to
${\bf Z}_2 \times {\bf Z}_2$, which then acts as above.
As an ${\bf H}$ orbifold, we omit one-loop twisted sectors
involving group elements that do not commute, but it is straightforward
to check that the one-loop twisted sectors which the
${\bf H}$ orbifold omits are the same ones that vanish in the
description of the ${\bf Z}_2 \times {\bf Z}_2$ orbifold above.
The other remaining difference is that there is a 
multiplicity of $| {\bf Z}_2 |^2$ states.
Thus, the one-loop partition function of
$[[W/{\bf Z}_2]/{\bf H}]$ is given by
\begin{displaymath}
\frac{1}{|{\bf H}|} | {\bf Z}_2 |^2 
\sum_{a,b \in {\bf Z}_2 \times {\bf Z}_2}
{\scriptstyle a} \square_b\,{'} \: = \:
\frac{ | {\bf Z}_2 |}{|{\bf H}|} 
\sum_{g,h \in {\bf H}, gh=hg}
{\scriptstyle g} \square_h
\end{displaymath}
which is exactly twice the one-loop partition function of 
$[W/{\bf H}] = X$, exactly matching previous presentations of this
gerbe.

It is now a completely straightforward exercise to check that
at genus $g$, the $g$-loop partition function of $[[W/{\bf Z}_2]/{\bf H}]$
is $| {\bf Z}_2 |^g$ times the $g$-loop partition function of
$[W/{\bf H}] = X$, exactly matching previous presentations of this
gerbe.

Finally, let us consider the fourth presentation of the trivial gerbe,
as a $U(1)$-gauged sigma model on $X \times U(1)$, where
the $U(1)$ action rotates the $U(1)$ twice, and leaves $X$ invariant.
Here we cannot make statements quite as clean or explicit
as for 
presentations by finite quotients, but some things can still be said.
For example, since the $U(1)$ action rotates the $U(1)$ twice,
the gauge field has holonomies $\exp(\int A dx)$
around the worldsheet, which lie in the trivially-acting ${\bf Z}_2$.
Thus, for each loop of the worldsheet, one might expect the partition function
to contain a factor of $2$, corresponding to field configurations
differing solely by that holonomy, so at genus $g$ one would
expect the partition function of the theory to have a factor of
$2^g$ relative to a sigma model on $X$, which agrees with
the calculations above for presentations by finite quotients.
See also \cite{edwzw}[section 3.1] for some related general statements.

\subsection{The topological A and B models on stacks}
\label{tftstx}

The reader should now find it very straightforward to
understand the topological A and B models on stacks.
After all, a sigma model on a stack, for a fixed presentation,
is just a gauged sigma model on an ordinary space,
and A and B twists of gauged sigma models are well-known.

The statement of presentation-independence of universality classes
of RG flow is slightly more interesting in this context.
Although we have primarily spoken about {\it Calabi-Yau} stacks
classifying universality classes of RG flow, 
we will see in several exmaples in \cite{glsm} that the same
statement applies to A-twisted TFT's for more general stacks.

In passing, we should also mention that the B model constraint,
namely that the
B model is only well-defined for Calabi-Yau targets due to an anomaly,
naturally generalizes here.  Recall that a Calabi-Yau stack has the 
property that any presentation as a global quotient by a group,
has the property that the covering space is Calabi-Yau and the
group preserves the holomorphic top form.  That is precisely the constraint
for the gauged B model to be well-defined with respect to the usual
anomaly condition.  Moreover, if any presentation has that form,
then the stack is necessarily Calabi-Yau.  Thus, the B model
on a stack satisfies the usual anomaly constraint precisely when
the target is a Calabi-Yau stack, partially justifying the notion of
``Calabi-Yau stack'' used in this paper.

We shall study the closed string A model in more detail in
\cite{glsm}, where we see a number of results for correlation
functions, encoded in quantum cohomology relations.
The open string B model we shall consider in more detail later in this paper.

\section{Closed string spectra}   \label{closedspectra}

\subsection{The basic conjecture}

There is a conjecture (see {\it e.g.} \cite{agv,cr})
that the closed string spectrum of the CFT associated
to a Calabi-Yau stack is given, additively,
by the de Rham cohomology of the associated inertia stack.
We will argue that this conjecture is correct.

For the case of stacks presentable as global quotients by finite
effectively-acting groups, this is known to agree with the usual
result for the massless spectrum of an orbifold, as we shall
review in the next section.

For the case of stacks presentable as global quotients by finite
{\it noneffectively}-acting groups, to our knowledge no serious
work exists in the physics literature, but we argued in \cite{nr} 
that this result
is still true, {\it i.e.} that the spectrum calculation for
ordinary orbifolds generalizes in the obvious way to noneffective orbifolds.  
We checked this by examining implications of partition function factors
for string propagator pole counting, and by examining alternative
attempts at spectrum calculations.  
A subtle, and very serious, potential difficulty revolves around
deformation theory:  with this result for the massless spectrum calculation,
there are significantly more physical deformations of the CFT
than there are mathematical deformations of the stack.
We will examine this mismatch in greater detail in section~\ref{defthy},
and especially in \cite{glsm}, where we will construct new abstract CFT's
that realize noneffective twist field vevs, and where we will see that
these abstract CFT's and these nonmathematical physical deformations
play a crucial role in understanding mirror symmetry.

For stacks {\it not} presentable as global quotients by discrete groups,
we have to work harder still.  For gauged sigma models where one
gauges a nonfinite group, it is well-known (from experience with
gauged linear sigma models, for example) that the complete massless
spectrum of the IR fixed point is not typically directly accessible in the UV
theory -- one cannot explicitly write down vertex operators,
as one does not have Lagrangian access to the CFT, and although
in some special cases ({\it e.g.} Landau-Ginzburg orbifolds)
there are tricks for computing the spectrum,
there are no general methods known that work for all cases.  
Furthermore, since no one understands whether V-manifolds\footnote{
Only some special presentations of stacks can be concretely realized
in physics; it is not clear at present whether V-manifolds are 
examples of such physically-relevant presentations, as discussed earlier.
This is the reason why we recast local orbifolds as global quotients.} can be
used to directly define a CFT, no formal calculation based on
V-manifolds can be used to give a physically sensible spectrum calculation.
Thus, we appear to have a problem, in that
no direct calculation exists that can 
explicitly confirm this conjecture.
Despite that fact, we do have some very strong evidence for this 
massless-spectrum conjecture.
Specifically, in examples \cite{glsm}
in which
the sigma model on a presentation takes the form of a
gauged {\it linear} sigma model,
we can compute massless spectra at Landau-Ginzburg
points (contingent on our results for massless spectra in quotients by
finite non-effectively-acting groups).
Although Landau-Ginzburg spectra
need not precisely match large radius spectra, at least they typically
have the same form.  In such examples we shall find in \cite{glsm}
that Landau-Ginzburg
spectra do have exactly the same form as that predicted by the 
massless-spectrum conjecture above for
large radius points, giving very strong evidence for that conjecture.

Another strong test comes from quantum cohomology calculations presented
in \cite{glsm}.  Some of the older methods that physicists have used
historically to derive quantum cohomology do not require first knowing
the additive part of the ring, so one can compute the additive part by
first computing the quantum cohomology.  In examples of the A model on
toric stacks discussed in \cite{glsm}, we shall see that the
quantum cohomology ring is a product structure on the de Rham cohomology
of the associated inertia stack, in complete agreement with the
massless spectrum conjecture described above.

In addition, there are indirect tests.
For example, we will see in
\cite{glsm} that Witten indices are computed by the Euler characteristic
of the associated inertia stack, giving further evidence for this conjecture.

Understanding the massless spectrum of IR fixed points of gauged sigma
models also gives us predictive power for local orbifolds.
In this paper, we have argued that local orbifolds can be understood
concretely in physics by recasting them as global quotients of
(larger) spaces by (bigger) groups.
The local orbifold itself should be understood as the conformal fixed
point of worldsheet renormalization group flow of such globally
gauged sigma models.  For the reasons discussed above,
a direct calculation of the massless spectrum of such a conformal fixed
point is sometimes tricky, but the massless spectrum conjecture described
here gives us predictive power.

In passing, it is important to note that the massless spectrum
conjecture above implicitly assumes in its form that universality
classes are classified by stacks.  After all, the inertia stack depends
only upon the stack, not upon any presentation thereof, so saying that
the massless spectrum is computed as the cohomology of the inertia
stack implicitly assumes that the IR physics depends upon stacks and
not presentations.  Although we believe this conjecture is true,
and provide a great deal of evidence to support it,
we nevertheless think it is important to emphasize that it is
a conjecture, since 
a direct explicit verification is in principle impossible
with existing technology.

Let us also again mention that this same statement about massless
spectra of strings on stacks has been implicitly assumed in some
recent mathematical work on stacks.  However, we must emphasize that
this has been merely a conjecture -- the mathematics papers in question
did not include a physically valid calculation of the spectrum,
but rather arrived at the conjecture via other means.
The only cases in which the physics
was thought through before this paper were the special cases of
global quotients by finite effectively-acting groups.

\subsection{Quotients by finite effectively-acting groups}

Let us take a moment to review why the massless spectrum of
a traditional orbifold is given by the cohomology of the
inertia stack.

In an orbifold $[X/G]$, where $G$ is finite and acts effectively
on $X$, recall the standard result that the closed string massless spectrum
is given additively by
\begin{displaymath}
\sum_{[g]} H^*\left(X^g, {\bf C}\right)^{Z(g)}
\end{displaymath}
where the sum is over conjugacy classes of elements in $G$,
$X^g$ denotes the subset of $X$ invariant under $g \in G$,
and $Z(g)$ denotes the centralizer of an element $g \in G$
representing some cohomology class.

Now, it is a standard fact that 
for any quotient stack $[X/G]$,
\begin{displaymath}
H^*\left( [X/G], {\bf C} \right) \: = \:
H^*\left(X, {\bf C}\right)^G
\end{displaymath}
{\it i.e.} the de Rham cohomology of $[X/G]$ is the same as the
$G$-invariant part of the de Rham cohomology of $X$.

Thus, we can rewrite the expression for the closed string massless spectrum
as
\begin{displaymath}
\sum_{[g]} H^*\left( [X^g/Z(g)], {\bf C} \right)
\: = \: H^*\left( \bigoplus_{[g]} [X^g/Z(g)] , {\bf C} \right)
\end{displaymath}

However, the inertia stack $I$ associated to the quotient stack $[X/G]$
is given by
\begin{displaymath}
I \: = \: \bigoplus_{[g]} [X^g/Z(g)]
\end{displaymath}
so we immediately see that the standard expression for closed string
massless spectra can be rewritten as the cohomology of the inertia stack.

Now, upon further reflection, the reader might be slightly disturbed
by this result, for the following reason.
The closed string massless spectrum of a sigma model on a space $X$
is given by the cohomology of $X$; however, for stacks,
we see here that the closed string massless spectrum is {\it not}
given by the cohomology of the target, but rather by the cohomology
of some other stack entirely.
The reason this does not contradict the claim that a string orbifold
is a sigma model on a quotient stack was discussed in 
\cite{meqs}.  In a nutshell, closer examination of the reasoning
behind massless spectrum calculations reveals that one takes the
cohomology of the zero-momentum part of the loop space of the target.
After all, mechanically we construct the massless spectrum
by acting on the Fock vacuum with operators corresponding to the
zero modes of a Fourier expansion of the worldsheet fields.
When the target is an ordinary manifold, the zero-momentum part of the
loop space is the same as the target, hence the massless spectrum
is the cohomology of the target.  When the target is a
quotient stack by a finite effectively-acting group, the
zero-momentum part of the loop space is {\it not} the target,
but rather the inertia stack associated to the target.
Thus, the fact that the massless spectrum is given by the cohomology
of the inertia stack, and not by the cohomology of the target stack,
is consistent with the description of string orbifolds as 
sigma models on quotient stacks.

\subsection{Quotients by finite noneffectively-acting groups}
\label{spectra:noneff}

The inertia stack of $[X/G]$ for $G$ finite but noneffectively-acting
has the same form as for $G$ finite and effectively-acting, namely
\begin{displaymath}
I \: = \: \bigoplus_{ [g] } [X^g / Z(g) ]
\end{displaymath}
We argued in \cite{nr} that for $G$ finite and noneffectively-acting,
the closed string massless spectrum has the same form as for
$G$ effectively-acting, namely
\begin{displaymath}
\bigoplus_{[g]} H^*\left(X^g; {\bf C}\right)^{Z(g)}
\end{displaymath}
and so the massless spectrum can be understood as cohomology of the
inertia stack, just as described in the previous section.

We should emphasize that it is not immediately obvious that the
massless spectrum for $G$ finite but noneffectively-acting is the
same as for $G$ effectively-acting; a significant portion of
\cite{nr} is spent checking technical details.

A more mathematical objection to this particular result for the spectrum
comes from thinking about deformation theory.  Mathematically,
a gerbe over a manifold $M$ has only as many deformations as $M$.
On the other hand, since the
physical moduli are determined by the massless spectrum, a
physical theory on, say, $[X/{\bf Z}_k]$ where the ${\bf Z}_k$ acts
trivially has $k$ times as many moduli as $X$.  Thus, according to the
result of the massless spectrum calculation of \cite{nr}, 
a physical theory on a gerbe will typically have many more moduli
than there are mathematical moduli of the gerbe.  We will resolve this
discrepancy in section~\ref{defthy}.  In a nutshell, those 
`extra' physical deformations take us to some new abstract CFT's,
of a form not previously studied, which we will be able to 
understand explicitly, and which play a crucial role in understanding
mirror symmetry for stacks.

\subsection{Quotients by nondiscrete groups}

A given stack may have presentations in the form of quotients by
both finite and nonfinite groups.  We have seen that massless spectra
of presentations as quotients by finite groups 
are computed by the inertia stack associated to the original stack,
so for consistency, {\it i.e.} for universality classes of gauged sigma
models to be classified by stacks and not presentations thereof,
one would like for massless spectra of presentations as quotients by
nonfinite groups to also be counted by cohomology of the inertia stack.
After all, the inertia stack is defined by the stack itself,
not by a presentation thereof.
This was part of the reasoning behind the conjecture that
massless spectra of IR fixed points of gauged sigma models should
be counted by cohomology of the inertia stack.

Unfortunately, a direct calculation of massless spectra
of gauged sigma models in which a nonfinite group is gauged is not
currently possible.  
In gauged linear sigma models, for example, it is well-known that
only a subset of the IR states
are accessible directly in the UV theory.
For gauged WZW models, a direct computation was attempted in
\cite{edwzw}[section 2.2], but the author was careful to state
that extra branches of semiclassical phase space, arising for
various degenerate flat connections on the worldsheet, could give
contributions to massless spectra, which he did not understand.

Nevertheless, 
although a direct computation is not currently possible,
there are very strong indirect tests of that statement.
For example, in \cite{glsm}, we will find that many Calabi-Yau gerbes,
which themselves cannot be presented as global quotients by finite
groups, have associated Landau-Ginzburg orbifold theories.
Although the large radius spectrum may not be directly calculable,
the Landau-Ginzburg orbifold spectrum is directly calculable using
standard methods.  Now, in general the Landau-Ginzburg orbifold spectrum
and the large radius massless spectrum need not agree, but in typical
cases they are closely related, and for the Landau-Ginzburg orbifolds
associated to Calabi-Yau gerbes studied in \cite{glsm},
we find that the Landau-Ginzburg orbifold spectrum is computed by
cohomology of the inertia stack.  Again, this is not a direct computation,
but it is a very strong indirect check of the assertion.

Another strong test involves the quantum cohomology computations
in \cite{glsm}.  The quantum cohomology ring that we derive in that reference
for gerbes on projective spaces turns out to be a product structure on
the cohomology of the inertia stack.

There are also some weaker tests that this conjecture satisfies.
For example, for toric Fano stacks, we see in
\cite{glsm} that the Witten index is
the Euler characteristic of the inertia stack.

In any event, given the indirect tests outlined above and described in
more detail in \cite{glsm}, we are led to believe that massless spectra
of IR fixed point theories are given by the cohomology of the inertia stack.

\subsection{Local orbifolds}

In section~\ref{locorb:mathexs} we discussed some simple examples of stacks
which are local orbifolds, {\it i.e.} stacks which are not globally
quotients by finite groups, but which have local quotient structures.
We can describe such stacks as quotients of larger spaces by nonfinite
groups, and if our conjecture that stacks classify universality classes
of gauged sigma models is correct, then physically we can now describe
local orbifolds -- as IR fixed points of certain other 
globally gauged sigma models.

Let us describe the inertia stack for these examples, to help the reader
understand how the inertia stack should be related to massless spectra.

The first example we discussed looked like $[T^4/{\bf Z}_2]$,
with some of the quotient points blown up.
The inertia stack of $[T^4/{\bf Z}_2]$ itself is given by
\begin{displaymath}
I \: = \: [T^4/{\bf Z}_2] \coprod_1^{16} B{\bf Z}_2
\end{displaymath}
where $B{\bf Z}_2 = [\mbox{point}/{\bf Z}_2]$ and we have one copy
of $B {\bf Z}_2$ for each fixed point of the ${\bf Z}_2$ action on $T^4$.
The cohomology of $I$ is given by the ${\bf Z}_2$-invariant part of the
cohomology of $T^4$, plus one ${\bf C}$ for each $B{\bf Z}_2$, precisely
duplicating twist field counting.  (Also recall that the grading in twisted
sectors is shifted, because of the boundary conditions on the
worldsheet fields, so that each ${\bf C}$ from a $B{\bf Z}_2$ contributes
in degree one, rather than degree zero.)

The local orbifold corresponding to $T^6/{\bf Z}_2$ with, say, $k$ singular
points blown up and the remaining points given stacky structures,
we labelled ${\cal Y}$, and the inertia stack of ${\cal Y}$ is given by
\begin{displaymath}
I \: = \: {\cal Y} \coprod_1^{16-k} B {\bf Z}_2
\end{displaymath}
Again, for each local ${\bf Z}_2$ quotient structure, we get a contribution
to the cohomology.  In the case $[T^4/{\bf Z}_2]$, we could understand
such contributions as twist fields in the traditional sense.
Now, since we can only understand the local orbifold physically
as the IR fixed point of a gauged sigma model with a nonfinite group gauged,
we cannot really speak of these $B{\bf Z}_2$ contributions as twist
fields in the usual sense, though intuitively that is what they are doing.

The second example we discussed in section~\ref{locorb:mathexs}
was a local orbifold ${\cal Y}$ based on $[T^6/{\bf Z}_3]$, in which some
of the ${\bf Z}_3$ singularities were blown up.
This example follows the same pattern as the example above.
The inertia stack of the basic $[T^6/{\bf Z}_3]$ orbifold is given by
\begin{displaymath}
I \: = \: [T^6/{\bf Z}_3] \coprod_1^{27} \left( B {\bf Z}_3 \coprod B {\bf Z}_3 \right)
\end{displaymath}
where we get a $B {\bf Z}_3 \coprod B {\bf Z}_3$ from each of the $27$
fixed points of the ${\bf Z}_3$ action on $T^6$.
The cohomology then is given by the ${\bf Z}_3$-invariant part of the
cohomology of $T^6$, plus a ${\bf C}^2$ from each $B{\bf Z}_3 \coprod
B {\bf Z}_3$.

For the local orbifold ${\cal Y}$, with $k$ points blown up,
the inertia stack is given by
\begin{displaymath}
I \: = \: {\cal Y} \coprod_1^{27-k} \left( B {\bf Z}_3 \coprod B {\bf Z}_3 \right)
\end{displaymath}
where again, each remaining stacky point contributes a 
$B {\bf Z}_3 \coprod B {\bf Z}_3$ to the inertia stack,
so that the cohomology gets contributions that closely resemble 
twist fields in ordinary (global) orbifolds.

\section{Deformation theory of stacks}   \label{defthy}

One of the most important puzzles in understanding whether stacks are
physically relevant to gauged sigma models lies in comparing
mathematical deformation theory of stacks to physical deformation theory
of CFT's.  The mathematical deformations of a stack encode only
the {\it untwisted} sector moduli of the corresponding CFT,
and omit twist field moduli.  Specifically,
for the stack $[X/G]$ for $G$ finite, the deformations of the stack are merely
the $G$-invariant deformations of $X$, whereas it is well-known that
$G$-gauged sigma models can have more moduli than that.
In a string compactification
on a smooth Calabi-Yau manifold, the mathematical deformations of
the Calabi-Yau manifold match the physical deformations of the corresponding
sigma model, which suggests that this is a serious problem.
In examples, this appears even worse.  For example, the physical theory
given by the standard ${\bf Z}_2$ orbifold of ${\bf C}^2$ has physical
deformations to sigma models on both deformations and resolutions of
the quotient space ${\bf C}^2/{\bf Z}_2$, whereas by contrast the stack
$[{\bf C}^2/{\bf Z}_2]$ is rigid, admitting neither complex structure
deformations nor K\"ahler resolutions.  In noneffective orbifolds we will
see that the closed string massless spectrum conjecture typically predicts
far more physical moduli than either mathematical moduli of the stack
or geometric moduli of the underlying space, leaving one with a serious
puzzle concerning the interpretation of those moduli.
As the massless spectrum in noneffective orbifolds has not been
thoroughly studied in the physics literature before, this difficulty
can even lead one to question it.

We will resolve these difficulties in this section, by noticing
that at the conformal fixed point of a gauged sigma model
only some of the marginal operators lead to physical theories
for which geometry is a good description, and the mathematical
deformation theory of stacks is precisely counting those marginal
operators.

\subsection{Effective orbifolds:  a potential contradiction and a resolution}

One of the most basic problems in any attempt to understand
string compactifications on stacks is to understand how
physical deformations of the CFT correspond to deformations of the
stack.  For example, we know physically that the orbifold $[{\bf C}^2/
{\bf Z}_2]$ CFT has marginal operators that deform it to
sigma models on 
complex structure deformations and K\"ahler resolutions
of the quotient space ${\bf C}^2/{\bf Z}_2$, which is one of the bases
for typical physical interpretations of string orbifolds.

If a string orbifold CFT is the same thing as a sigma model on a stack,
then from our experience with sigma models on smooth Calabi-Yau's,
we expect those physical marginal operators
to correspond to deformations of the stack $[ {\bf C}^2/{\bf Z}_2]$.
In particular, the string orbifold CFT for $[{\bf C}^2/{\bf Z}_2]$
lies in a family of CFT's, which at other points describe sigma models
on deformations of ${\bf C}^2/{\bf Z}_2$.  If the string orbifold
is a sigma model on $[{\bf C}^2/{\bf Z}_2]$, then from this description
one would expect there must exist a family whose fibers are
mostly deformations of ${\bf C}^2/{\bf Z}_2$, but with
the singular quotient space ${\bf C}^2/{\bf Z}_2$ replaced by the
quotient stack $[ {\bf C}^2/{\bf Z}_2 ]$.

However, we have a problem.  There is no such family relating
the quotient stack to deformations of the quotient space.

It is easy to show that the stack $[{\bf C}^2/{\bf Z}_2]$ has
no infinitesimal complex structure moduli.
From the usual Kodaira-Spencer-based notions, applied to stacks,
deformations of the complex structure of a smooth Deligne-Mumford
stack ${\cal X}$ are
counted by elements of $H^1( {\cal X}, T {\cal X} )$,
just as complex structure deformations of a smooth Calabi-Yau
manifold $X$ are counted by $H^1(X, TX )$.
However, 
\begin{displaymath}
H^1\left( [ {\bf C}^2/{\bf Z}_2] , T [ {\bf C}^2 / {\bf Z}_2 ] \right) \: = \:
H^1\left( {\bf C}^2, T {\bf C}^2 \right)^{ {\bf Z}_2 }
\end{displaymath}
and ${\bf C}^2$ has no complex structure deformations,
much less ${\bf Z}_2$-invariant complex structure deformations.
This problem has nothing to do with compactness -- for example,
repeating the same calculation, we find that the only possible
complex structure deformations of the stack $[T^4/{\bf Z}_2]$
are those of the torus $T^4$ which are invariant under the ${\bf Z}_2$,
{\it i.e.} marginal operators arising in the untwisted sector.
We see in these examples that there are no complex structure deformations
of the quotient stacks corresponding to the marginal operators in twisted 
sectors, hence, we have a serious problem.

Not only are complex structure deformations a problem,
but so are K\"ahler parameters.  Although we can blow up
the singular space ${\bf C}^2/{\bf Z}_2$ to a Calabi-Yau, 
attempts to blow up
the stack $[ {\bf C}^2/{\bf Z}_2 ]$ give stacks with nontrivial
canonical bundle, which cannot be Calabi-Yau.
(See section~\ref{ex:nonCY} for an example of a stack constructed
as an attempt to blow up $[ {\bf C}^2/{\bf Z}_2 ]$.)
The mathematical problem is ultimately simply that the stack
$[ {\bf C}^2/{\bf Z}_2]$ is everywhere smooth, and although we
can blow up singularities without making the canonical bundle nontrivial,
any blowup of a smooth point makes the canonical bundle nontrivial.

In other words, not only do we have a problem matching complex structure
moduli of the stack $[ {\bf C}^2/{\bf Z}_2]$ to the CFT marginal operators,
but we also have a problem matching K\"ahler parameters of the stack
to CFT marginal operators.

One potential (and failed) workaround is as follows:
if a string on a gerbe
were somehow the same as a string on the underlying space, perhaps with
a $B$ field, then perhaps there exists a family relating
the quotient stack $[ {\bf C}^2/{\bf Z}_2 ]$ to some stacks over
deformations of ${\bf C}^2/{\bf Z}_2$, but with gerby structures.
That is essentially the structure seen above in the tentative
`blow up' of $[{\bf C}^2/{\bf Z}_2]$.
However, as we have seen implicitly throughout this paper and \cite{nr},
a string on a gerbe is not the same thing as a string on the underlying
space, even with a $B$ field.  They have different closed string massless
spectra, different open string boundary states, different open string massless
states, and so forth.

Another possible partial fix to this problem would be
to modify the definition of ``Calabi-Yau stack.''
If a stack could have nontrivial canonical bundle while sigma models
on the stack still had spacetime supersymmetry, then perhaps we could
at least match K\"ahler parameters to some of the CFT marginal operators.
Then our `blow up' of $[ {\bf C}^2/{\bf Z}_2]$ would be Calabi-Yau.
However, in our experience, attempts at alternative definitions
are not workable.  For example, as outlined in section~\ref{cystxdefn},
for open string B model spectra to be well-behaved, one really
wants the canonical bundle to be trivial.  Similarly, an analysis of
the physics of blowups of the stack (see section~\ref{ex:nonCYphysics})
reveals that physical theories on attempted blowups of the stack are not
spacetime supersymmetric.  Thus, we look elsewhere for the resolution
to this puzzle in matching stack deformations to CFT marginal operators.

We resolve this dilemma as follows.
First, recall that in a nonlinear sigma model
on a smooth Calabi-Yau manifold at large radius, 
where a geometric interpretation
of the CFT is sensible, both K\"ahler and complex structure deformations
result in other Calabi-Yau's at large radius.  Mechanically,
giving a vev to such a marginal operator $h$ amounts to adding
a term of the form 
\begin{displaymath}
\int_{\Sigma} \{ G_+, [ G_-, h] \}
\end{displaymath}
to the action, and this simplifies to give terms of the form
\begin{displaymath}
\int_{\Sigma} h_{i \overline{\jmath}} \partial X^i \cdot 
\partial X^{\overline{\jmath}} \: + \: \cdots
\end{displaymath}
This manifestly just describes a deformation of the metric in the
nonlinear sigma model, as the leading term above can be combined with
the kinetic term for the bosons $X^i$ to give a new kinetic term of the
form
\begin{displaymath}
\int_{\Sigma} \left( g_{i \overline{\jmath}} \: + \:
h_{i \overline{\jmath}} \right)
\partial X^i \cdot
\partial X^{\overline{\jmath}} 
\end{displaymath}
for $g_{i \overline{\jmath}}$ the original metric on the Calabi-Yau.
In other words, for smooth
Calabi-Yau manifold targets, the physical moduli yield other CFT's in which
a geometric interpretation is sensible.

In an orbifold by a finite freely-acting group, however,
many of the physical deformations do {\it not} give CFT's in which
a geometric interpretation is really sensible.  
Recall from section~\ref{sigmacoupling}
that we are able to give a pseudo-geometric interpretation to orbifolds
because sigma model perturbation theory is weakly coupled when the atlas
is at large radius.  If we give a vev to an untwisted sector modulus,
{\it i.e.} a $G$-invariant deformation of the covering space,
then we deform to a new CFT which also has a sensible large radius 
interpretation.  However, if we give a vev to a twist field modulus,
for example in $[ {\bf C}^2 / {\bf Z}_2 ]$, then the new CFT describes
a sigma model on a space with a very small rational curve, which receives
strong worldsheet instanton corrections.  
The new CFT is still a small deformation away from a weakly coupled
sigma model, namely the orbifold theory, but this new CFT can really no
longer be well-described by geometry, because of worldsheet instantons.
Mechnically, if we follow the same procedure as above,
by adding a term of the form
\begin{displaymath}
\int_{\Sigma} \{ G_+, [ G_-, h] \}
\end{displaymath}
to the action, where $h$ is now a twisted-sector marginal operator,
then the usual selection rule dictates that when we expand out
the exponential of this term in any given correlation function,
only a few powers can contribute, not all of the terms of the
exponential.  The resulting physical theory is no longer described
by anything as simple as a nonlinear sigma model with a deformed
metric, but rather now looks like a nonlinear sigma model in which
correlation functions are deformed by the addition of a factor containing
certain selected powers in the Taylor series expansion of the exponential
of the term above.  This is some abstract CFT, no longer purely of the form
of a nonlinear sigma model on some space,
reflecting the fact that worldsheet instanton
corrections are large, and so geometry is no longer a good description
of the physical theory.  

Strictly speaking, the only orbifold marginal operators that we can
add to the action that will result in a new theory that still has
the same form as a nonlinear sigma model are untwisted
sector marginal operators, which
describe $G$-invariant deformations of the covering space.
Any twisted sector marginal operators will result in the strange form
described above, an abstract CFT whose interpretation can only be obtained
by deforming to a limit in which worldsheet instanton corrections are small
and geometry is a good guide.

Thus, the only marginal operators that one should really expect to
have a mathematical interpretation are the $G$-invariant deformations
of the covering space $X$, for $G$-gauged sigma models on $X$ in which
$G$ is finite.  As it happens, the mathematical deformations of the
stack $[X/G]$, for $G$ finite, are precisely those, the $G$-invariant
deformations of $X$.

Our resolution of the puzzling mismatch between physical and mathematical
deformations of stacks is that the mathematical deformations encode
deformations from theories with good geometric interpretations
({\it i.e.} small worldsheet instanton corrections), to other
theories which also have good geometric interpretations.  
For a smooth Calabi-Yau
manifold target, all of the physical deformations of a weakly-coupled
sigma model result in nearby theories with good geometric interpretations,
{\it i.e.} descriptions as weakly-coupled nonlinear sigma models.
For an orbifold, on the other hand, only the untwisted sector moduli
yield new theories with good geometric interpretations; the twisted
sector moduli, which yield theories in which geometry is strongly corrected
by worldsheet instantons, do not appear as mathematical deformations of the 
stack.

There is an alternative resolution of this mismatch,
that stems from ideas that we should take D-brane probe arguments literally,
and that the ``correct'' stringy geometry of the $[ {\bf C}^2/{\bf Z}_2 ]$
orbifold is a resolution $\widetilde{ {\bf C}^2 / {\bf Z}_2 }$ of the
quotient space.  We shall see that this method does not work so well
for noneffective orbifolds, but as some readers may find this
line of thought fruitful, let us review the idea.

In this alternative resolution of the deformation mismatch,
one says that the physical deformations of the orbifold
should really be interpreted as deformations of the classical vacua
of the sigma model, which is the same as the target when the target
is a space, but may be slightly different for more general targets. 
In order to make this intuition precise, let us make the ansatz
that the space of classical vacua of the closed string should be
represented by the Quot scheme $\mbox{Quot}_1$ of the target.
(Recall this Quot scheme essentially counts skyscraper sheaves.)
When ${\cal X}$ is an actual space, $\mbox{Quot}_1 {\cal X} = {\cal X}$,
but for orbifolds, for example,
\begin{displaymath}
\mbox{Quot}_1 [ {\bf C}^2/{\bf Z}_2 ] \: = \:
\widetilde{ {\bf C}^2 / {\bf Z}_2 }
\end{displaymath}
Thus, if we identify CFT marginal operators with deformations of
$\mbox{Quot}_1$ of the stack, rather than with deformations of the stack
itself, we see that our difficulty is resolved.

This method is not perfect.  For example,
for a gerbe over a space $X$, the relevant component of
$\mbox{Quot}_1$ of the gerbe is just a copy of $X$,
the same as seen by D-brane probes.
However, we have argued that the correct massless spectrum calculation
for a gerbe on $X$ typically yields several copies of the moduli of $X$.
Thus, this alternative resolution of the deformation mismatch only
makes sense for finite effectively-acting groups.

Furthermore, $\mbox{Quot}_1$ can sometimes have some other odd features.
We only consider Deligne-Mumford stacks in this paper,
{\it i.e.} stacks in which the stabilizers of any fixed point are finite,
but if one considers the Artin stack
 $[{\bf C}^2/{\bf C}^{\times}]$,
where $(x,y) \mapsto (\lambda x, \lambda y)$ for $\lambda \in {\bf C}^{\times}$,
its $\mbox{Quot}_1$ is the  disjoint union of
${\bf P}^1$ plus a single point.

Although this alternative resolution of the deformation mismatch
puzzle is not perfect, it may provide some solace for those
readers who are particularly religious about their
D-brane probe interpretations.

\subsection{Calabi-Yau gerbes (noneffective quotients)}

Many Calabi-Yau gerbes seem to suffer from the same potential
contradiction as string orbifolds:  there are not enough complex
structure moduli to match all of the marginal operators in the CFT
that ought to describe deformations of complex structure.

For example, for a trivial ${\bf Z}_k$ gerbe over a space $X$,
namely $[X/{\bf Z}_k] = X \times B {\bf Z}_k$,
we have argued in section~\ref{spectra:noneff} that the massless spectrum
is $k$ copies of the cohomology of $X$.  The mathematical infinitesimal
moduli of the algebraic stack corresponding to this gerbe, however,
contain only one copy of the moduli of $X$.
If those extra physical moduli really do exist, if our spectrum calculation
is correct, then what can those extra moduli correspond to?

Our resolution of this puzzle is the same as for effectively-acting
groups:  the mathematical moduli correspond to deformations of the stack
which result in physical theories having well-behaved mathematical
interpretations.  The `extra' physical moduli result in physical
theories, abstract CFT's, which are not weakly-coupled nonlinear sigma models,
and so 
do not have clean mathematical interpretations.

A skeptic might wonder if instead we have managed to miscalculate
the physical massless spectrum, despite the large number of arguments
and calculations we have already provided.  However, these 
physical moduli can be seen explicitly -- the new abstract CFT's one
gets by deforming along such extra marginal operators can often be
described algebraically.

Begin with a Landau-Ginzburg model corresponding to a Calabi-Yau hypersurface,
so that the superpotential is the hypersurface polynomial.
As is well-known,
a marginal deformation of the theory corresponds to a deformation
of that superpotential by terms which do not change the degree of
homogeneity of the polynomial.

*123*

Now, construct a trivial ${\bf Z}_k$ orbifold of that Landau-Ginzburg
model.  As discussed previously in 
\cite{nr}
this is equivalent to adding a field $\Upsilon$ that takes values
in $k$th roots of unity.  According to our massless spectrum calculation,
we now have $k$ times as many moduli in the physical theory as before,
given by multiplying any vertex operator corresponding to a modulus
of the original theory by a power of $\Upsilon$.
Giving a vev to such a twisted sector modulus is equivalent to
adding a term to the Landau-Ginzburg superpotential which has a factor
of $\Upsilon$ to some power, since such terms are just supersymmetry
transformations of the relevant vertex operator.
Thus, we can see these new physical deformations very explicitly,
as {\it e.g.} Landau-Ginzburg superpotential terms with factors
of $\Upsilon$, the field valued in roots of unity.
As discussed in \cite{nr}, this is completely equivalent to conformal
perturbation theory for a twist field -- the sum over roots of unity in the
path integral measure enforces the orbifold selection rule, for example.

Ordinarily, giving a vev to a twist field is somewhat messy,
as the twist field introduces a branch cut, and as this changes the moding
of worldsheet fields, the resulting marginal operators are no longer
quite so simple to express.  In the present case, however,
since the twist field in question corresponds to a group element that
acts trivially, the moding of worldsheet fields does not change,
and so an algebraic description of the process of giving a vev to a
twist field, as we have outlined above, becomes possible.

These resulting abstract CFT's no longer seem to have a clean
relationship to geometry -- we do not think one can understand them as
weakly-coupled nonlinear sigma models on spaces.  Thus, there is
no good reason why mathematical deformation theory {\it should} see them,
and indeed it does not.

We will return to these abstract CFT's and study them more extensively
in \cite{glsm}, where they will be independently re-derived from a 
completely different
direction.  Here and in \cite{nr}, we have derived Landau-Ginzburg models 
with fields
taking values in roots of unity from considering physical deformations
of noneffective orbifold theories.  In \cite{glsm} we will find that the
same sort of Landau-Ginzburg theories appear when one builds mirrors
to stacks.  The fact that we are seeing these same physical theories
appear in a different context, is an excellent check that our
analysis is consistent.

In the previous section, we outlined how ideas stemming from D-brane
probes could give an alternative understanding of the mismatch
between physical and mathematical moduli.  For noneffective orbifolds,
however, that idea breaks down.  Motivated by D-brane probe ideas,
we made the ansatz that the physical moduli should be counted
by $\mbox{Quot}_1$ of the stack, which for orbifolds by finite
effectively-acting groups, gave us the correct answer.
However, for a Calabi-Yau gerbe ${\cal G}$ over a Calabi-Yau manifold $X$,
it can be shown\footnote{The structure sheaf of the gerbe,
as well as any quotient thereof, is of weight zero, a pullback from
the manifold $X$.  The only way to get ideals not from $X$, is for the
corresponding quotient of the structure sheaf to have nonzero weight,
but that can never happen.} that
\begin{displaymath}
\mbox{Quot}_1 {\cal G} \: = \: X
\end{displaymath}
Unfortunately, the number of physical moduli associated to a gerbe
is usually much larger than the number of moduli of the underlying manifold,
as we have discussed.
Thus, this alternative resolution, 
which worked for orbifolds by finite effectively-acting
groups, fails for noneffectively-acting groups.

\subsection{B fields at orbifold points}

An important part of the string duality story is the idea that
``string orbifolds describe strings with B fields at quotient singularities''
\cite{edstr95,paul95}.  In particular, this is how string theory distinguishes
singular points in moduli space at which the low-energy effective theory gets a
nonperturbatively enhanced gauge symmetry, corresponding to badly-behaved
CFT's, from points in moduli space corresponding to string orbifolds, at which
the CFT's are well-behaved.  Although in principle both can be associated
to quotients, physically they differ in the B field.

To be specific, let us focus on $[ {\bf C}^2/{\bf Z}_2 ]$, which in any
event is the only case in which the physics is understood.
To be precise, the ``value of the B field'' referred to above is the
monodromy of the B field around the exceptional divisor of
the minimal resolution of ${\bf C}^2/{\bf Z}_2$.
Of course, that exceptional divisor has shrunk to zero size 
at the points in moduli space where a nonperturbatively enhanced gauge symmetry
might occur.  Nevertheless, despite the fact that the exceptional divisor
has zero size, one can still talk about the monodromy of the B field.
This is really because such discussions assume one is
doing sigma model perturbation theory in the
curvature of the resolution, and so we are 
talking about a regime in the moduli space
where worldsheet instanton corrections are strong, so speaking of geometry
at all is misleading in that expansion parameter.  
The fact that we physically appear to have a 
nonzero monodromy
of a B field on a divisor that is no longer present, is an artifact 
of the worldsheet instanton corrections, a signature of the fact that
we are doing perturbation theory with a parameter that is no longer small.

Perhaps the most nearly correct way to understand this 
B field business is to say that as the exceptional divisor
shrinks in size, the limiting value of the B field monodromy is nonzero,
as seen in perturbation theory on the curvature of the resolution
$\widetilde{ {\bf C}^2/{\bf Z}_2 }$.

In perturbation theory in the curvature of the covering space ${\bf C}^2$,
the approach taken in this paper, 
in which the string orbifold is weakly coupled, 
this ``B field at orbifold points''
business is invisible, and relating these two perspectives is an
old technical problem.

A possible understanding of this phenomenon at the orbifold point,
in perturbation theory in the curvature of the covering space, was
pointed out in \cite{meqs}.  
One sensible-seeming way to understand the
blowup of $[ {\bf C}^2/{\bf Z}_2 ]$ is as the quotient
\begin{equation}  \label{blowup}
\left[ \frac{\mbox{Bl}_1 {\bf C}^2 }{ {\bf Z}_2 } \right]
\end{equation}
where the ${\bf Z}_2$ is extended over the exceptional divisor.
This stack naturally lies over the resolution
$\widetilde{ {\bf C}^2/{\bf Z}_2 }$,
and has the structure of a ${\bf Z}_2$ gerbe over the exceptional divisor.
Given the fact that a sheaf on a gerbe is a twisted sheaf on the underlying
space, twisted in the sense of B fields, one can plausibly speculate
that a closed string on a gerbe should be related to a closed string
on the underlying space, but with a nonzero B field, whose holonomy
is determined by the gerbe.  In such an event, in the present example,
since we have a ${\bf Z}_2$ gerbe over the exceptional divisor,
it would be very plausible that a string on this stack would be
a string on the resolution $\widetilde{ {\bf C}^2/{\bf Z}_2 }$,
but with a nonzero holonomy on the exceptional divisor.
Thus, blowing up the stack $[ {\bf C}^2/{\bf Z}_2 ]$, in the sense above,
would lead to physics seeing a string on $\widetilde{ {\bf C}^2/{\bf Z}_2 }$,
with nonzero B field holonomy, exactly as seen physically.

Unfortunately, after further analysis, this proposal does not work out.
There are several reasons why this fails:
\begin{enumerate}
\item The stack given above in equation~(\ref{blowup}) is not a Calabi-Yau
stack, as discussed in section~\ref{ex:nonCY},
and so we should not consider string compactifications on it,
at least for the purposes of this paper. 
\item In order for the ${\bf Z}_2$ gerbe on the exceptional divisor
to uniquely determine a B field holonomy, we must specify a map
${\bf Z}_2 \rightarrow U(1)$.  Now, although there are only two such maps,
neither emerges physically in any fashion.  One might guess that for a 
nonobvious reason, the nontrivial one should be taken, but then, as soon
as one tries to generalize the procedure above to other quotient
singularities, one gets more complicated situations where there is even
less reason to make one particular choice.
\item Perhaps the most important objection to the proposal above is that,
unlike sheaves, a closed string propagating on a gerbe is not the same as
a closed string on a space with a nonzero B field.
In this paper, we have now seen numerous examples of strings on gerbes,
and judging from one-loop partition functions, 
massless spectra, {\it etc},
the theory of a closed string on a gerbe is very different from that of
a closed string on a space with a nonzero B field.
\end{enumerate}

Thus, we do not at present have a derivation of 
``B fields at orbifold points'' in the language of stacks.
We suspect such a derivation may not be possible, for the same
reasons that mathematical deformations of stacks only see those 
physical deformations that yield theories in which geometry is a useful
tool.  Since the blowup moduli are twist fields, which yield
CFT's in which worldsheet instanton corrections are strong and
geometry a poor approximation, understanding ``B fields at orbifold points''
in terms of the geometry of stacks may not be possible.

Instead, let us observe that the results of this paper give us
an alternate perspective on this issue.
Instead of thinking of ``B fields at orbifold singularities'' as the
fundamental reason why string orbifold CFT's are well-behaved,
perhaps the ``B field'' in such cases is a consequence of quantum corrections,
and there is another more nearly fundamental reason why the CFT
is well-behaved.

In particular, there are certainly other examples in which this
standard lore does not make much sense.  For example, in nonsupersymmetric
orbifolds one example often studied is the quotient
$[ {\bf C} / {\bf Z}_n ]$, where the generator of ${\bf Z}_n$ acts
on the holomorphic coordinate by multiplication by phases.
From the older perspective, this case is confusing for a variety of
reasons:
\begin{enumerate}
\item As algebraic varieties, the quotient space ${\bf C}/{\bf Z}_n$
is indistinguishable from the complex line ${\bf C}$.
In more detail, the algebraic functions on ${\bf C}/{\bf Z}_n$
are the ${\bf Z}_n$ invariant part of the coordinate ring
${\bf C}[x]$, but since $x$ is acted upon by $n$th roots of unity,
the invariant ring is given by ${\bf C}[x^n]$, which is isomorphic
as a ring to ${\bf C}[x]$.  Thus, as algebraic varieties,
${\bf C}/{\bf Z}_n$ and ${\bf C}$ are indistinguishable.
Physically, the two cases are very different -- 
one is a nonsupersymmetric orbifold, the other is a supersymmetric
free field theory.
Now, by contrast, as algebraic stacks, $[ {\bf C} / {\bf Z}_n ]$ is
very different from ${\bf C}$, so the fact that the physical theories
are distinct could be viewed as evidence for the physical relevance
for stacks.  On the other hand,
a skeptic might point out that this could be because differential
geometry is pertinent to physics,
and certainly
in differential geometry ${\bf C}/{\bf Z}_n$ is distinct from
${\bf C}$.  However, there are other puzzles....
\item There is no blowup mode associated with ${\bf C}/{\bf Z}_n$,
no obvious place to encode a ``B field at an orbifold singularity.''
If the fundamental reason why string orbifolds are well-behaved
CFT's, unlike sigma models on singular spaces, is that B field,
then we appear to have a contradiction, since a B field cannot be
encoded here, and yet the CFT is well-behaved.
\end{enumerate}

Analogous puzzles appear in other limits.
For example, the quotient ${\bf C}^4/{\bf Z}_2$ is a terminal singularity
 -- although it admits blowup modes, there are no blowup modes that
preserve the Calabi-Yau structure.  The string orbifold
$[ {\bf C}^4/{\bf Z}_2]$ is a well-behaved CFT,
but the corresponding singular space cannot be resolved to a smooth
Calabi-Yau, and again, no place to encode B fields.

Perhaps the resolution of these puzzles 
is that the well-behavedness of the CFT is fundamentally
a consequence of some other aspect of the physics,
perhaps related to the mathematical fact that the stack is smooth,
and not fundamentally because of some sort of ``B field at the
orbifold singularity.''  Perhaps this B field phenomenon is merely
a result of quantum corrections in certain special cases, 
a reflection of some other more nearly
fundamental result, related by internal consistency to other
phenomena but not so fundamental in its own right.

\subsection{Summary}

In this section we have addressed a basic problem for both the
physics and mathematics of string orbifolds, namely,
what does it mean to give a vacuum expectation value to a twist field?

Physically, one can work in conformal perturbation theory to formally
define a deformed CFT in terms of correlation functions computed
in the original CFT, but that begs the question of the
interpretation of moduli.  If the marginal operator lies in a twisted
sector, then the resulting CFT's are not clearly nonlinear sigma models
on anything, but are just merely abstract CFT's.
For string orbifolds by finite effective
group actions, this issue has, in the past, been addressed indirectly,
but we would argue that a direct thorough understanding of this matter
does not exist.

The very same issue arises mathematically.
Deformation theory of stacks, as currently defined by mathematicians,
sees precisely {\it untwisted} sector moduli -- it does not see
twisted sector moduli at all.

We have argued that giving vevs to twist fields results in 
physical theories that either do not have geometric interpretations,
or for which geometry is a poor approximation,
so that the mathematical notion of deformation theory is consistently
picking out only those physical deformations for which one could
expect a good geometric interpretation.

However, the reader might wonder if that is the entire
story.  For example, in orbifolds by finite effectively-acting
groups, although giving a vev to a twist field results in a theory
with strong worldsheet instanton corrections, one might speculate
on the existence of some sort of ``stringy'' 
generalization of deformation
theory which could realize the process of giving a vev to such
twist fields. 
For orbifolds by finite noneffectively-acting groups,
one might wonder if, despite our work, there might
actually exist some geometric interpretation of the abstract CFT's
we find (discussed in greater detail in \cite{glsm}).
If those new abstract CFT's, in which fields valued in roots of unity
appear, do have a geometric interpretation, then there should be
some stringy generalization of mathematical
deformation theory which realizes the process
of giving a vev to a twist field.

We shall leave these questions for future work.

\section{D-branes on stacks}   \label{Dbranegen}

In this section, we collect a few miscellaneous facts about
D-branes in gauged sigma models.
We begin by very briefly reviewing D-branes in noneffective orbifolds,
and their mathematical interpretation.
Then, we discuss D-brane probes in orbifolds, and the oft-quoted
claim that since D-brane probes perceive a resolution of the
quotient space, ``string orbifolds describe strings on 
resolutions, not stacks.''

\subsection{D-branes in noneffective orbifolds}  \label{Dbranenoneff}

D-branes in noneffective orbifolds were discussed in \cite{nr}.
For completeness, let us briefly review the result here.

Suppose we are given a $G$-gauged sigma model, where a 
normal subgroup $K$ of $G$ acts trivially on the space.
When defining D-branes in the $G$-gauged sigma model on this space,
in general $K$ can act nontrivially on the Chan-Paton factors,
even though $K$ acts trivially on the space.
This is consistent with the Cardy condition, as discussed in \cite{nr};
one needs merely to be careful to distinguish boundary states
with insertions of twist fields in $K$ from boundary states with
no twist fields at all.

Mathematically, B-branes in gauged sigma models with noneffective
gaugings correspond to sheaves on gerbes.  Now, as we shall review
in section~\ref{sheavesgerbesspaces}, sheaves on gerbes are the same
thing as (twisted) sheaves on underlying spaces.  
The reader can deduce that fact from the description above:
an equivariant sheaf in a noneffective quotient is a projectively
equivariant sheaf over the corresponding effective quotient.

\subsection{D-branes in orbifolds:  a potential contradiction
and its resolution}    \label{dbranemckay}

Several years ago, it was observed in \cite{dgm} that the classical
Higgs moduli space of D-branes in orbifolds is a resolution of the
quotient space, a result that has led some to claim that 
string orbifold CFT's describe closed strings on resolutions of
quotient spaces, and not quotient stacks.
This fact has often been cited as a contradiction, a reason to believe
that stacks cannot be relevant to physics.

This interpretation of D-branes in orbifolds
seems rather fishy to the authors for a variety
of reasons.  One reason is that it is well-known that one should
work with $G$-equivariant objects on covering spaces when dealing
with orbifolds, and objects on resolutions of quotient spaces are
simply {\it not} the same thing as $G$-equivariant objects on covering spaces.
(By contrast, an object on the stack
$[X/G]$ is precisely a $G$-equivariant
object on $X$.)
Another reason to be suspicious of such D-brane
interpretations is that in complex dimension four,
resolutions of quotient spaces are often not Calabi-Yau, yet they
still appear as classical Higgs moduli spaces.
The resolution $\widetilde{ {\bf C}^4/{\bf Z}_2 }$ is a well-known
example of this phenomenon.

Let us point out that the D-brane probe results of \cite{dgm}
have an alternative
understanding within the framework of stacks, that is more logically
coherent.
This explanation was originally described in
\cite{meqs}, based on ideas first presented in \cite{tomasme} and
recently expanded in \cite{dks}, and works as follows:
\begin{itemize}
\item Re-reading \cite{dgm}, one finds that the exceptional divisors
in the classical Higgs moduli spaces are composed of nilpotent Higgs
vevs, a fact which should already make a careful reader pause before
intepreting D-brane probe results too literally.
\item Nilpotent Higgs vevs are modelled by structure sheaves of nonreduced
schemes, and in the case at hand, $G$-invariant nonreduced schemes
or equivalently nonreduced schemes on the quotient stack.
\end{itemize}
Typically nonreduced schemes are unstable, but orbifolds provide
an exception to the rule.

In other words, there is an alternative interpretation of the D-brane
probe results, that is consistent with stacks, and does not require
one to try to believe nonsensical statements such as
``$G$-equivariant objects on a cover are the same as objects
on a resolution of a quotient space.''
In the context of the open string B model, where we typically
describe D-branes with sheaves, stable sheaves include some
objects (structure sheaves of nonreduced schemes) which are not stable
on a smooth space.  In essence, in an orbifold there are stable
D-brane configurations which are not present on smooth spaces.
Thus, although D-brane probes are well-understood for smooth spaces,
one should be careful in interpreting their on stacks, 
where additional objects are present.

A skeptic might nevertheless argue that one should take 
D-brane probe results completely literally in all cases,
despite the problems with such an interpretation, and despite
the subtleties observed above.
Such a skeptic would argue that
the ``correct underlying stringy geometry'' of an orbifold,
however that should be defined, really is a resolution of a quotient
space, and not a stack.  Perhaps there is some sense in which
at very small distance scales, the `correct' geometry is that
of a resolution of a quotient space; however, in such an event,
we see no reason to believe that the geometry seen by open and closed
strings need necessarily match, so perhaps at the same time that
open strings `see' a resolution of a quotient space,
the closed strings see a quotient stack.

In the next few subsections, we shall describe in more detail
how the D-brane probe results can be understood within the framework
of stacks, for the special case of the orbifold $[ {\bf C}^2/{\bf Z}_2]$.
(See \cite{dks} for more detail.)

\subsubsection{Exceptional divisors are nilpotent Higgs fields}

For simplicity, we shall consider two D0 branes, both supported
at the origin of ${\bf C}^2$, hence described by the direct
sum of two skyscraper sheaves at the origin.
Consider the usual ${\bf Z}_2$ orbifold of ${\bf C}^2$, which inverts
both coordinates.

In order to define the ${\bf Z}_2$ orbifold, we need to choose
a ${\bf Z}_2$-action on the Chan-Paton factors,
{\it i.e.},
a ${\bf Z}_2$-equivariant structure on the corresponding
pair of skyscraper sheaves.
In such a simple case, the equivariant structure is 
really just a choice of two-dimensional representation
of ${\bf Z}_2$.  We shall use the regular representation.

The fields on the D0 branes in the orbifold are ${\bf Z}_2$-invariants.
Thus, given 2 D0 branes as above, we have two fields, call them
$X$ and $Y$, which are the Higgs fields on the D0 brane worldvolume,
and arise from open strings connecting the D0 branes to themselves.
$X$ and $Y$ are both $2 \times 2$ matrices.
In the orbifold theory, we take the ${\bf Z}_2$ invariants,
and it's easy to check that that means $X$ and $Y$ must have the form
\begin{displaymath}
X \: = \: \left[ \begin{array}{cc}
                 0 & x_1 \\
                 x_0 & 0 \end{array} \right], \:
Y \: = \: \left[ \begin{array}{cc}
                 0 & y_1 \\
                 y_0 & 0  \end{array} \right]
\end{displaymath}
We can work in a complexified theory and think of $x_{0,1}$, $y_{0,1}$
as all being complex numbers.

The original $U(2)$ 
gauge symmetry is reduced, by the orbifold projection,
to $U(1)^2$.
The matrices above transform as adjoints under this
gauge symmetry.  It is easy to check that one of the $U(1)$'s decouples,
and under the other $U(1)$, $x_0$, $y_0$ have the same charge
(call it $+1$), and $x_1$, $y_1$ have the opposite charge
(call it $-1$).

So far we have described the Higgs fields on the D0 brane worldvolume,
arising from open strings connecting the D0 branes to themselves.

In order to describe classical vacua of the theory,
these fields must satisfy additional constraints\footnote{Strictly
speaking, these constraints emerge from a triplet of D-terms, given
the amount of supersymmetry present; however, we shall refer to them
as F- and D-term constraints, in reference to \cite{dgm}.}:
\begin{enumerate}
\item F term constraints:  $[X,Y] = 0$.
It is straightforward to check that the condition that the two
matrices commute reduces to the single equation $x_1 y_0 = x_0 y_1$.
\item D term constraints:
\begin{displaymath}
| x_0 |^2 \: + \: | y_0 |^2 \: - \: | x_1 |^2 \: - \: | y_1 |^2 \: = \: r
\end{displaymath}
where $r$ is a constant.
This is, of course, part of a symplectic quotient corresponding to
the nontrivial $U(1)$.
\end{enumerate}

After modding out the remaining $U(1)$, or,
if one prefers, performing a GIT quotient wherein
$(x_0, x_1, y_0, y_1)$ have weights $(+1, -1, +1, -1)$,
the classical moduli space of Higgs fields is given by
\begin{displaymath}
\{ ( x_1 y_0 = y_1 x_0) \subset {\bf C}^4 \} // {\bf C}^{\times}
\end{displaymath}
It can be shown\footnote{Describe $\widetilde{ {\bf C}^2 / {\bf Z}_2 }$ as
$ {\bf C}^3 // {\bf C}^{\times}$,
where if we label coordinates on ${\bf C}^3$ by $(x,y,p)$,
the coordinates have weights $(+1,+1,-2)$ under the ${\bf C}^{\times}$.
Then identify $x_0 = x$, $x_1=xp$, $y_0=y$, and $y_1=yp$.} 
that this quotient is the same as the minimal
resolution of the quotient space ${\bf C}^2 / {\bf Z}_2$.

We can also outline the result using basic linear algebra.
Since the original matrices $X$ and $Y$ commute, they're simultaneously
diagonalizable.  It is easy to check that since they're ${\bf Z}_2$
invariant, the (nonzero) eigenvalues come in ${\bf Z}_2$ pairs.
If we map the pair $(X,Y)$ to the point on $\widetilde{ {\bf C}^2/
{\bf Z}_2 }$ determined by the eigenvalues, then for nonzero
eigenvalues, we have an isomorphism.
The zero eigenvalues are degenerate -- they are mapping out the
exceptional divisor.

In this language, it is clear that the exceptional divisor
in $\widetilde{ {\bf C}^2 / {\bf Z}_2 }$ is arising from
nilpotent matrices $X$, $Y$.
In particular, if $r > 0$, then the matrices corresponding to
the exceptional divisor are given by
\begin{displaymath}
X \: = \: \left[ \begin{array}{cc}
                 0 & 0 \\
                 x_0 & 0  \end{array} \right], \:
Y \: = \: \left[ \begin{array}{cc}
                 0 & 0 \\
                 y_0 & 0 \end{array} \right]
\end{displaymath}
where $x_0$, $y_0$ are homogeneous coordinates on the ${\bf P}^1$.
The corresponding length 2 ideal on ${\bf C}[x,y]$ is given by
$(x^2, x_0 y -  y_0 x)$.
We have seen the corresponding sheaves previously in this text:
$D_x$ is the sheaf corresponding to the case $y_0=0$,
and $D_y$ is the sheaf corresponding to the case $x_0=0$.

We have only discussed the special case $[ {\bf C}^2/{\bf Z}_2]$,
but as should be clear from the linear-algebra-based discussion
above, the same ideas work in generality -- D-branes on orbifolds
see resolutions of quotient spaces because the gauge theory classical moduli
space admits nilpotent Higgs vevs.  
See \cite{dgm} for a much more general discussion.

\subsubsection{McKay:  exceptional divisors are nonreduced
schemes}   \label{mckay}

In the previous section, we showed that the exceptional divisors
in moduli spaces of D-branes in orbifolds are determined by
nilpotent Higgs vevs.  As described in much more detail
in \cite{dks} (see \cite{melec} for a review),
mathematically we can describe such D-brane configurations using
{\it e.g.} structure sheaves of nonreduced schemes.
In \cite{dks} we formulated a precise map between D-branes with
Higgs vevs and sheaves, that matches open string spectra to
Ext groups, and for nilpotent Higgs vevs, the corresponding sheaves
are typically related to nonreduced schemes.

Rather than review \cite{dks}, let us instead review an alternative
way of understanding the results of \cite{dks}, namely the
McKay correspondence, which gives consistent results.

The version of the McKay correspondence that is applicable
here is due to Bridgeland-King-Reid \cite{bkr}, who described
McKay at the level of an equivalence of derived categories of sheaves.
If one starts with a skyscraper sheaf on the exceptional divisor
of $\widetilde{ {\bf C}^2/{\bf Z}_2 }$,
then just as discussed in \cite[section 6.2]{kps} for skyscraper
sheaves on $\widetilde{ {\bf C}^3/{\bf Z}_3 }$,
the image under the McKay functor is a ${\bf Z}_2$-equivariant
nonreduced scheme on ${\bf C}^2$, with support at the origin
(the ${\bf Z}_2$ fixed point), and with scheme structure determined
by the location of the skyscraper sheaf on the exceptional divisor.

Clearly, to be consistent with \cite{dgm}, one would like
for nonreduced schemes to be related to nilpotent Higgs fields,
and that is exactly what was found in \cite{dks}.

Also, in followup work to \cite{dgm}, it was noticed that the
resolutions could also be obtained as Hilbert schemes of points,
{\it i.e.} moduli spaces of $G$-equivariant nonreduced schemes
(see {\it e.g.} \cite{mohri}), as relevant for another version
of the McKay correspondence \cite{itonakajima}.  This is yet more
supporting evidence for the statement.

In any event, we see that the nilpotent Higgs vevs computed physically
have a mathematical understanding in terms of certain sheaves.
It turns out that the sheaves that correspond to nilpotent Higgs vevs,
are stable in orbifolds, but unstable on smooth spaces.
Thus, we are implicitly discussing a class of D-branes which are
physically more relevant to orbifolds than to smooth spaces,
and it is this class of orbifold-specific D-brane configurations
that were sensed in the work of \cite{dgm}.

\section{Open string B model on stacks}   \label{openB}

In this section we shall consider the open string B model
when the target is a Calabi-Yau stack.
It is now believed (see \cite{medc,mikedc,paulalb} for early work)
that boundary states in the open string B model (on a Calabi-Yau
space) should be counted by derived categories of coherent sheaves on that
space.  We will argue that the same result holds true when the
target is a Calabi-Yau stack, instead of a space.

We begin by calculating open string spectra between two
sets of D-branes, working through the analogue for stacks
of calculations in \cite{orig,kps,cks,melec}.  
Specifically, recall that
massless boundary Ramond sector states of open strings
connecting D-branes wrapped on complex submanifolds
of Calabi-Yau's, with holomorphic gauge bundles, should
be counted by Ext groups.
More precisely, if the D-branes are wrapped on the
complex submanifolds $i: S \hookrightarrow X$ and
$j: T \hookrightarrow X$ of a Calabi-Yau $X$,
with holomorphic vector `bundles' ${\cal E} \otimes \sqrt{K_S^{\vee}}$, 
${\cal F} \otimes \sqrt{K_T^{\vee}}$,
respectively, then massless boundary Ramond sector states
are in one-to-one correspondence with elements of
either
\begin{displaymath}
\mbox{Ext}^*_X\left( i_* {\cal E}, j_* {\cal F} \right)
\end{displaymath}
or
\begin{displaymath}
\mbox{Ext}^*_X\left( j_* {\cal F}, i_* {\cal E} \right)
\end{displaymath}
(depending upon the open string orientation).

We shall argue that D-branes wrap substacks,
and that if we have D-branes wrapped on
two smooth closed substacks $i: {\cal S} \hookrightarrow {\cal X}$ and
$j: {\cal T} \hookrightarrow {\cal X}$ of a Calabi-Yau stack ${\cal X}$,
with holomorphic vector `bundles' 
${\cal E} \otimes \sqrt{K_{{\cal S}}^{\vee}}$, 
${\cal F} \otimes \sqrt{K_{ {\cal T} }^{\vee}}$,
respectively,  then massless boundary Ramond sector states
are in one-to-one correspondence with elements of
either
\begin{displaymath}
\mbox{Ext}^*_{ {\cal X} }\left( i_* {\cal E}, j_* {\cal F} \right)
\end{displaymath}
or
\begin{displaymath}
\mbox{Ext}^*_{ {\cal X} }\left( j_* {\cal F}, i_* {\cal E} \right)
\end{displaymath}
(depending upon the open string orientation),

After working through this computation and checks thereof,
we will outline how the formal justification for counting
off-shell B model boundary states with derived categories also
goes through when the target is a Calabi-Yau stack,
resulting in a classification by derived categories on stacks.

We will assume the reader is familiar with the results of
section~\ref{Dbranenoneff} where D-branes in noneffective
orbifolds are discussed.

A word on notation.  We shall denote ordinary spaces with
standard Roman fonts, as {\it e.g.} $S$, $T$, $X$,
whereas stacks will be denoted with calligraphic fonts,
as {\it e.g.} ${\cal S}$, ${\cal T}$, ${\cal X}$.

\subsection{Open string spectra}

Before calculating spectra, let us make an observation concerning
what a D-brane can wrap.  Let ${\cal S}$ be a substack of the stack
${\cal X}$.
Since ${\cal X}$ is a Deligne-Mumford stack, it has an atlas,
call it $X$.  The substack ${\cal S}$ also necessarily
has an atlas $S$, given by $S = {\cal S} \times_{ {\cal X} } X$,
and since by definition of substack the inclusion $i: {\cal S} \rightarrow
{\cal X}$ is a representable injective map, the projection
${\cal S} \times_{{\cal X}} X \rightarrow X$ is injective,
{\it i.e.}, the atlas for ${\cal S}$ is a submanifold of the atlas for
${\cal X}$.  In fact, if we present the stack ${\cal X}$ as
$[X/G]$ for some group action $G$, then ${\cal S} = [S/G]$.
Thus, a substack of $[X/G]$ is really a $G$-invariant submanifold of $X$.
Since in a $G$-gauged sigma model on $X$, the D-branes are located
on $G$-invariant submanifolds of $X$, we see that
D-branes wrap substacks.

Next, let us turn to the computation of massless spectra of the open strings.
We shall begin with parallel coincident branes,
both wrapped on a closed smooth substack ${\cal S} = [S/G]$ 
of the stack ${\cal X}$.

In order to calculate open string spectra for D-branes
on stacks, we shall generalize the ansatz of \cite{dougmoore,kps}.
Recall \cite{dougmoore,kps} computed open string spectra in orbifolds
by finite effectively-acting groups
by first computing spectra on a covering space, and then
taking invariants.
We will follow essentially the same prescription.
Specifically, we construct $G$-invariant boundary vertex 
operators on the space $S$
(or equivalently, since ${\cal S} = [S/G]$, one could say we construct
boundary vertex operators on the substack ${\cal S}$).

As in \cite{orig,kps,cks}, there are important subtleties in the
boundary conditions:
\begin{itemize}
\item If $TX|_S$ does not split holomorphically as $TS \oplus {\cal N}_{S/X}$,
then statements such as $\theta_i=0$ are not sensible globally.
\item If the gauge bundles have nonzero curvature, then the boundary
conditions are twisted.
\end{itemize}
For the moment, we shall assume that $TX|_S$ splits holomorphically
and that the gauge bundles have trivial curvature, to simplify the
analysis.

Thus, we can write the ($G$-invariant) massless boundary
Ramond sector states in the same form as in \cite{orig,kps,cks},
namely
\begin{displaymath}
b^{\alpha \beta j_1 \cdots j_m}_{\overline{\imath}_1 \cdots \overline{
\imath}_n}(\phi_0) \, \eta^{\overline{\imath}_1} \cdots \eta^{
\overline{\imath}_n} \, \theta_{j_1} \cdots \theta_{j_m}
\end{displaymath}
where $\phi$ are scalars sweeping out the atlas $S$,
the $\eta$ are interpreted as one-forms on $S$,
and the $\theta$ 
couple to the normal bundle
${\cal N}_{ S/X}$.
The bundles ${\cal E}$, ${\cal F}$ on the stack ${\cal S}$ define
bundles on the atlas $S$, which we shall denote with
${\cal E}$, ${\cal F}$ also.  The Chan-Paton indices couple
to the `bundles'\footnote{The twisting by $\sqrt{K_S^{\vee}}$ is a 
reflection of the Freed-Witten anomaly, as discussed in \cite{orig}.}
${\cal E}\otimes\sqrt{K_S^{\vee}}$, 
${\cal F}\otimes\sqrt{K_S^{\vee}}$ on the atlas.

In order to interpret these vertex operators, we need to be a little
careful about the BRST operator.  Since we have gauged $G$,
the square of the BRST operator need only vanish up to a gauge transformation.
However, since we are acting on $G$-invariant states, which do not have
gauge transformations, the BRST operator acts as $\overline{\partial}$,
just as in previous work.

Thus, just as in previous work, these vertex operators
are naturally interpreted as elements of
the sheaf cohomology groups
\begin{displaymath}
H^n\left(S, {\cal E}^{\vee} \otimes {\cal F} \otimes \Lambda^m 
{\cal N}_{ S / X }
\right)
\end{displaymath}
on the space $S$.
Note that at this point, everything is being described on an honest
{\it space} ({\it i.e.} $S$, the atlas of the stack ${\cal S}$).

Just as in \cite{orig,kps,cks}, we have made simplifying assumptions involving
twisting of the boundary conditions.
Before we consider those subtleties, however, let us pause for a
moment to reinterpret the work above.
First, the normal bundle ${\cal N}_{ S / X }$ 
has canonical isomorphisms
$p_1^* {\cal N}_{ S/X } \cong p_2^* {\cal N}_{
S / X }$ making it not only into a bundle
over the stack ${\cal S}$, but in fact making it into the 
normal bundle\footnote{Even more can be said.  The total space
of the bundle ${\cal N}_{{\cal S}/{\cal X}}$ over the stack ${\cal S}$ 
is also a stack,
and the total space of the bundle ${\cal N}_{S/X}$ is
an atlas for the stack ${\cal N}_{{\cal S}/{\cal X}}$.} 
${\cal N}_{{\cal S}/{\cal X}}$ between the stacks ${\cal S}$ and ${\cal X}$.
Second, the tangent bundle has analogous isomorphisms,
so it can also be interpreted as the tangent bundle to the stack ${\cal S}$.
Finally, the elements of the sheaf cohomology groups on the space
$S$ above with the $G$-invariance property are precisely the
elements of the  
sheaf cohomology groups on
the stack ${\cal S}$:
\begin{eqnarray*}
H^n\left({\cal S}, {\cal E}^{\vee} \otimes {\cal F} \otimes
\Lambda^m {\cal N}_{{\cal S}/{\cal X}} \right)
& = &
H^n\left(S, {\cal E}^{\vee} \otimes {\cal F} \otimes \Lambda^m {\cal N}_{
S/X} \right)^{ p_1^* = p_2^* } \\
& = &
H^n\left(S, {\cal E}^{\vee} \otimes {\cal F} \otimes \Lambda^m {\cal N}_{
S/X} \right)^{G} 
\end{eqnarray*}
just as, in the special case of orbifolds \cite{kps}, sheaf cohomology
on the quotient stack $[S/G]$ is the same as the $G$-invariant
part of sheaf cohomology on $S$.
(This result, however, is specific to presentations as quotient stacks;
for alternative presentations of a given stack, the obvious analogue
of this statement is more complicated, as we discuss in 
section~\ref{ext:altpres}.)
Thus, we can summarize our results so far
much more compactly by saying that, using the obvious generalization
of the ansatz in \cite{dougmoore,kps}, open string spectra between
D-branes on a stack ${\cal S}$ are counted by sheaf cohomology on the
stack ${\cal S}$.

Notice that if we had taken the results of \cite{orig} and
simply replaced every occurrence of the word ``space'' with the
word ``stack,'' we would arrive at the same result.

As in \cite{orig,kps,cks}, this analysis is slightly too naive,
in that we have assumed the tangent bundle to the stack ${\cal X}$
splits holomorphically into $T{\cal S} \oplus {\cal N}_{{\cal S}/{\cal X}}$, 
and that
we have ignored twisting of boundary conditions induced by
Chan-Paton curvature \cite{abooetal}.
Taking these subtleties into account, 
we are led to look for a spectral sequence, just as in \cite{orig,kps,cks},
and indeed, we shall see momentarily that 
a spectral sequence of exactly the desired form exists,
relating the sheaf cohomology groups above to Ext groups on the stack 
${\cal X}$.

The analysis now follows exactly the same 
form as that in previous work \cite{orig,kps,cks} -- taking into
account the two subtleties above has the effect of physically
realizing a spectral sequence directly in BRST cohomology,
(for Chan-Paton factors describing line bundles, and for open strings
connection a D-brane to itself).
Rather than drag the reader through the same analysis again, we shall
merely summarize relevant details, and refer the interested reader
to our earlier work \cite{orig,kps,cks}.

Analogues of the spectral sequences used in previous papers
\cite{orig,kps,cks} exist (which also reflect the
Freed-Witten anomaly, whose relevance here is exactly as
discussed previously in \cite{orig,kps,cks}).  Specifically:
\begin{enumerate}
\item Parallel coincident branes on ${\cal S} \hookrightarrow {\cal X}$.
Let ${\cal S}$ be a closed substack of a Calabi-Yau stack ${\cal X}$.
Let ${\cal E}$, ${\cal F}$ be vector bundles on ${\cal S}$.
There is a spectral sequence
\begin{displaymath}
H^n\left({\cal S}, {\cal E}^{\vee} \otimes {\cal F} \otimes
\Lambda^m {\cal N}_{{\cal S}/{\cal X}} \right) \: \Longrightarrow \:
\mbox{Ext}^{n+m}_{ {\cal X} }\left( i_* {\cal E}, i_* {\cal F} \right)
\end{displaymath}
\item Parallel branes on submanifolds of different dimension.
Let ${\cal S}$ be a closed substack of a Calabi-Yau stack ${\cal X}$,
and let ${\cal T}$ be a closed substack of ${\cal S}$.  Let ${\cal E}$,
${\cal F}$ be vector bundles on ${\cal S}$, ${\cal T}$, respectively.
Then there is a spectral sequence
\begin{displaymath}
H^n\left({\cal T}, {\cal E}^{\vee}|_{ {\cal T} } \otimes {\cal F} \otimes
\Lambda^m {\cal N}_{{\cal S}/{\cal X}}|_{ {\cal T} } \right) \: \Longrightarrow \:
\mbox{Ext}^{n+m}_{ {\cal X} } \left( i_* {\cal E}, j_* {\cal F} \right)
\end{displaymath}
\item General intersections.
Let ${\cal S}$, ${\cal T}$ be closed substacks of a Calabi-Yau stack ${\cal X}$.
Let ${\cal E}$, ${\cal F}$ be vector bundles on ${\cal S}$, ${\cal T}$, respectively.
There are spectral sequences
\begin{eqnarray*}
H^p\left({\cal S}\cap {\cal T}, {\cal E}^{\vee}|_{{\cal S} \cap {\cal T}} \otimes {\cal F}|_{{\cal S}\cap {\cal T}}
\otimes \Lambda^{q-m} \tilde{N} \otimes \Lambda^{top} {\cal N}_{{\cal S}\cap {\cal T}/{\cal T}}
\right) & \Longrightarrow &
\mbox{Ext}^{p+q}_{ {\cal X} }\left( i_* {\cal E}, j_* {\cal F} \right) \\
H^p\left( {\cal S} \cap {\cal T}, {\cal E}^{\vee}|_{{\cal S} \cap {\cal T}} \otimes {\cal F}|_{{\cal S} \cap {\cal T}}
\otimes \Lambda^{q-n} \tilde{N} \otimes \Lambda^{top} {\cal N}_{{\cal S}\cap {\cal T}/{\cal S}}
\right) & \Longrightarrow &
\mbox{Ext}^{p+q}_{ {\cal X} }\left(j_*{\cal F}, i_* {\cal E}\right)
\end{eqnarray*}
where 
\begin{displaymath}
\tilde{N} \: = \: T{\cal X}|_{{\cal S} \cap {\cal T}}/\left( T{\cal S}|_{{\cal S} \cap {\cal T}} + T{\cal T}|_{{\cal S} \cap {\cal T}}
\right) 
\end{displaymath}
and $m = \mbox{rk } {\cal N}_{{\cal S}\cap {\cal T}/{\cal T}}$,
$n = \mbox{rk } {\cal N}_{{\cal S}\cap {\cal T}/{\cal S}}$.
\end{enumerate}
Note that for stacks, ${\cal S} \cap {\cal T}$ is defined as the fiber product
${\cal S} \times_{ {\cal X} } {\cal T}$.
(These spectral sequences can be proven in the same way as in the
appendix of \cite{kps}.)

Just as in \cite{orig,kps,cks}, the spectral sequences above are realized
directly in worldsheet BRST cohomology, in exactly the same manner
as in \cite{orig,kps,cks}.  
Let us briefly review the special case that the D-branes are
wrapped on the same substack, {\it i.e.} ${\cal S} = {\cal T}$,
and that they have the same bundle on their worldvolumes,
{\it i.e.} ${\cal E} = {\cal F}$, and that the Chan-Paton factors describe
line bundles.
The differential 
\begin{displaymath}
d_2: \: H^0({\cal E}^{\vee} \otimes {\cal E} \otimes
{\cal N}_{S/X})^G \: \longrightarrow \: H^2({\cal E}^{\vee} \otimes {\cal E})^G
\end{displaymath}
can be described as the composition
\begin{displaymath}
H^0\left(S, {\cal E}^{\vee} \otimes {\cal E} \otimes {\cal N}_{S/X}
\right)^G \: \stackrel{\delta}{\longrightarrow} \:
H^1\left(S, {\cal E}^{\vee} \otimes {\cal E} \otimes TS \right)^G
\: \stackrel{eval}{\longrightarrow} \: 
H^2\left(S, {\cal E}^{\vee} \otimes {\cal E} \right)^G
\end{displaymath}
where the map $\delta$ is the coboundary map in the long
exact sequence associated to the short exact sequence
\begin{displaymath}
0 \: \longrightarrow \: TS \: \longrightarrow \: TX|_S \:
\longrightarrow \: {\cal N}_{S/X} \: \longrightarrow \: 0
\end{displaymath}
and the evaluation map encodes the Chan-Paton-induced twisting
of the boundary conditions \cite{abooetal,aboo2}.

Just as in \cite{orig,kps,cks}, we can build vertex operators by starting
with normal-bundle-valued differential forms and lifting to $TX|_S$-valued
differential forms.  The resulting differential forms may no longer be
closed under $\overline{\partial}$ -- in which case their BRST images
are encoded by the map $\delta$ above, and taking into account the
boundary conditions \cite{abooetal,aboo2}, we see that a normal-bundle-valued
differential form defines a BRST-closed operator if and only if it is
annihilated by $d_2$.

We shall not describe other cases in detail here; suffice it to say,
we conjecture that in other cases, the spectral sequences above are
realized directly in BRST cohomology, as happened in the case above.

Thus, taking into account subtleties in the boundary conditions,
as in \cite{orig,kps,cks}, we see explicitly 
that the spectrum of (BRST-invariant) massless boundary Ramond sector states
on open strings between D-branes wrapped on substacks ${\cal S}$, ${\cal T}$
of a Calabi-Yau stack ${\cal X}$ is counted by either
\begin{displaymath}
\mbox{Ext}^*_{ {\cal X} }\left( i_* {\cal E}, j_* {\cal F} \right)
\end{displaymath}
or
\begin{displaymath}
\mbox{Ext}^*_{ {\cal X} }\left( j_* {\cal F}, i_* {\cal E} \right)
\end{displaymath}
depending upon the open string orientation.

Note that in retrospect, if we had taken the results of
\cite{orig} and everywhere replaced the word ``space'' with ``stack,''
we would have obtained the same result.

Also note that since our results on open string spectra are
phrased directly in terms of stacks, they are necessarily
independent of presentation of the stack -- if we present a given
stack in different ways and calculate open string spectra, although
the details of the calculations will differ, the results are the
same.  We check this explicitly in section~\ref{check:openspectra} in
an example involving
fractional branes on $[ {\bf C}^2/{\bf Z}_n]$ (comparing
results in \cite{kps}[section 3.3.2]) and using the alternate
presentation of $[ {\bf C}^2/{\bf Z}_n ]$ discussed in section~\ref{multpres}.

So far we have generalized the ansatz of \cite{dougmoore,kps} in the obvious
way to compute open string spectra for D-branes on stacks.
Now, stacks include both ordinary spaces and orbifolds as special
cases, but a less obvious fact is that we have also implicitly
reproduced the flat $B$ field computations of \cite{cks}.
In the next few sections we will
check explicitly that the general analysis just described completely
subsumes the results of \cite{orig,kps,cks}.

\subsection{Aside:  alternative presentations}  \label{ext:altpres}

In the previous section we described how to compute open string
spectra on a Calabi-Yau stack, presented as a global quotient
$[X/G]$, for some (not necessarily finite) group $G$.
However, although all Deligne-Mumford stacks can be (non-uniquely)
presented in
this form, there are other presentations not of this form,
as discussed previously in section~\ref{transverseCY}.   

Although in this paper we concentrate on presentations of the
form $[X/G]$, we shall take a few moments here to illustrate some
of the technical complications of the more general case.
Let us illustrate these complications in the case of the
calculation of open string spectra between two D-branes wrapped
on the same substack ${\cal S}$.

As before, one begins by computing open string spectra on the covering
space, and with assumptions of splitting of $TX|_S$ and triviality of
gauge bundles, one finds that the open string spectra on the covering
space are counted by the sheaf cohomology groups
\begin{displaymath}
H^n\left(S, {\cal E}^{\vee} \otimes {\cal F} \otimes \Lambda^m 
{\cal N}_{ S / X }
\right)
\end{displaymath}
on the space $S$.

There is a naive way to try to generalize the notion of taking
$G$-invariants to more general presentations. 
Specifically, if we let $p_{1,2}$ denote the projection maps
\begin{displaymath}
p_{1,2}: \: S \times_{ {\cal S} } S \: \longrightarrow \: S
\end{displaymath}
then we impose the condition that if $b$ is any bundle-valued form
representing a cohomology group above, we will only consider those
$b$ such that $p_1^* b = p_2^* b$.
This is precisely the $G$-invariance constraint in the special
case\footnote{For a quotient stack ${\cal S} = [S/G]$, 
$S \times_{ {\cal S} } S = S \times G$,
one of the projection maps is the (forgetful) projection $S \times G 
\rightarrow S$,
and the other projection map describes the $G$-action on $S$.}
of quotient stacks, and so one would naively guess that
also for alternative presentations of the stacks,
\begin{equation}    \label{try1}
H^n\left({\cal S}, {\cal E}^{\vee} \otimes {\cal F} \otimes
\Lambda^m {\cal N}_{{\cal S}/{\cal X}} \right)
\: = \:
H^n\left(S, {\cal E}^{\vee} \otimes {\cal F} \otimes \Lambda^m {\cal N}_{
S/X} \right)^{ p_1^* = p_2^* }
\end{equation}

However, although the statement above does generalize the notion of
taking $G$-invariants, equation~(\ref{try1}) is not correctly mathematically.
The correct description is more complicated.
If we define
\begin{displaymath}
X_n \: = \: \underbrace{ X \times_{ {\cal X} } \cdots \times_{ {\cal X} }
X }_{n+1}
\end{displaymath}
and let ${\cal E}_n$, {\it etc} be the pullback of the sheaf ${\cal E}$
to the space $X_n$, then for example sheaf cohomology groups of
${\cal E}$ are derived from a spectral sequence
\begin{displaymath}
E_2^{p,n}: \: H^p(X_n, {\cal E}_n) \: \Longrightarrow \: H^{p+n}({\cal X}, {\cal E})
\end{displaymath}
Equation~(\ref{try1}) essentially only realizes one term in the spectral
sequence.
Ext groups are calculated in the obvious analogue.

\subsection{Serre duality}    \label{serre}

As discussed in {\it e.g.} \cite{orig}, Serre duality always
has the property of mapping string spectra back into themselves.
In the case of open strings on Calabi-Yau spaces, Serre duality
exchanges the groups
\begin{displaymath}
\mbox{Ext}^n_X\left( i_* {\cal E}, j_* {\cal F} \right)
\end{displaymath}
with the groups
\begin{displaymath}
\mbox{Ext}^{n-p}_X\left( j_* {\cal F}, i_* {\cal E} \right)
\end{displaymath}
effectively flipping the open string orientation.

For Ext groups of sheaves on Deligne-Mumford stacks ${\cal X}$, Serre duality
acts exactly as one would naively guess \cite{tonyandrewunpub}:
\begin{displaymath}
\mbox{Ext}^p_{ {\cal X} }\left( {\cal S}, {\cal T} \right) \: \cong \:
\mbox{Ext}^{n-p}_{ {\cal X} }\left( {\cal T}, {\cal S} \otimes
K_{ {\cal X} } \right)^*
\end{displaymath}
where $n$ is the dimension of ${\cal X}$, and ${\cal S}$, ${\cal T}$
are sheaves on ${\cal X}$.
In particular, when the canonical bundle of the stack is trivial,
Serre duality acts on Ext groups precisely as a flip of open string
orientation, exactly as discussed for Calabi-Yau spaces.

This fact gives independent evidence that the `right' definition
of Calabi-Yau stack is a stack with trivial canonical bundle,
which is indeed the definition we chose in section~\ref{cystxdefn}.
If Serre duality did not map open string spectra back into themselves,
then our definition of Calabi-Yau stack would be much more suspect.
See section~\ref{cystxdefn} for further discussion and 
justification of what it means
for a Deligne-Mumford stack to be Calabi-Yau.

\subsection{Some examples}

For our first example, consider the case that the stack
${\cal X}$ is an 
ordinary smooth Calabi-Yau, call it $X$.

To specialize the general language above to this case,
use the fact that a closed substack
of a smooth Calabi-Yau (viewed as a stack) is an honest space\footnote{
The inclusion map $i: {\cal S} \rightarrow {\cal X}$ is a representable closed
immersion, by definition of closed substack.  
Now, for a map $f: {\cal F} \rightarrow {\cal G}$ between
stacks to be representable means \cite{gomez} that for all maps $Y \rightarrow {\cal G}$,
where $Y$ is an honest space, the fiber product ${\cal F} \times_{{\cal G}} Y$
is an honest space.  In this case, for $i: {\cal S} \rightarrow {\cal X}$ to be representable
in the case that ${\cal X}$ is a space $X$ means that ${\cal S} \times_{ {\cal X} } X$ is an honest
space, but ${\cal S} \times_{ {\cal X}} X = {\cal S}$, hence ${\cal S}$ is an honest space.},
not a stack.  So, in the language of the general analysis,
the sub`stacks' ${\cal S}$ and ${\cal T}$ are actually submanifolds of $X$.

In particular, it should now be clear that in the special case
that ${\cal X}$ is an ordinary smooth Calabi-Yau, the general analysis
in the previous section reproduces, on the nose, the results
for spectra of open strings between D-branes on ordinary Calabi-Yau's
as described in \cite{orig}.

In passing, we observe that it is completely trivial
to specialize our analysis for general $G$-gauged sigma models
to orbifolds by finite effectively-acting groups.
The derivation goes over essentially verbatim,
and we immediately recover the results of \cite{kps} as a special
case.

We should also mention that
understanding string orbifolds as sigma models on
quotient stacks has some specific conceptual advantages.
For example, fractional branes are describable as
sheaves on quotient stacks, but they cannot be understood as sheaves
on quotient spaces \cite{kps} -- so for sheaf-theoretic descriptions
of fractional branes, one is forced to use quotient stacks.

\subsection{Example:  Calabi-Yau gerbes}

The examples in the last two sections could be derived from stacks
in a very obvious way:  spaces are special cases of stacks,
and global orbifolds are mathematically closely related with quotient stacks.

Having rederived the results of \cite{orig,kps} as special cases
of the current work, let us now turn our attention to
\cite{cks}.  The twisted
sheaves described in \cite{cks}, corresponding to D-branes in
nontrivial $B$ field backgrounds,
are the same as sheaves on certain stacks, where the stacks are
simply the gerbes corresponding to the $B$ field in question,
a fact we shall review in more detail shortly.

The fact that the twisted sheaves of \cite{cks} are the same as sheaves
on gerbes leads one to suspect that the results of \cite{cks} could be
rederived by considering the open string B model on gerbes.
Unfortunately, such an analysis assumes that the closed string
B model on the gerbe is the same physical theory as a closed
string on the underlying manifold with a $B$ field, which is not true,
as we shall review shortly.

\subsubsection{Sheaves on gerbes are twisted sheaves on spaces}
\label{sheavesgerbesspaces}

Next, let us consider sheaves on gerbes.
We shall outline how a sheaf on a gerbe over a space $X$
is the same thing as a twisted sheaf on $X$.
Consider a $G$-gerbe on $X$
with an atlas given by a principal $K$-bundle $P$ on $X$,
for some $K$, as above, so that the gerbe itself is $[P/A]$,
where $A$ is an extension of $K$ by $G$.
A sheaf on $[P/A]$ is, from our general discussion of quotient stacks,
the same thing as an $A$-equivariant sheaf on $P$.

Now, since $P$ is a principal $K$-bundle,
a $K$-equivariant sheaf on $P$ is the same thing as an ordinary
sheaf on $X$.

In the present case, however, we do not quite have a $K$-equivariant
sheaf, but rather an $A$-equivariant sheaf, where $A$ is an extension
of $K$ by $G$.  Since $G$ acts trivially on $P$, the effect of imposing
an $A$-equivariant structure is to add (base-preserving) sheaf automorphisms.
As a result, an $A$-equivariant sheaf on $P$ is the
same thing as a twisted sheaf \cite{cks} on $X$, twisted by some multiple
of the element
of $H^2(X, C^{\infty}(G))$ defining the gerbe.
Recall a twisted bundle on a space $X$ is a bundle whose transition
functions $g_{ij}$ do not quite close on triple overlaps,
but rather only close up to a \v{C}ech 2-cocycle:
\begin{equation}   \label{gtwist}
g_{ij} g_{jk} g_{ki} \: = \: \alpha_{ijk}
\end{equation}
for some $(\alpha_{ijk})$ defining an element of $H^2(X, C^{\infty}(G))$.
Twisted sheaves can be defined similarly; see \cite{cks} for a recent
discussion.

It is important to note at this point that from sheaves on gerbes we
recover not only twisted sheaves on the base space $X$ that are
twisted by the element of $H^2(X, C^{\infty}(G))$ defining the gerbe,
but also sheaves twisted by all multiples of that element.
For example, as a special case, we can recover untwisted, ordinary,
sheaves on the space $X$.  In terms of the description of the gerbe
as a global quotient $[P/A]$, all of these possible results are encoded
in the choice of $A$-equivariant structure on the sheaf.  In other words,
different $A$-equivariant structures describe sheaves on $X$ with
different degrees of twisting.

Such twisted bundles and sheaves arise physically when describing
D-branes in flat nontrivial $B$ field backgrounds \cite{freeded}.
Recall this is because gauge transformations of the $B$ field
are only consistent on open string worldsheets if the Chan-Paton
factors receive an affine translation:
\begin{eqnarray*}
B & \mapsto & B \: + \: d \Lambda\\
A & \mapsto & A \: - \: \Lambda
\end{eqnarray*}
If the $B$ field has nonzero transition functions, then the
Chan-Paton gauge field picks up affine translations between 
coordinte charts.  One effect is that the transition functions
of the bundle are twisted, as in equation~(\ref{gtwist}).

These twisted sheaves have recently played an important
role in other aspects of string theory.  For example,
the recent derivation \cite{medt,dtrev} of Douglas's projectivization
of orbifold group actions on D-branes, in discrete torsion backgrounds
\cite{dougdt},
used the fact that D-brane `bundles' are twisted as above.

Let us consider a more specific example, that of ${\bf Z}_n$-gerbes
on a Calabi-Yau $X$.
As discussed previously in section~\ref{flatgerbes}, we shall
describe the gerbe in the form $[P/G]$ where $G$ is a nontrivial 
extension of ${\bf C}^{\times}$ by ${\bf Z}_n$, isomorphic to ${\bf C}^{\times}$
(but denoted differently for clarity), and $P$ is the total space of
a principal ${\bf C}^{\times}$ bundle over $X$ with the property
that the mod $n$ reduction of its first Chern class is the element
of $H^2(X, {\bf Z}_n)$ classifying the gerbe.  The possible sheaf twistings
corresponding to such gerbes are classified by characters $\chi:
{\bf Z}_n \rightarrow {\bf C}^{\times}$, as we shall see in more detail
momentarily.

Let ${\cal S}$ be a sheaf on $[P/G]$, {\it i.e.} a $G$-equivariant
sheaf on $P$.

The subgroup ${\bf Z}_n \subseteq G$ acts trivially on $P$, so 
${\bf Z}_n$ acts on ${\cal S}$ fiber-by-fiber.  In other words, the 
$G$-equivariant structure on ${\cal S}$ induces a representation 
$\rho: {\bf Z}_n \rightarrow \mbox{Aut}({\cal S})$ of ${\bf Z}_n$ in the
group of base-preserving sheaf automorphisms of ${\cal S}$.

On the other hand, every character $\chi: {\bf Z}_n \rightarrow {\bf C}^{\times}$
gives a natural action of ${\bf Z}_n$ on ${\cal S}$ by sheaf automorphisms,
namely, $g \in {\bf Z}_n$ acts on ${\cal S}$ by multiplication by the
number $\chi(g)$ in the fibers.

As the $G$-equivariant structure on ${\cal S}$ is varied, the induced
representations $\rho$ vary over precisely the automorphisms generated
by the characters $\chi$.  Also, the representation $\rho$ induced
by the $G$-equivariant structure will match $\chi$ precisely when
the $G$-equivariant sheaf ${\cal S}$ on $P$ is a $\chi$-twisted sheaf
on $X$.

Thus, by varying the possible $G$-equivariant structures on ${\cal S}$,
we map out possible twistings of sheaves on the base space $X$.

\subsubsection{Results of Caldararu-Katz-Sharpe are not a special case}

Given that a twisted sheaf on a manifold is the same thing as a
sheaf on a gerbe over the manifold, it is tempting to try to
rederive the results of \cite{cks} from considering the open string
B model on gerbes.  Let us set this up more precisely, then
we shall discover a basic problem.

First, when is a gerbe a Calabi-Yau stack?
Flat gerbes on Calabi-Yau manifolds are all Calabi-Yau stacks.
Non-flat gerbes on Calabi-Yau manifolds are not Calabi-Yau stacks,
however.
Thus, for the purposes of string compactifications on Calabi-Yau
stacks, we shall only consider flat gerbes.

Next, what are the possible substacks of a flat gerbe?
Using the same methods as for spaces and global orbifolds,
it is easy to see that a substack of a flat gerbe is itself
another flat gerbe.  Moreover, the normal bundles to the substacks
are the same as the normal bundles appearing in \cite{cks}.
In particular, the degree-zero twisted sheaves on the gerbe are pullbacks
of ordinary sheaves from the underlying space, and the normal bundle is
an exampel of one such degree-zero sheaf.

Thus, most of the components we need to derive \cite{cks} from 
the open string B model on gerbes seem to be in place.

Unfortunately, one significant component is lacking.
The paper \cite{cks} discussed D-branes in CFT's corresponding to
closed strings on manifolds with B fields.  Here, we are
implicitly discussing D-branes in CFT's corresponding to gerbes.
In order to honestly rederive \cite{cks}, we would need for
the two closed string theories to match.
It is clear from the results of this paper, however, that they do not:
massless spectra of gerbes are not the same as massless spectra of
the underlying spaces, partition functions of the closed string theories
differ, and so forth.  We have repeatedly seen throughout this paper
that the CFT of a closed string on a gerbe is very different from the
CFT of a closed string on a manifold with a flat $B$ field.

Thus, although the counting of D-branes in string theories on gerbes
matches the counting of D-branes in string theories on manifolds
with $B$ fields, otherwise the closed string theories are very different,
and so we cannot honestly rederive the results of \cite{cks}
by considering the open string B model on gerbes.

\subsection{Check:  multiple presentations have the same spectra}  
\label{check:openspectra}

In this subsection we shall work through an explicit example to
show that when we calculate open string spectra, the results
are independent of the presentation of the stack.
This is guaranteed by the fact that the open string spectra
can be expressed just in terms of the stacks, but an explicit
example may help convince any remaining skeptical readers.

In particular, let us take the Calabi-Yau stack to be
$[ {\bf C}^2/{\bf Z}_n]$.  We are going to repeat the
calculation of \cite{kps}[section 3.3.2], using the alternative
presentation of $[ {\bf C}^2/{\bf Z}_n ]$ discussed in 
section~\ref{multpres}.
Although the details of the calculations will differ,
the final results for open string spectra will be the same.

Let us take a moment to recall the results of \cite{kps}[section 3.3.2].
In that example, on the covering space,
we have two sets of D0-branes, both sitting at the fixed
point of the ${\bf Z}_n$ action.  
These D0-branes will be fractional branes.
Let $\rho_i$ denote the $i$th one-dimensional irreducible representation
of ${\bf Z}_n$, in conventions in which $\rho_0$ is the trivial
representation.  Fix the ${\bf Z}_n$-equivariant structure
on the first set of D0-branes by taking $a_i$ D0-branes to be
in the $\rho_i$ representation, and fix the equivariant structure
on the second set of D0-branes by taking $b_i$ D0-branes to be in the
$\rho_i$ representation.  Let ${\cal E}$, ${\cal F}$ denote the
corresponding sheaves over the origin of ${\bf C}^2$,
so that ${\cal E} = {\cal O}^{ \oplus a_0 +a_1 + \cdots + a_{n-1} }$
and similarly for ${\cal F}$.  Using the fact that the induced
equivariant structure on the normal bundle ${\cal O} \oplus {\cal O}$
is given by $\rho_1 \oplus \rho_{n-1}$, and that $\rho_i^{\vee} = 
\rho_{n-i}$, we computed that
\begin{displaymath}
\mbox{dim } \mbox{Ext}^p_{ [ {\bf C}^2/{\bf Z}_n ] } 
\left( i_* {\cal E}, i_* {\cal F}
\right) \: = \:
\left\{ \begin{array}{ll}
        a_0 b_0 + a_1 b_1 + a_2 b_2 + \cdots + a_{n-1} b_{n-1} & p=0, 2 \\
        a_0 b_{n-1} + a_1 b_0 + a_2 b_1 + \cdots + 
        a_{n-1} b_{n-2} + & \, \\
        \: \: \: 
        + a_{n-1} b_0 + a_0 b_1 + a_1 b_2 + \cdots + a_{n-2} b_{n-1} & p=1 
        \end{array} \right.
\end{displaymath}

Now, let us recall the alternative presentation of $[ {\bf C}^2/{\bf Z}_n]$
discussed earlier.  Since we shall be manipulating sheaves, we shall
describe it as a quotient by a reductive algebraic group instead of
a compact Lie group.  
Define 
\begin{displaymath}
X \: = \: \frac{ {\bf C}^2 \times {\bf C}^{\times} }{ {\bf Z}_n }
\end{displaymath}
where the generator of ${\bf Z}_n$ acts as
\begin{displaymath}
\left(x,y,t\right) \: \mapsto \: \left( \alpha x,
\alpha^{n-1} y, \alpha t \right)
\end{displaymath}
for $\alpha = \exp(2 \pi i / n)$.
Then, we can write
\begin{displaymath}
[ {\bf C}^2/{\bf Z}_n ] \: = \: [ X / {\bf C}^{\times} ]
\end{displaymath}
as discussed in section~\ref{multpres}.

In terms of $[X/{\bf C}^{\times}]$, a fractional D0-brane is
a D-brane on $X$ whose support lies on the image of $0 \times {\bf C}^{\times}
\subset {\bf C}^2 \times {\bf C}^{\times}$ in the quotient.
(In particular, since the support must be invariant under both
the ${\bf Z}_n$ and ${\bf C}^{\times}$ quotients, it must cover all of
the ${\bf C}^{\times}$.)
The different types of fractional D0-branes are defined by
different ${\bf Z}_n$-equivariant structures on the structure sheaf
of $0 \times {\bf C}^{\times}$.  Since the ${\bf Z}_n$ group action
on $X$ is free, those equivariant structures can be interpreted on
the quotient manifold, and are simply flat ${\bf Z}_n$-valued holonomies of
the $U(1)$ gauge field around the ${\bf C}^{\times}$.

Following \cite{orig,kps}, since the gauge bundles are trivial
(modulo the Freed-Witten anomaly), the spectral sequence that played
an important role there trivializes, and computing the spectrum
reduces to computing sheaf cohomology groups.

Let us begin by working on ${\bf C}^2 \times {\bf C}^{\times}$,
then apply the techniques of \cite{kps} to compute Ext groups
on the ${\bf Z}_n$ quotient to get Ext groups on the space $X$,
and then finally take ${\bf C}^{\times}$ invariants to get
Ext groups on the stack $[X/{\bf C}^{\times}]$.

Following \cite{orig,kps}, since the gauge bundles are trivial
(modulo the Freed-Witten anomaly), the spectral sequence that played
an important role there trivializes, and computing the spectrum
reduces to computing sheaf cohomology groups.

Since the support has dimension greater than zero,
in principle we need to compute both degree zero and degree one sheaf
cohomology groups.
However,
\begin{eqnarray*}
H^0\left( {\bf C}^{\times}, {\cal O} \right) & = & {\bf C}[x,x^{-1}] \\
H^1\left( {\bf C}^{\times}, {\cal O} \right) & = & 0 
\end{eqnarray*}
so our computation reduces to degree zero sheaf cohomology.

Let ${\cal E}$ denote the bundle over $0 \times {\bf C}^{\times}$
corresponding to the first set of D-branes, namely
\begin{displaymath}
{\cal O}^{\oplus a_0 + a_1 + \cdots + a_{n-1} }
\end{displaymath}
and similarly for ${\cal F}$.  We will use ${\cal E}$ and ${\cal F}$
to denote the same bundle on $0 \times {\bf C}^{\times}$,
on its image in $X$, and on the image of that image in $[X/{\bf C}^{\times}]$.

Then, on ${\bf C}^2 \times {\bf C}^{\times}$,
\begin{eqnarray*}
\mbox{Ext}^0_{ {\bf C}^2 \times {\bf C}^{\times} } \left( 
i_* {\cal E}, i_* {\cal F} \right) & = &
H^0\left( 0 \times {\bf C}^{\times}, {\cal E}^{\vee} \otimes {\cal F} \right)\\
& = & {\bf C}[x,x^{-1}]^{\left( a_0 + \cdots + a_{n-1} \right) \left(
b_0 + \cdots + b_{n-1} \right) }
\end{eqnarray*}
The group $\mbox{Ext}^1\left(i_* {\cal E}, i_* {\cal F} \right)$
gets contributions from the sheaf cohomology groups
\begin{displaymath}
\begin{array}{c}
H^0\left( 0 \times {\bf C}^{\times},
{\cal E}^{\vee} \otimes {\cal F} \otimes {\cal N} \right) \\
H^1\left( 0 \times {\bf C}^{\times}, {\cal E}^{\vee} \otimes
{\cal F} \right)
\end{array}
\end{displaymath}
where ${\cal N} = {\cal O} \oplus {\cal O}$ 
is the normal bundle,
but as degree one sheaf cohomology vanishes,
we are left with
\begin{eqnarray*}
\mbox{Ext}^1_{ {\bf C}^2 \times {\bf C}^{\times} }\left(
i_* {\cal E}, i_* {\cal  F} \right) & = &
H^0\left( 0 \times {\bf C}^{\times}, {\cal E}^{\vee} \otimes
{\cal F} \otimes {\cal N} \right) \\
& = & {\bf C}[x,x^{-1}]^{ 2 \left( a_0 + \cdots + a_{n-1} \right) \left(
b_0 + \cdots + b_{n-1} \right) }
\end{eqnarray*}
Finally,
\begin{eqnarray*}
\mbox{Ext}^2_{ {\bf C}^2 \times {\bf C}^{\times} }\left(
i_* {\cal E}, i_* {\cal  F} \right) & = &
H^0\left( 0 \times {\bf C}^{\times}, {\cal E}^{\vee} \otimes {\cal F} \otimes
\Lambda^2 {\cal N} \right) \\
& = & {\bf C}[x,x^{-1}]^{\left( a_0 + \cdots + a_{n-1} \right) \left(
b_0 + \cdots + b_{n-1} \right) }
\end{eqnarray*}
Notice that, because the support is noncompact, the
Ext groups are not merely vector spaces, but are products of
polynomial rings.

So far we have computed Ext groups between the preimages of the
fractional branes on ${\bf C}^2 \times {\bf C}^{\times}$.
Next let us work out the Ext groups between their preimages on
$X = \left( {\bf C}^2 \times {\bf C}^{\times} \right)/{\bf Z}_n$,
and then finally we shall compute Ext groups on $[X/{\bf C}^{\times}]$,
at which point we shall recover the same result as on
$[{\bf C}^2/{\bf Z}_n]$.

Since $X$ is a global quotient by a finite group, we can apply the
same methods as in \cite{kps}, namely, given the Ext groups on the
covering space ${\bf C}^2 \times {\bf C}^{\times}$,
we take ${\bf Z}_n$ invariants, and so we recover
\begin{eqnarray*}
\mbox{Ext}^0_X\left( i_* {\cal E}, i_* {\cal F} \right) & = &
{\bf C}[x,x^{-1}]^{a_0 b_0 + a_1 b_1 + a_2 b_2 + \cdots + a_{n-1} b_{n-1}} \\
\mbox{Ext}^1_X\left( i_* {\cal E}, i_* {\cal F} \right) & = &
{\bf C}[x,x^{-1}]^{a_0 b_{n-1} + a_1 b_0 + a_2 b_1 + \cdots + a_{n-1} b_{n-2}
+ a_{n-1} b_0 + a_0 b_1 + \cdots + a_{n-2} b_{n-1} } \\
\mbox{Ext}^2_X\left( i_* {\cal E}, i_* {\cal F} \right) & = &
{\bf C}[x,x^{-1}]^{a_0 b_0 + a_1 b_1 + a_2 b_2 + \cdots + a_{n-1} b_{n-1}}
\end{eqnarray*}

Now that we have computed Ext groups on $X$, we can compute Ext groups
on $[X/{\bf C}^{\times}]$ by taking ${\bf C}^{\times}$-invariants.
The group ${\bf C}^{\times}$ acts on the polynomial ring generator $x$
only, so taking ${\bf C}^{\times}$ invariants means taking only the
constant part of the polynomial rings above.
In other words, the Ext groups on $[X/{\bf C}^{\times}]$
are merely finite-dimensional complex vector spaces, of dimension
\begin{displaymath}
\mbox{dim } \mbox{Ext}^p_{ [X/{\bf C}^{\times}] } \left( i_* {\cal E},
i_* {\cal F} \right) \: = \: \left\{ \begin{array}{ll}
        a_0 b_0 + a_1 b_1 + a_2 b_2 + \cdots + a_{n-1} b_{n-1} & p=0, 2 \\
        a_0 b_{n-1} + a_1 b_0 + a_2 b_1 + \cdots + 
        a_{n-1} b_{n-2} + & \, \\
        \: \: \: 
        + a_{n-1} b_0 + a_0 b_1 + a_1 b_2 + \cdots + a_{n-2} b_{n-1} & p=1 
        \end{array} \right.
\end{displaymath}
which exactly matches the Ext group computations in
\cite{kps}[section 3.3.2], outlined above.

Thus, we see in this example that when we compute open string
spectra between two sets of D-branes on the orbifold,
represented on the one hand as $[{\bf C}^2 / {\bf Z}_n ]$
and on the other hand as $[ X / {\bf C}^{\times} ]$,
although the presentations and calculational details are different, 
we do indeed get the
same result.  In principle, since open string B model spectra
can be defined in terms of the corresponding stack,
the result is guaranteed to be presentation-independent,
but hopefully this explicit computation will help convince
any skeptical readers.

\subsection{Derived categories and Calabi-Yau stacks}

In principle, the same formal methods that allow one to argue
that the off-shell states of the open string B model on
a Calabi-Yau manifold are classified by the derived category
of coherent sheaves on that manifold, also apply to stacks,
yielding a classification of off-shell B-branes in gauged sigma models
by the derived category of coherent sheaves on a Calabi-Yau stack.
In this section we shall briefly review how this proceeds.

We can associate off-shell B model states, for the B model on a 
Calabi-Yau stack, to objects in a derived category, in much
the same way as on a Calabi-Yau space.
We will merely outline the details here;
see \cite{medc,paulalb,melec} for a more complete treatment
of the details, for Calabi-Yau spaces.

Recall that given an object in the derived category of coherent sheaves
on a Calabi-Yau manifold, we pick a representative given by a complex
of locally-free sheaves.  We associate the odd elements, say,
with branes, and the even elements with antibranes, wrapped on the
entire Calabi-Yau manifold.  Maps between the branes and antibranes
are interpreted as tachyon vevs, and a physical analysis of 
necessary conditions to preserve boundary ${\cal N}=2$ supersymmetry
on the worldsheet leads to the constraint defining a complex,
that the composition of maps must vanish.
Quasi-isomorphisms are believed to be realized physically via
boundary renormalization group flow.

In principle the same analysis applies here.
First, pick a presentation $[X/G]$ of the stack.
The derived category of coherent sheaves on the stack can be represented
on this presentation as the derived category of $G$-equivariant sheaves
on $X$.  Objects in this derived category are represented by complexes
of $G$-equivariant sheaves, in which all of the maps are also
$G$-equivariant.

Proceeding as for spaces, to any object in the derived category
pick a representative consisting of locally-free sheaves,
and associate the elements of that complex with branes and antibranes
wrapped on $X$, as above.
Tachyons between branes and antibranes,
corresponding to sheaves ${\cal E}$, ${\cal F}$,
are elements of
\begin{displaymath}
\mbox{Ext}^0_X\left({\cal E}, {\cal F}\right)^G \: = \:
\mbox{Hom}_X\left({\cal E},{\cal F}\right)^G
\end{displaymath}
as discussed in the previous section on massless spectra in the open string
B model.  A $G$-invariant element of $\mbox{Hom}({\cal E},{\cal F})$
is the same thing as a $G$-equivariant map from ${\cal E}$ to ${\cal F}$,
so we see that the tachyons in this theory are naturally the maps appearing
in the complexes.

If we give a tachyon a vev, by adding a boundary term to the action
as in \cite{paulalb}, then also just as in \cite{paulalb} the
BRST operator is modified to the form
\begin{displaymath}
Q \: = \: \overline{\partial} \: + \: \phi
\end{displaymath}
We need to be slightly careful about notation here.
In the gauged sigma model, the BRST operator only squares to zero up to
a gauge transformation, but to simplify the analysis, the BRST operator
above will be assumed to only act on $G$-invariant states,
on which it would ordinarily have the form $\overline{\partial}$.
Deforming the action by giving the tachyon a vev $\phi$ modifies the
BRST operator, and the $\phi$ above should be interpreted as an operator
inserting $\phi$ at suitable points inside a $G$-invariant combination
of states.

As in \cite{paulalb,melec}, demanding that $Q^2=0$ is a necessary condition
to preserve boundary ${\cal N}=2$ supersymmetry, and in the presence of
multiple tachyon vevs, that amounts to a pair of constraints,
saying that the tachyons are holomorphic, and that their composition vanishes.
This is physically how the condition for a complex of maps, rather than
just a set of maps, arises.

By giving vevs to more general boundary states, and demanding that
$Q^2=0$, we can also recover the Bondal-Kapranov enhanced triangulated
categories formalism.

The BRST-invariant open string states of this nonconformal
topological field theory can be
shown to be counted by Ext groups between derived category elements,
{\it i.e.} cohomologies of RHom's,
in a fashion precisely analogous to that described in
\cite{melec,paulalb}.

Let $\left( {\cal E}_{\cdot}, \phi^{ {\cal E} }_{ \cdot } \right)$
be a complex of locally-free sheaves ${\cal E}_{\cdot}$ on the stack with
tachyons $\phi^{ {\cal E} }_{\cdot}$, as described above,
representing one object of the derived category,
and let $\left( {\cal F}_{\cdot}, \phi^{ {\cal F} }_{ \cdot } \right)$
be a similar complex representing another object.
The boundary BRST operator has the form
\begin{displaymath}
Q \: = \: \overline{\partial} \: + \: \sum_n \phi^{ {\cal E} }_n
\: - \: \sum_n \phi^{ {\cal F} }_n
\end{displaymath}
and acts on bundle-valued differential forms of the form
\begin{displaymath}
b^{\alpha \beta }_{\overline{\imath}_1 \cdots \overline{\imath}_n}
\eta^{\overline{\imath}_1} \cdots \eta^{\overline{\imath}_n}
\end{displaymath}
where the bundles are $\oplus_n {\cal E}_n$
and $\oplus_n {\cal F}_n$.
Note that effectively we are describing differential forms valued
in complexes.  Also note that there are no $\theta$'s here
because the D-branes are supported on all of the stack, not a substack.

The BRST cohomology of these bundle-valued differential forms
is precisely 
\begin{displaymath}
\mbox{Ext}^*_{ {\cal X} }\left( {\cal E}_{\cdot}, {\cal F}_{\cdot} \right)
\: = \:
H^* {\bf R}\mbox{Hom}_{ {\cal X} }\left( {\cal E}_{\cdot}, {\cal F}_{\cdot}
\right)
\end{displaymath}
It can be proven that this BRST cohomology matches the cohomology of
RHom's.  The proof is just a $G$-equivariant version of
\cite{melec}[appendix B], and as the extension is trivial,
we omit the details.

\section{Conclusions}

In this paper we have set down the basics of what it means to compactify
a string on a stack.  In a nutshell, every stack has a presentation
of the form $[X/G]$, to which one associates a $G$-gauged sigma model on
$X$.  Such presentations are not unique, and the corresponding gauged
sigma models are often very different physical theories; stacks
classify universality classes of worldsheet renormalization group flow
of gauged sigma models, not gauged sigma models themselves.
Demonstrating such a claim directly and explicitly is not possible
with existing technology, but here and in \cite{nr,glsm} we have
described a wide variety of strong indirect tests of this assertion.

We have discussed open and closed string spectra in gauged sigma models.
Closed string spectra were only previously known for gaugings of
finite effectively-acting groups; for finite noneffectively-acting groups,
the discussion is a bit subtle, and for nondiscrete groups,
it is in principle impossible to directly calculate the massless spectrum,
but here and in \cite{glsm} we have given several strong indirect arguments
that confirm a conjecture for the closed string spectra.

We have also discussed D-branes and the derived categories program for
stacks.  One of the most commonly quoted objections to understanding
gauged sigma models in terms of stacks involves D-brane probes,
an issue we address.  We also get a nice physical description of
D-branes in noneffective gaugings (gerbes).  Specifically, even if a group
acts trivially on the underlying space, it can still act nontrivially
on the Chan-Paton factors, which is a physical reflection of the
mathematical statement that sheaves on gerbes are (twisted) sheaves
on the underlying space.

One of the most important difficulties in understanding the consistency
of this program, {\it i.e.} whether IR physics is presentation-independent,
revolves around deformation theory, an issue first pointed out in
\cite{meqs}.  Physically, gauged sigma models have marginal operators
describing deformations which cannot be understood in terms of stacks,
leading to many easy puzzles.  For example, in effective orbifolds,
the standard ${\bf Z}_2$ orbifold of ${\bf C}^2$ can be deformed physically
to sigma models on deformations and resolutions of the quotient
space ${\bf C}^2/{\bf Z}_2$.  The stack $[ {\bf C}^2/{\bf Z}_2 ]$,
by contrast, is rigid, admitting neither complex structure deformations
nor K\"ahler resolutions to Calabi-Yau's.  In noneffective gaugings this
puzzle becomes even more tricky:  since the marginal operators are determined
as part of the massless spectrum, and the massless spectrum was
subtle and not well understood, opportunities for confusion abound,
some of which can be tentatively addressed by modifying proposals for
massless spectrum calculations.  In particular, the massless spectrum
that we arrive at in this paper predicts many more physical moduli than
mathematical moduli, and unlike effective orbifolds, here the extra
physical moduli do not have any obvious geometric interpretations.

We resolve these difficulties by realizing that the existing mathematical
deformation theory of stacks is describing those physical deformations which
yield theories with geometric interpretations and small quantum corrections.
In the case of the $[ {\bf C}^2/{\bf Z}_2]$ orbifold, the physical marginal
operators result in sigma models with small rational curves,
and so receive large quantum corrections.  In the case of the
noneffective gaugings, the `extra' marginal operators lead to abstract
CFT's which do not seem to have any geometric interpretation.
We provide a completely explicit description of many of these
abstract CFT's, which is the reason we believe this conclusion.

One application of this work is to local orbifolds, which are
special cases of stacks.
Such orbifolds historically have had a description in terms of ``V-manifolds,''
that is, local orbifold coordinate charts.  However, it has never
been clear whether such a description in terms of local coordinate
charts could be used to build a CFT; all existing physical technology
is only applicable to globally gauged theories.
We are able to side-step this issue by using presentations of local orbifolds
as global quotients of (larger) spaces by (bigger) groups, which can
be concretely realized in physics, unlike V-manifolds.

Another application of this work is to physical fields valued
in roots of unity, whose existence and properties were first pointed out
in \cite{nr}.  They play an important role in understanding 
noneffective gaugings and the surrounding CFT moduli spaces.

In \cite{glsm} we shall discuss toric stacks and gauged linear sigma models
for toric stacks.  We will also discuss mirror symmetry for 
stacks, and will see that the physical fields valued in roots of unity
mentioned above will play an important role, for multiple independent
reasons, in understanding mirror symmetry for stacks.

\section{Acknowledgements}

We would like to thank A.~Adams, J.~Distler, S.~Katz, J.~McGreevy, and
R.~Plesser for useful conversations.  We would also like to thank the
Aspen Center for Physics for hospitality while this work was being
done and the UPenn Math-Physics group for the excellent conditions for
collaboration it provided during several stages of this work.
T.P. was partially supported by NSF grants DMS 0403884 and FRG
0139799.

\appendix

\section{Principal ${\bf C}^{\times}$ bundles and
Calabi-Yau manifolds}   \label{ppalcxCY}

Many Calabi-Yau manifolds can be constructed as total spaces of
principal ${\bf C}^{\times}$ bundles.  Specifically, 
suppose we have an algebraic variety $X$ and a line bundle
$L \rightarrow X$.  Let $Z$ be the total space of the
${\bf C}^{\times}$ bundle $L^{\times} \rightarrow X$,
obtained by removing the zero section of $L$.
Then $Z$ will be a Calabi-Yau manifold if and only if
$K_X$ is isomorphic to some integral power of $L$.

Before describing the proof, let us describe two important special
cases of this result.
\begin{enumerate}
\item First, consider the case that $L = K_X$.
Then, it is well-known that the total space of $L$ is a Calabi-Yau manifold,
and simply restricting to the total space of $L^{\times}$ does not
change that fact.
\item Next, suppose that $X$ itself is a Calabi-Yau manifold.
Then, the total space of {\it every} principal ${\bf C}^{\times}$ bundle
over $X$ is also Calabi-Yau, for the simple reason that for any
$L$, $L^{\otimes 0} = {\cal O}$, so if $K_X = {\cal O}$,
then there exists an integral multiple (namely, zero) of $L$ for
all $L$ matching $K_X$.
\end{enumerate}

The second corollary is somewhat counterintuitive.
After all, it is well-known that the total space of a line bundle $L$
over a Calabi-Yau is not typically itself Calabi-Yau, unless the line
bundle is trivial.  Nevertheless, the total space of
{\it any} principal ${\bf C}^{\times}$
bundle over a Calabi-Yau manifold is itself Calabi-Yau, regardless of whether
or not the bundle is trivial, unlike line bundles.
One way to see this is as follows\footnote{One of the authors (E.S.)
would like to thank S.~Katz for providing the argument we review here.}.
Let $L^{\times}$ be the total space of a principal ${\bf C}^{\times}$ bundle
over some Calabi-Yau $M$, with projection $\pi: L^{\times} \rightarrow M$.
Then, we have the short exact sequence
\begin{displaymath}
0 \: \longrightarrow \: T_{\pi} \: \longrightarrow \: TL^{\times} \:
\longrightarrow \: \pi^* TM \: \longrightarrow \: 0
\end{displaymath}
so
\begin{displaymath}
\mbox{det } TL^{\times} \: = \: \pi^* \left( \mbox{det } TM \right)
\otimes
\left( \mbox{det } T_{\pi} \right)
\end{displaymath}
However, the line bundle $T_{\pi}$ is trivializable, because of the
${\bf C}^{\times}$ action.  In more detail, the
${\bf C}^{\times}$ action gives a map
\begin{displaymath}
\psi_t: \: L^{\times} \: \longrightarrow \: L^{\times}
\end{displaymath}
that sends $x$ to $t \cdot x$.
In particular, $\frac{d}{dt} \psi_t |_{t=0}$ is a nowhere-zero vector
field on $L^{\times}$, that lies in $T_{\pi}$.  Thus, the line bundle
$T_{\pi}$ has a global nowhere-zero section, hence
\begin{displaymath}
\mbox{det } TL^{\times} \: = \: \pi^* \mbox{det } TM
\end{displaymath}
so if $M$ is Calabi-Yau, then so is the total space of $L^{\times}$.

Now, let us give the proof of the full theorem.
Let $X$, $L$, and $Z$ be as above.
Write $p: Z \rightarrow X$ for the natural projection.
The tangent sequence for the map $p$ reads,
\begin{displaymath}
0 \: \longrightarrow \: T_{Z/X} \: \longrightarrow \: TX \:
\stackrel{ dp }{\longrightarrow} p^* TX \: \longrightarrow \: 0
\end{displaymath}
Since $Z$ is contained in the total space of $L$ we have $T_{Z/X} = p^* L$.
But the tautological section of $p^* L$ does not vanish anywhere,
and therefore $T_{Z/X} = p^* L \cong {\cal O}_Z$.
In particular, $K_{Z/X} \cong {\cal O}_Z$ and $K_X = p^* K_X \otimes
K_{Z/X} = p^*K_X$.  Thus, $Z$ will be a Calabi-Yau manifold if and only
if $p^* K_X$ is trivial on $Z$.
As already explained $p^* L$ is trivial on $Z$ and so it is sufficient
to have $K_X \cong L^{\otimes k}$ for some $k \in {\bf Z}$.
Conversely, if we compactify $Z$ to the ${\bf P}^1$-bundle
$P \equiv {\bf P} \left( L \otimes {\cal O}_X \right) \rightarrow X$
we see that a line bundle $M \in \mbox{Pic}(X)$ will pull back to a trivial
line bundle on $Z$ if and only if there are integers $a$ and $b$, so that
the pullback of $L^{\otimes a}\otimes M$ is isomorphic to 
${\cal O}_P(b)$, where ${\cal O}_P(1)$ is the unique line bundle on $P$
whose pushforward to $X$ is isomorphic to $L^{-1} \oplus {\cal O}_X$.
On the other hand, ${\cal O}_P(b)$ has degree $b$ along the fibers of
$P \rightarrow X$, and so ${\cal O}_P(b)$ can be a pullback of a line
bundle on $X$ only when $b=0$.  This implies that the pullback of
$L^{\otimes a} \otimes M$ is trivial on $P$, which by the see-saw theorem
yields that $L^{\otimes a} \otimes M$ is trivial on $X$.  Therefore
$K_Z$ will be trivial as an algebraic vector bundle if and only if
$K_X$ is isomorphic to some integral power of $L$.

In passing, note that this result is tricky to see with
gauged linear sigma models, because those most naturally describe
${\bf C}$-valued fields, not ${\bf C}^{\times}$-valued fields.

\end{document}